\documentclass{aa}  

\usepackage{graphicx}
\usepackage{txfonts}

\usepackage{hyperref}

\begin{document}

   \title{Average solar active region}
   \subtitle{I. Intensities, velocities, and the photospheric magnetic field}
   \titlerunning{Average solar active region I.}

    \author{
        Michal {\v S}vanda\inst{1, 2}
        \and
        Jan Jur{\v c}{\' a}k\inst{1}
        \and
        Markus Schmassmann\inst{3}
    }
    \institute{
        Astronomical Institute of the Czech Academy of Sciences, Fri\v{c}ova 298, CZ-25165, Ond{\v r}ejov, Czech Republic\\
        \email{svanda@asu.cas.cz}
        \and
        Astronomical Institute, Faculty of Mathematics and Physics, Charles University, V Hole\v{s}ovi\v{c}k\'ach 2, CZ-18200, Prague, Czech Republic
        \and
        Institut f{\" u}r Sonnenphysik (KIS), Georges-K\"ohler-Allee 401A, D-79110 Freiburg im Breisgau, Germany
    }

   \date{Received ---; accepted ---}

  \abstract

  {Solar active regions (ARs) are key manifestations of the Sun’s magnetic activity, displaying diverse spatial and temporal characteristics. Their formation and evolution play a crucial role in understanding the solar dynamo and space weather. While individual ARs exhibit significant variability, ensemble averaging offers a method to extract their typical properties and evolution.}
  {This study aims to construct an average bipolar AR using ensemble averaging of observational data. By normalizing ARs in space and time, we seek to identify general trends in the evolution of magnetic flux, velocity fields, and atmospheric emissions, providing insights into the underlying physical mechanisms governing AR development.}
  {We analysed a sample of bipolar ARs observed by the Helioseismic and Magnetic Imager and Atmospheric Imaging Assembly aboard the Solar Dynamics Observatory. The ARs were selected based on strict criteria, ensuring clear polarity separation and emergence within 60\degr{} of the solar central meridian. Normalisation procedures were applied to align ARs spatially and temporally before computing an ensemble average of various observables, including line-of-sight magnetograms, Dopplergrams, and multi-wavelength intensity maps.}
  {The average AR exhibits a well-defined evolutionary pattern, with flux emergence followed by peak activity and subsequent decay. The leading polarity retains coherence longer than the trailing one, consistent with previous studies. Surface flow maps revealed a diverging outflow near the emergence site before the emerged AR is clearly visible in magnetograms. Atmospheric emission variations indicate enhanced heating above the AR in later phases, possibly due to persistent reconnection events. The ensemble averaging approach highlights systematic features of AR evolution that are often obscured by individual-case variability.}
  {}


   \maketitle
%

\section{Introduction}

The Sun is a magnetically active star that exhibits a broad spectrum of magnetic phenomena across varying spatial scales and field strengths. Central to these phenomena are the active regions (ARs), which are defined by the emergence of sunspots on the solar surface and appear as darkened features in continuum white-light observations. These features are caused by overly concentrated magnetic fields, which partly inhibit the heat flow from the solar interior. However, it has been found that not all ARs exhibit sunspots when the magnetic field is less concentrated than the threshold \citep[e.g.][]{2012ApJ...757L...8L,2018A&A...611L...4J,2018A&A...620A.104S}, leading to the formation of sunspots. Nevertheless, other activity phenomena, such as active filaments, flares, and plages are present in spotless ARs. Active regions predominantly occur as east-west aligned pairs of opposite magnetic polarity, with sizes covering about 10 to 100~Mm and lifetimes spanning days to weeks. Their characteristics, such as tilt angles and spatial distribution, are intimately tied to the Sun’s magnetic dynamo and have been found to correlate with the strength and progression of solar cycles \citep[e.g.][]{2010A&A...518A...7D,2021A&A...653A..27J}.  

The historical understanding of ARs revolved around the paradigm of magnetically isolated flux tubes that formed at the base of the solar convection zone. These structures were believed to rise buoyantly due to magnetic forces and subsequently pierce the solar surface, forming ARs. Early studies employing the `thin flux tube approximation', treated flux tubes as one-dimensional entities where buoyancy, magnetic tension, and the Coriolis force governed their dynamics. The advance of computing technologies allowed for the use of much more sophisticated models that solve the full set of magnetohydrodynamic equations in 3D, and significantly refined the picture \citep[e.g.][]{Hotta2020}.  

For example, helioseismic analyses, such as those by \cite{2013ApJ...762..131B}, \cite{2022MNRAS.517.2775K},  and \cite{2024ApJ...965..186M}, have constrained the amplitude and speed of flows beneath the solar surface before flux emergence. These results suggest that flux emergence may be influenced more by convection and turbulent flows in the upper convection zone than by buoyancy alone. It would seem that, at least in the near-surface layers of the solar convection zone, the turbulence influences the rise of the magnetic flux tubes significantly, challenging earlier assumptions that flux tubes rise almost intact from the tachocline towards the surface. 

The typical evolution of an AR was described in a review by \cite{vanDriel-Gesztelyi2015}. In short, from the observational point of view, the evolution may be split into five consecutive phases: emergence, growth, maximum activity, decay, and dispersal. During the emergence and growth phases, the AR gains magnetic flux, changing the initially horizontal magnetic field into a configuration with highly vertical fields in the cores of the sunspots and inclined fields in their outskirts \citep[e.g.][]{2024A&A...691A.119K}. Around the maximum phase, rapid phenomena such as flares typically start to appear, especially when AR undergoes a large change in the sunspot areas \citep{2013AdSpR..52.1561C}. During the decay, the AR starts to lose coherence and loses flux mainly due to the convective motions in the close vicinity, for example, via moving magnetic features \citep{1973SoPh...28...61H}. Constraining the details of the process is still a goal of current research \citep[e.g.][]{2024A&A...686A..75Z}. 

The time it takes an AR to move from emergence to decay varies. In general, the duration is at least a few days (not counting very short-lived, very small ARs), and the lifetime seems to be longer for larger ARs \citep[see an overview in][]{Schrijver_Zwaan_2000}. Also, the sizes of ARs that might be measured as a separation distance of two polarities ranges from about 10~Mm to more than 100~Mm. 

It is believed that the formation mechanism of all ARs is the same. Numerical models seeking the proper description of AR formation \citep[e.g.][among others]{2010ApJ...720..233C,2012ApJ...753L..13S,Hotta2020} are based on applications of principles of magnetohydrodynamics. The fact that a variety of sizes and lifetimes and even magnetic field topologies are observed together with studies emphasizing the effects of the near-surface turbulence (buffeting by convection) suggests that the final appearance is due to chance. The differing details of topology in various ARs may be viewed as random realisations of the background process that is common to all ARs. 

In statistics, retrieving the typical values from the noisy realisation of a random variable involves averaging. For the independent realisations, the noise level decreases proportionally to the square root of the number of realisations. The original statistical formulation of averaging is based on representation by numbers. A similar approach can be used in the case of more complex objects than numbers, such as for (presumably) independent realisations of the physical phenomena. Such an approach is usually called `ensemble averaging'. Ensemble averaging has recently been used in many studies in solar physics, for example, in the study of a typical flow structure of the solar supergranules \citep[e.g.][and many more]{2010ApJ...725L..47D,2015SoPh..290.1547D,2016A&A...596A..66L,2019A&A...623A..98F,2021A&A...646A.184K,2024NatAs...8.1088H}, the moat flow around isolated sunspots \citep[][]{2014ApJ...790..135S}, the persistent downflows in the supergranular lanes \citep[][]{2021A&A...647A.178R}, penumbral filaments \citep{2013A&A...557A..25T}, and the subtle effects of solar-induced geomagnetic storms on human infrastructure and natural phenomena \citep[e.g.][]{2020JSWSC..10...26S,2024JSWSC..14...37S}. Ensemble averaging has even been used in the case of ARs already \citep[][]{2014ApJ...786...19B,2016A&A...595A.107S,2019A&A...628A..37B}, especially when dealing with the pre-emergence signatures of average motions \citep[e.g.][]{2019A&A...625A..53S,2020A&A...640A.116S}.

The variety of sizes and lifetimes of ARs makes a straight ensemble averaging difficult. Realising this issue, and contrary to published studies, in this work we introduce a `normalisation' of ARs in the sample before the ensemble averaging is performed. The normalisation is motivated by the core of the theoretical description of the AR emergence, which expects that an $\Omega$-shaped flux tube raises through the solar surface, thereby creating two separate poles of opposite polarity and forming sunspots. We hope that the normalisation of AR sizes and lifetimes will help reveal the evolution of the flux-tube core and average out details of the magnetic topology, for which the vigorous turbulence is (presumably) responsible. 

\section{Data}
Our study is based entirely on the archival space-borne data, namely those obtained by instruments aboard the Solar Dynamics Observatory \citep[SDO;][]{SDO}. The majority of the presented results rely on line-of-sight data captured by both Helioseismic and Magnetic Imager \citep[HMI;][]{HMI1, HMI2} and Atmospheric Imaging Assembly \citep[AIA;][]{Lemen12}. Data products from both instruments consist of images recording various physical quantities over full-disc of the Sun, both instruments recorded the observations with a similar spatial resolution (pixel size of about 0.5\arcsec{}) and resulted in a series of various data products that are available with different time cadences. 

The data were downloaded from JSOC archives\footnote{\url{jsoc.stanford.edu}}. We did not download full-disc frames, instead, we used the JSOC tools to extract only patches around selected ARs, which were determined by Carrington coordinates. We also utilised JSOC tools to map the extractions using Postel's (azimuthal equidistant) projection to minimise geometrical distortions caused by projection effects as the ARs moved across the solar disc. The final extracted frames had a unified dimension of $768\times 768$ pixels with a pixel size corresponding to the original HMI observations (about 0.5\arcsec{}). The cadence of downloaded frames was chosen to be 12~minutes, which is the highest common cadence available for all the data products used. 

We only focused on bipolar ARs with clearly separated polarities over most of their evolution that formed sunspots within both polarities and that emerged within 60 degrees from the central meridian, preferably on the eastern hemisphere. The preference of the eastern hemisphere ensured that the length of the data series was limited rather by the evolution of the AR and not by the rotation of the Sun. Such a selection was done visually by using the SolarMonitor\footnote{\url{www.solarmonitor.org}}, namely the line-of-sight magnetograms and intensitygrams. Manual browsing with a daily sampling using SolarMonitor also allowed us to determine the date range that was considered for the data download. To cover at least part of the pre-emergence phase, we also requested data covering two days before an bipolar patch was seen by eye clearly above the noise in the line-of-sight magnetogram at the position of the new AR. Then we followed the AR until it decayed or disappeared at the western limb. 

For the purpose of this study, we downloaded HMI intensitygrams (series \texttt{hmi.Ic\_45s}), line-of-sight magnetograms (\texttt{hmi.m\_45s}), and Dopplergrams \texttt{hmi.v\_45s}, as well as three channels (1600, 304, and 171\,\AA) from AIA observations (\texttt{aia.lev1\_uv\_24s} and \texttt{aia.lev1\_uv\_12s}). Furthermore, we downloaded Milne-Eddington inverted HMI vector magnetograms \citep[\texttt{hmi.B\_720s},][]{2011SoPh..273..267B} with components transformed to local Cartesian coordinate system using the JSOC tools. Example fields of view of the finally considered sample of ARs are given in Fig.~\ref{fig:kolaz_int} and~\ref{fig:kolaz_mag}.

\section{Observables processing}
\label{sect:processing}
All the primary data products were downloaded from JSOC using a script in Python, utilising the \textsc{SunPy} \citep[][]{sunpy_community2020} and \textsc{drms} \citep[][]{Glogowski2019} packages. The script took an input file in a JSON format that contained the Carrington coordinates of the position of the given AR from JSOC, the time range of the requested data, and the requested frame dimensions and resolutions (the values of the key parameters are given in Table~\ref{tab:noaas} in the appendix). Then the script constructed the query for JSOC for each of the time intervals and downloaded the data. 

After the data were downloaded, the individual series of observables were processed separately for each considered AR. All the files were automatically inspected for missing pixels (this information was given in the files' headers in the MISSVALS record), and frames with missing pixels were discarded. The remaining frames were loaded into datacubes with segments 6-hours long (31 frames separated by 12 minutes). Missing frames were linearly interpolated. In the case of the Dopplergrams, we fit a quadratic surface to each frame and subtract it to remove the profile of the differential rotation and compensate for the motion of the satellite, effectively setting the line-of-sight velocity of the quiet Sun to zero. Such a procedure is common in helioseismology \citep[this step is explicitly mentioned in][]{2007SoPh..241...27S}. We also removed limb darkening from the intensitygrams by subtracting a limb-darkening model using a formula from \cite{astrophysical_quantities_2000}. There were typically over 30 datacubes covering the evolution of one AR in each observable. 

We then computed averages (over the whole time span) for each AR and for each observable. These averages were inspected visually and used to reject ARs that visually did not fulfil the requirements on an isolated bipolar AR (they demonstrated the existence of multipolar nature or showed the presence of another AR very close by with overlapping magnetic fields). This corresponds to the requiring the AR to be of B, C, D, E, or F \cite{1990SoPh..125..251M} classification type with the open sunspot distribution. This is the only point where human intervention in the data processing chain was required. The consecutive processing steps were performed completely automatically emphasizing the robustness, which made some of the steps complicated to properly account for peculiarities that appeared during the data processing. 

A local correlation tracking technique \citep[LCT;][]{1986ApOpt..25..392N} was applied to intensitygram datacubes using its implementation in Python \citep[\textsc{pyflowmaps} package by][]{pyflowmaps}. The spatial resolution of the HMI intensitygrams is sufficient to resolve the granules, and therefore these granules served as tracers of the surface apparent motions. The LCT was applied with a full width half maximum of 5\arcsec{}, which corresponded to 10 pixels, or about 4~Mm, and the flows were averaged over 6 hours. Therefore we obtained a series of surface flow fields with this temporal sampling. 

In the end, for each AR in the sample, we had a homogeneous series of intensitygrams, line-of-sight magnetograms, components of the vector magnetic field in the local Cartesian system, line-of-sight Dopplergrams, and intensity images in three AIA channels: 1600~\AA, 304~\AA, and 171~\AA\, which probed different layers of the solar atmosphere. The direct observables were accompanied by LCT horizontal flow maps.

\section{Active region stacking}
Each AR is unique. From the global point of view, they differ by the duration of their full evolution and their spatial extent. When assuming that all ARs are formed by the same process and it is only the random realisation that makes them different, one needs to normalise the individual ARs to prepare them for stacking so that they are co-aligned in both space and time. Only then may ensemble averaging reveal common features. Without normalisation, the features would average out. 

The stacking consisted of three major steps. The first two aimed at automatic detection of the polarity positions and utilised image processing of both intensitygrams and magnetograms. Initially, we used line-of-sight magnetograms in the process, but the projection effects especially when the AR was located at large heliocentric angles were severe, often even changing the sign of the magnetic field. Therefore, we decided to use the vertical (in the local Cartesian frame of reference) component of the vector magnetic field, $B_Z$, to mimic the line-of-sight magnetograms, $B_\mathrm{LOS}$. The third step performed the actual ensemble averaging of the ARs.

\subsection{Detection of polarities}

The chosen cut-out from the full-disc HMI observation was large enough to occasionally capture also compact magnetic regions belonging to other ARs or constituting leftovers after previous sunspot activity. It was thus important to limit the investigation only to the bipoles of the AR in question. To do so, we computed the average magnetogram and intensitygram frame by averaging the datacubes in time. The raw average magnetogram was used to guess the signs of both the leading and the trailing polarities. This task was achieved by computing the average of the magnetic flux separately in the western and eastern halves of the field of view. The sign prevailing in the western half was taken as an expectation for the leading polarity, and the sign prevailing in the eastern half was taken as an expectation of the trailing polarity. We have to note that in the investigated sample of ARs, we never encountered the situation that the same sign was identified in both halves of the average frame.

\begin{figure}
    \includegraphics[width=0.49\textwidth]{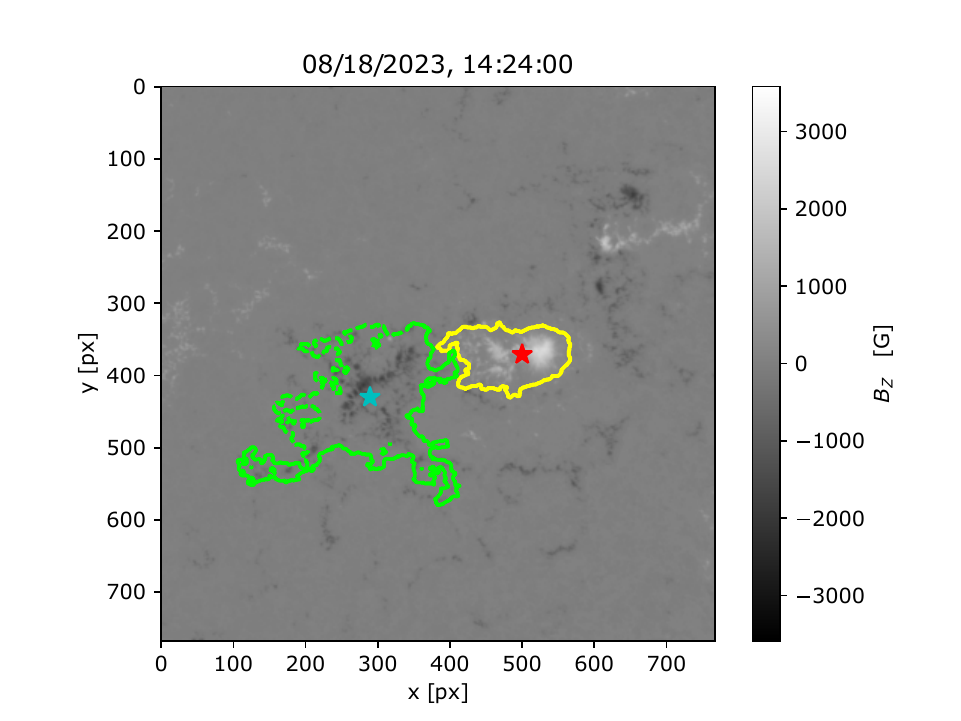} 
\caption{Demonstrative frame showing the performance of the automatic procedure to detect the magnetic polarities. The isolines outline compact regions of the positive (yellow) and negative (green) polarity. The red star indicates the center of mass of the positive polarity within the outlined region, and the blue star is the same but for the negative polarity. }
\label{fig:frame}
\end{figure}

The average magnetogram was thresholded so that pixels having an absolute value of the vertical magnetic field below 20~G were replaced by zeros. Further, we constructed a map containing only pixels with the positive magnetic field and a map with only the negative magnetic field. Then we applied a peak-detection routine from \textsc{SciKit} \citep{2014PeerJ...2..453V} package. This routine locally identifies all pixels that have a value larger than their surroundings of a given size. In our case, we requested to search for pixels having locally the largest value in the surrounding area of a radius of 50~px (about 20~Mm). This choice, obtained by a trial-and-error approach, gave robust detections in the case of the presence of noise spikes that could also be possibly identified as peaks. The peak search was performed separately for maps of the negative and positive magnetic fields. We also searched for peaks in the negatively taken average intensitygram frame, the results indicated the locations in intensitygrams where sunspot activity appeared throughout the time series. 

The procedure marked the local peaks in magnetograms and intensitygrams over the whole downloaded field of view. Not all the peaks in magnetograms had corresponding peaks in intensitygrams. In other words, not in all magnetic patches did sunspots appear. Therefore, we paired the peaks identified in the intensitygrams with the closest suitable peaks in the magnetograms. For each peak in an intensitygram, we searched for the closest peaks of the positive and the negative polarity of the magnetic field. The Euclidian distance between peak pairs was weighted by employing the information about the expected sign of the leading and trailing polarities. That is, for the peaks in the intensitygrams located in the western half of the average frame, peaks having magnetic signs agreeing with the expected leading polarity were weighted twice as strongly as the peaks having magnetic signs of the expected trailing polarity. The algorithm was constructed this way in order to be robust, e.g. in the case when a new flux emerged later during the evolution of the AR or another AR appeared in the field of view. Such peculiar cases might lead to ``jumps'' in the identification of the position of the leading or trailing polarity of the studied AR. 

The initial by-eye manual search aimed for bipolar ARs, so only two poles of the AR should be passed to the consequent analysis. By testing, we found that all ARs in our sample were tracked on Carrington coordinates corresponding to the point somewhere in the middle of the sunspot activity, the proper motion did not play a significant role. Therefore, we may safely assume that indeed, the leading polarity should be located in the western half of the average frame, whereas the trailing polarity has an opposite sign in the eastern half of the average frame. Those two should correspond to some of the peak pairs identified in both intensitygrams and magnetograms. 

It turned out impractical to choose the intensity peak that was the closest (in the Euclidian sense) to the image centre, which failed for instance in the case of a later emergence of the new flux. We constructed a custom metric combining two factors. One factor is based on the Euclidian distance from the image centre. It is expressed as the distance of the peak point from the image edge normalised by the half-width of the frame. Hence, the peak points located at the edge of the frame will get a value of 0, and the peak right in the middle of the frame will reach the value of 1.0. The second factor of the metric is the relative intensity of the magnetic field at the peak point with respect to the average magnetogram maximum (respecting the sign of the magnetic field in the peak point). These two factors were then summed up and the peak points with the highest total score (one for the negative magnetic field, one for the positive magnetic field) were chosen to represent the positions of the leading and trailing polarities. This way, the algorithm chose the deepest polarity closest to the frame centre. Possibly misidentified polarities from the surrounding ARs are rejected by the distance criterion, whereas the possible parasitic (e.g. newly emerged) polarities at the centre are rejected by the strength criterion. 

The boundaries of the leading and trailing polarity were determined by the flood-filling algorithm \citep{flood_filling} utilising the average magnetograms. This served as an initial estimate for the consecutive identification of the location of magnetic patches belonging to the AR in the individual frames of the series. The tolerance limit of the flooding was low (20~G, the root-mean-squared value of the average magnetogram in the quiet-Sun region was about 1~G), so that the flooding took the whole compact region of the given polarity, and the considered region was rather large. The flooding also avoided pixels of the opposite polarity. The flooded regions were used as initial masks to keep only magnetogram patches belonging to the leading and trailing polarity of the AR in the frame-by-frame processing, all pixels outside the flooded regions were set to zero. 

\subsection{Frame-by-frame segmentation}
Having the average masks for the leading and trailing polarities ready, we further proceeded in a frame-by-frame manner going through the magnetogram datacube (see an example in Fig~\ref{fig:frame}). 

\begin{figure*}[!ht]
    \includegraphics[width=0.33\textwidth]{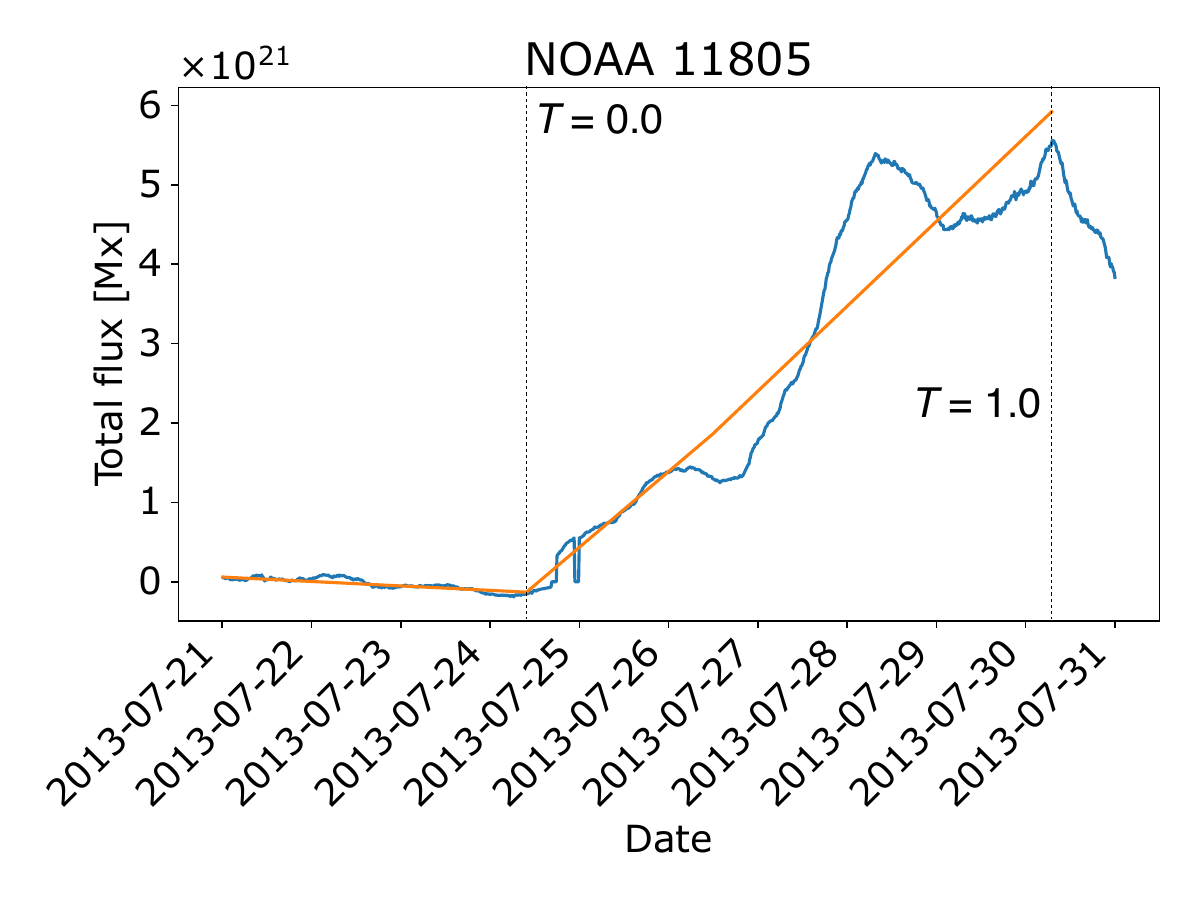} 
    \includegraphics[width=0.33\textwidth]{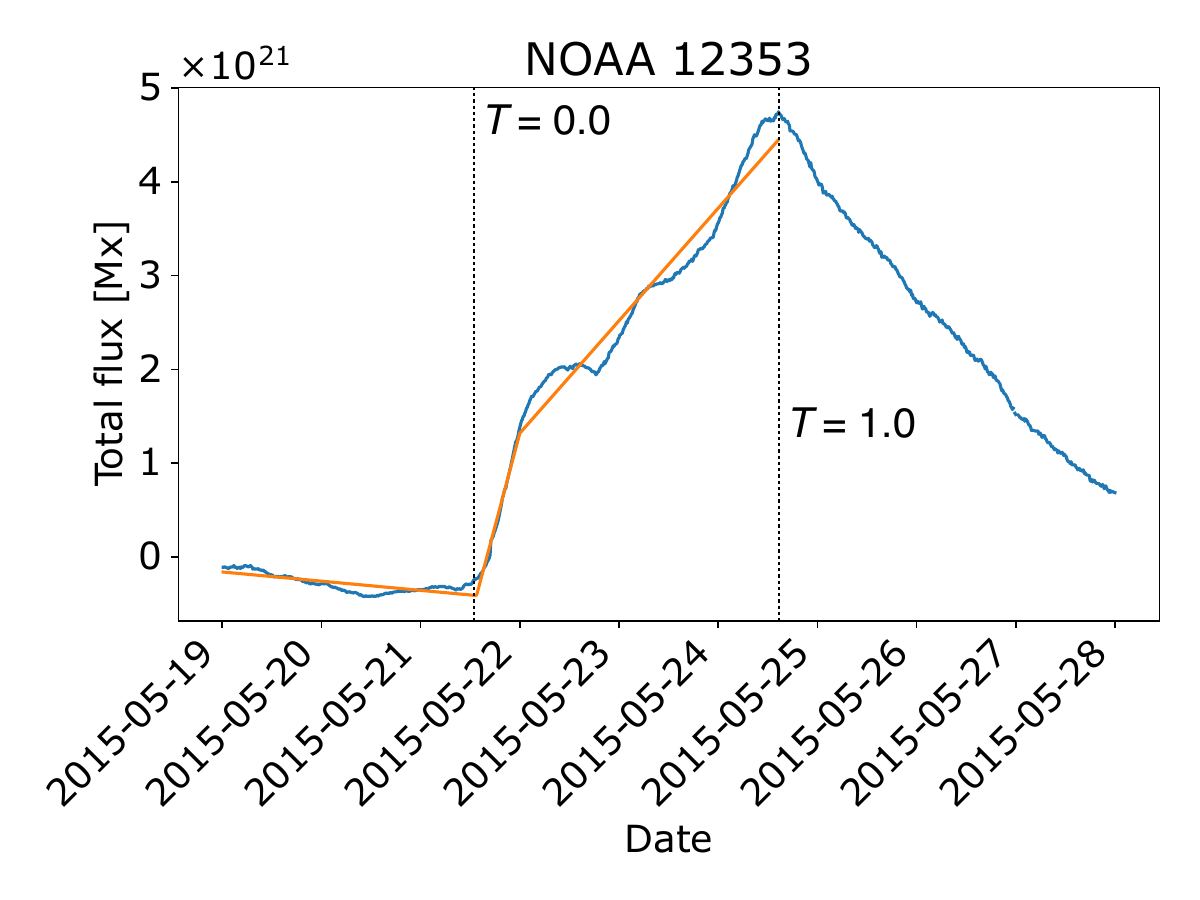} 
    \includegraphics[width=0.33\textwidth]{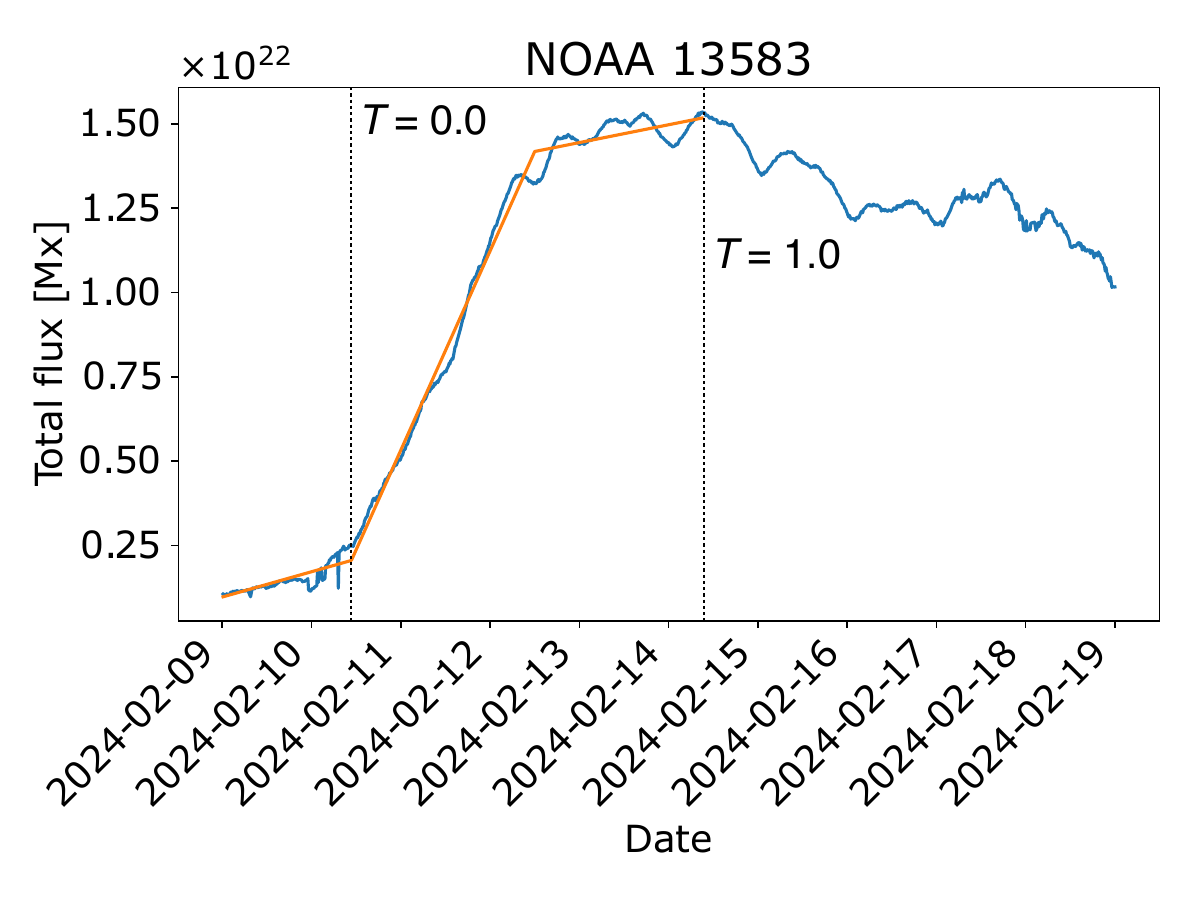} 
\caption{Total magnetic flux evolution for three chosen ARs and corresponding fits with the three-segment broken lines. The segments are fitted from the first frame until the maximum magnetic flux. The first broken point was chosen as a representative time for the beginning of the steep increase of the magnetic flux and therefore as the moment that the emergence starts (it corresponds to the normalised time $T=0$). The time of reaching the flux maximum corresponds to the normalised time $T=1$.}
\label{fig:broken}
\end{figure*}

The frames masked with the average polarity masks were investigated separately for both leading and trailing polarities. First of all, for the following, we only considered trailing or leading polarity if there were at least five pixels with the absolute vertical magnetic field above 500~G in it. If that was the case, we smoothed the magnetograms separated into leading and trailing polarities by a 30-pixel Gaussian and searched for the position of the maximum of each polarity. The Gaussian smoothing made the magnetic field continuous and suppressed small-scale variations. Such a property is needed for a proper application of the flood-filling algorithm. 

Then we applied the flood-filling algorithm to identify the boundaries of each magnetic polarity in the current frame. This created two polarity masks for each frame, which were used to determine the centre of mass of each polarity, taking the non-smoothed vertical magnetic field as the mass-weighting function. The coordinates of these barycentres were considered as representative positions for each magnetic polarity. 

Should both polarity centres be properly identified, we used the polarity masks to compute the signed magnetic flux of each polarity, and also computed their representative positions for the investigation of the proper motion and tilt. Should one of the polarities be so weak that there were not at least 5 pixels above 500~G, we used the average polarity masks and computed only the signed fluxes for both leading and trailing polarities. For such frames, we did not determine the polarity positions and other quantities based on knowledge of the polarity positions. 

Going frame by frame allowed us to construct time series of the polarity positions. Their change with respect to the initial position allowed us to compute their proper motion in time and consequently also the zonal and meridional velocities. Furthermore, we obtained a series of tilts. The missing values in these series were linearly interpolated (without extrapolations) and filtered from discontinuities (e.g. on the edges of the interpolated segments) using a Savitzky-Golay filter of the 2nd order with a window length of 10.2~hours (51 samples in the time domain) to ensure smoothness. 

\subsection{Normalisation}
Under the term `normalisation' we understand the mapping of all ARs both in space and time onto a predefined fixed space-time coordinate grid so that the chosen key points overlap for all ARs. Those key points were represented by the gravity centres of the polarities in the spatial domain and by the start of the flux emergence and its maximum in the time domain. 

\begin{figure*}
    \includegraphics[width=0.33\textwidth]{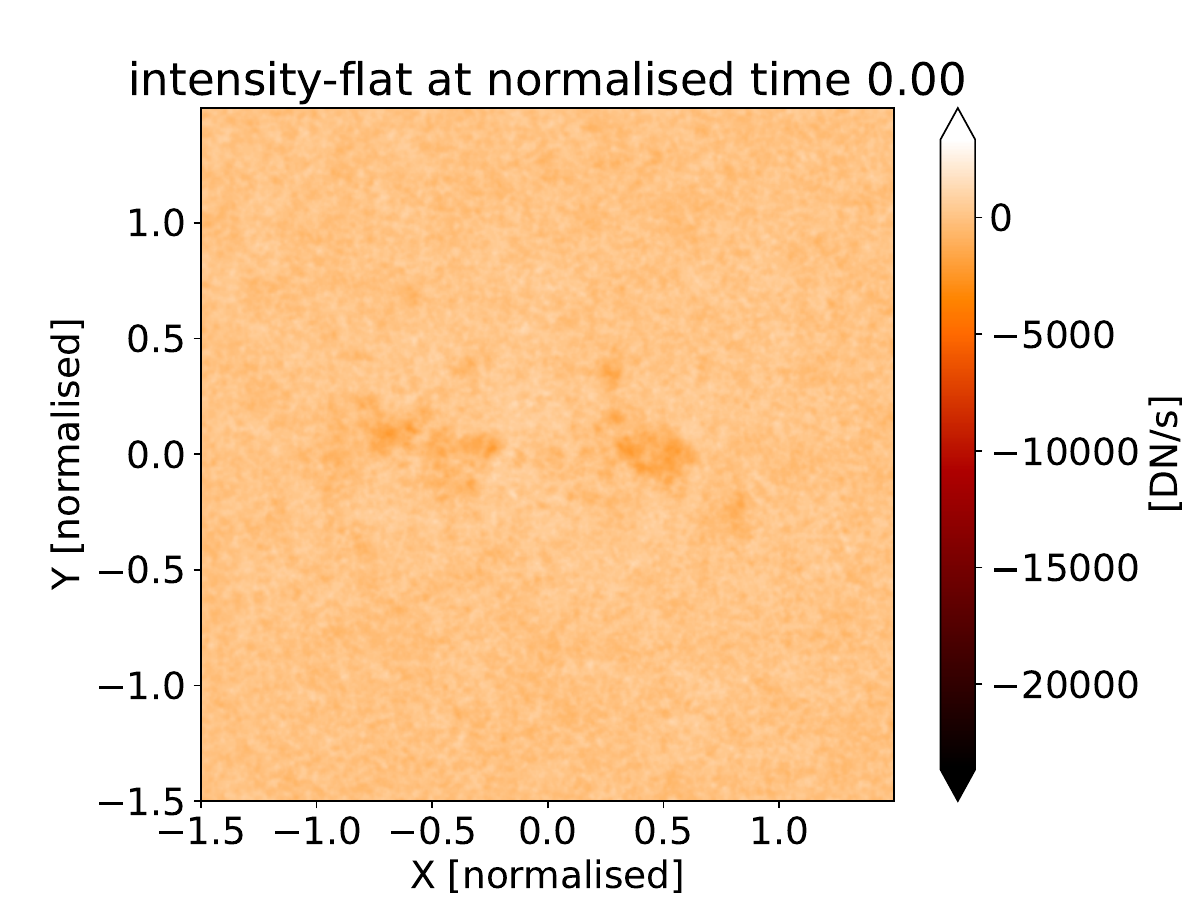} 
    \includegraphics[width=0.33\textwidth]{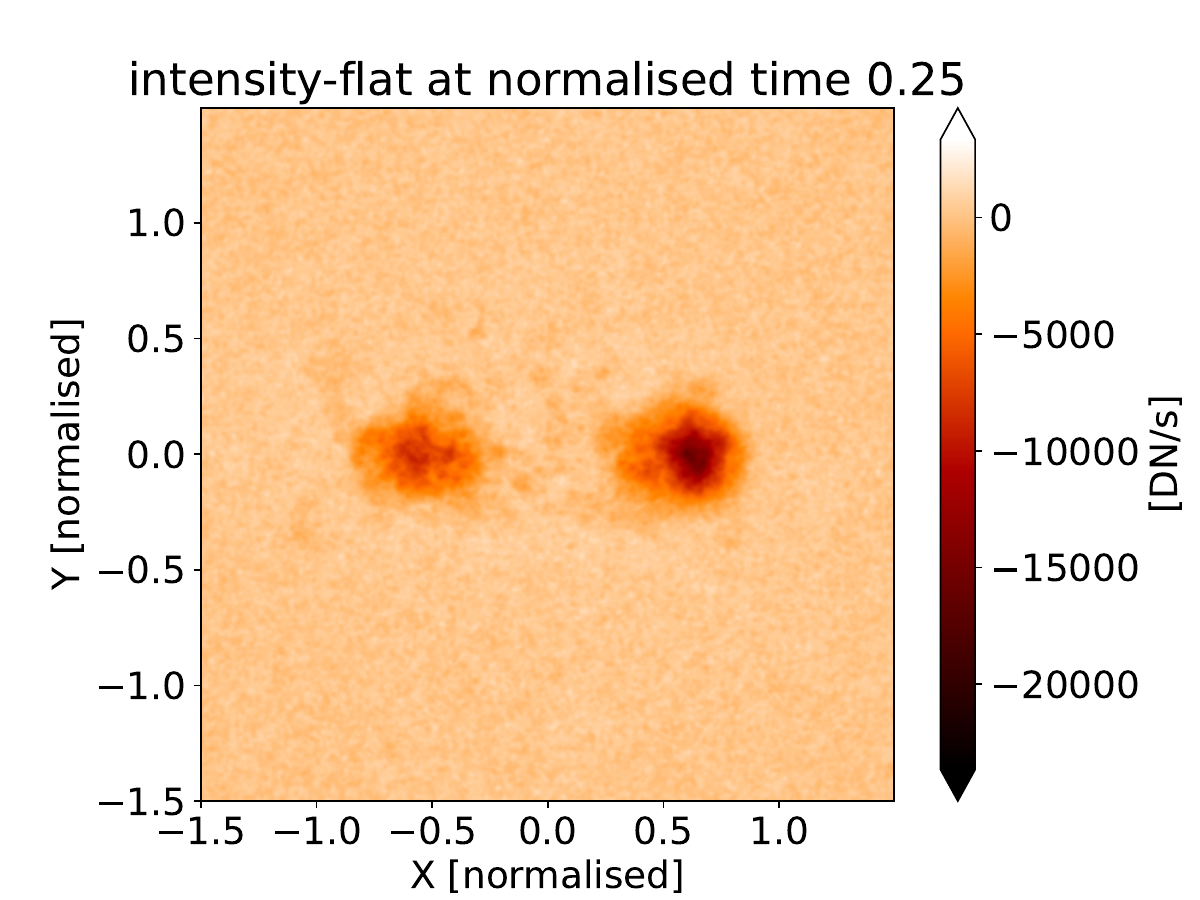} 
    \includegraphics[width=0.33\textwidth]{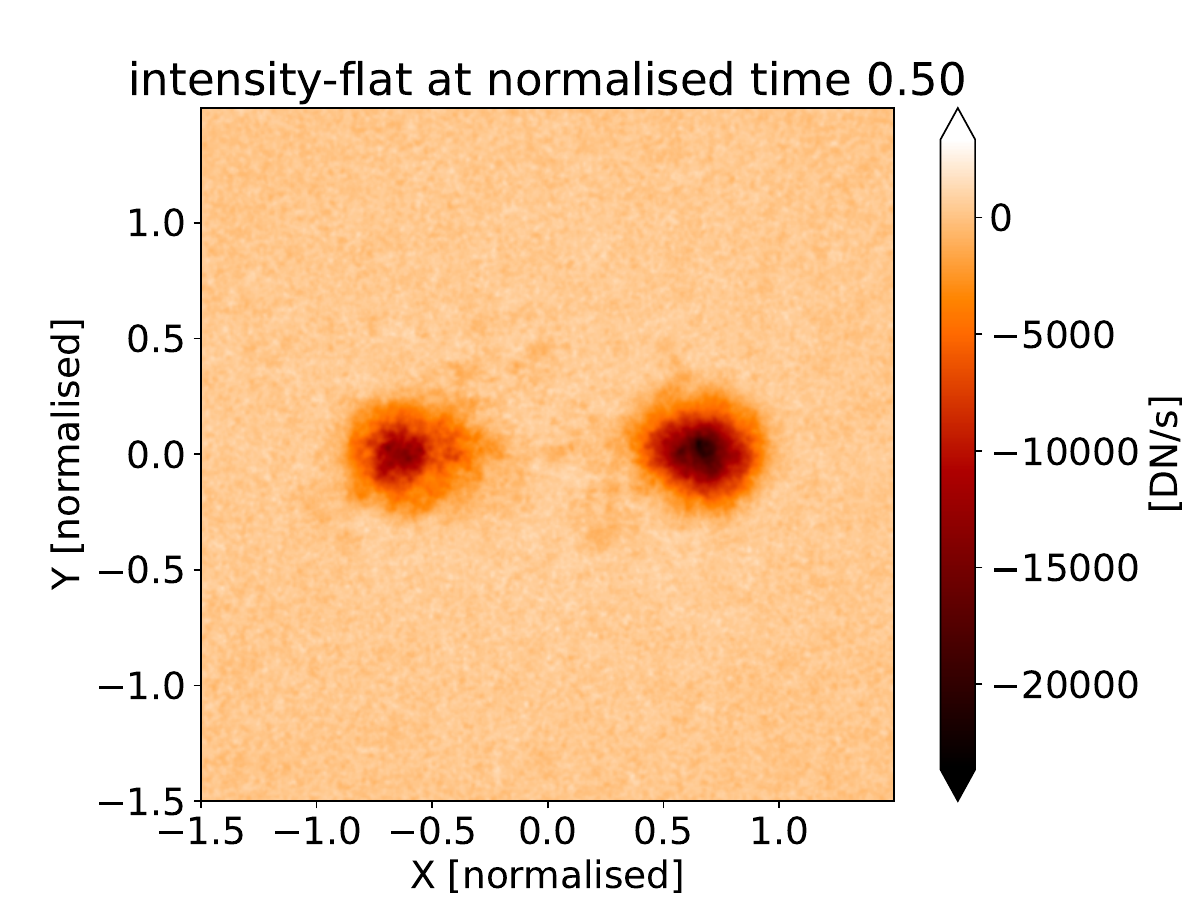} 
    \includegraphics[width=0.33\textwidth]{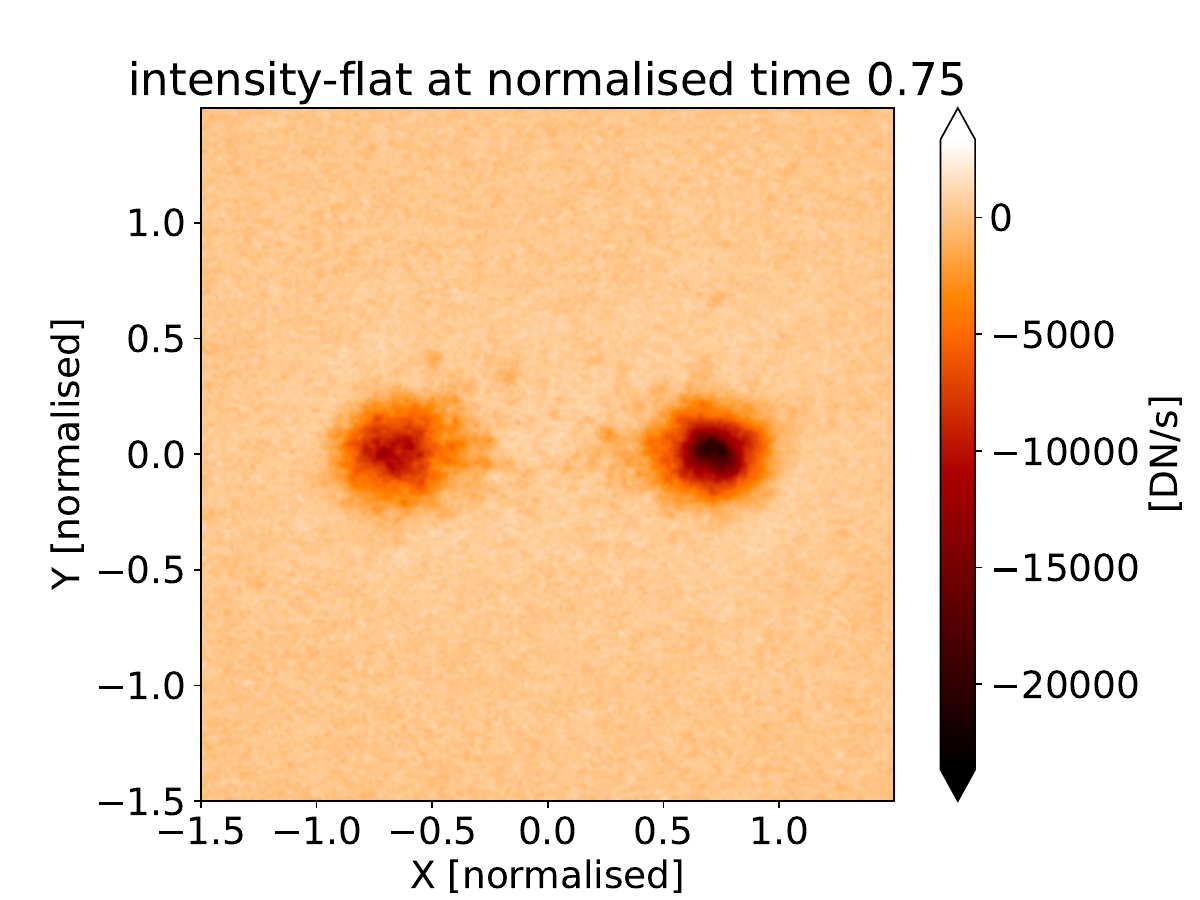} 
    \includegraphics[width=0.33\textwidth]{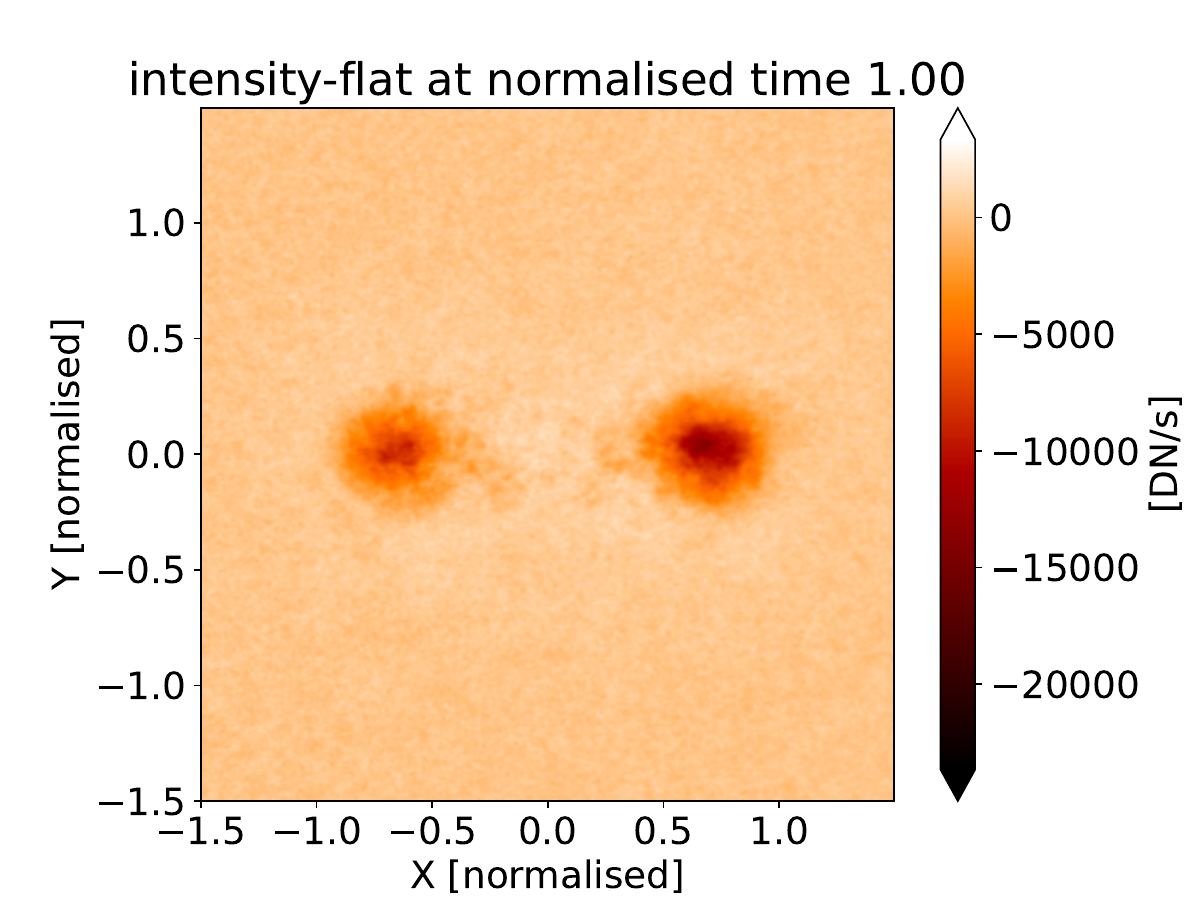} 
    \includegraphics[width=0.33\textwidth]{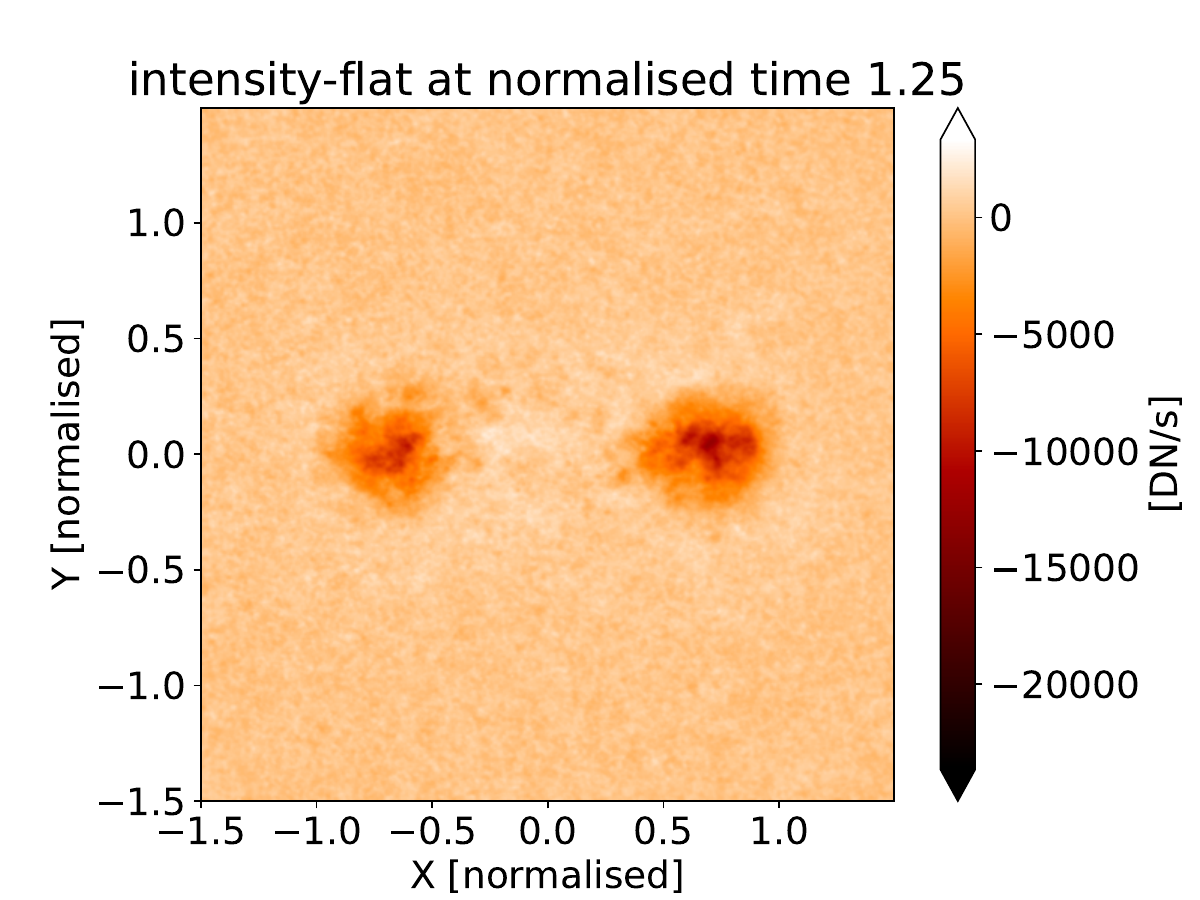} 
\caption{Snapshots of the HMI average intensity of the average normalised AR. Several normalised times are shown, where $T=0$ corresponds to the emergence time (see Fig.~\ref{fig:broken}) and $T=1$ represents the time at which the flux maximum was reached. Gravity centres of the polarities are located on coordinates $\pm 0.5$ on the horizontal axis and $0$ on the vertical axis. The intensity is relative with respect to the quiet-Sun intensity. A corresponding movie is available online (intensity.mp4) from \url{https://zenodo.org/records/15656676}.}
\label{fig:intensity}
\end{figure*}

\begin{figure*}
    \includegraphics[width=0.33\textwidth]{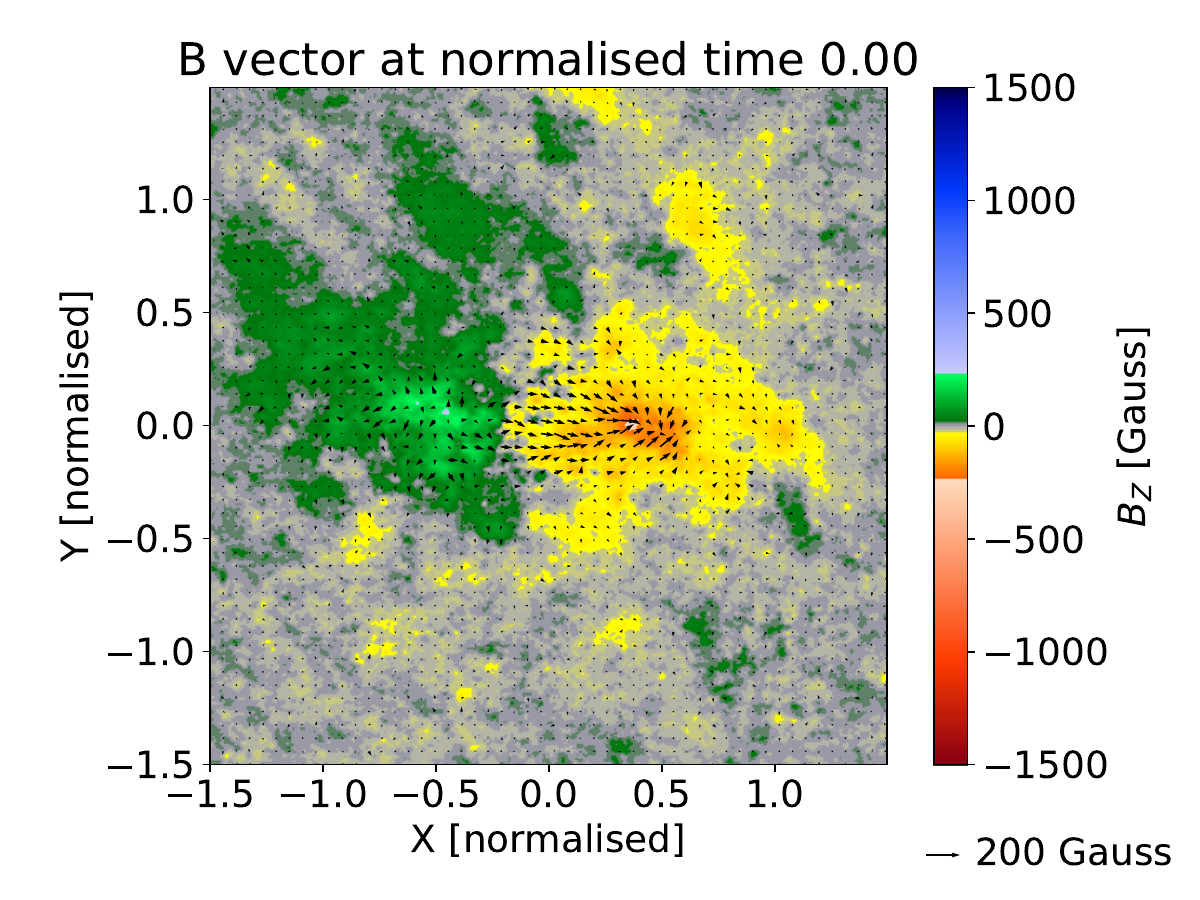} 
    \includegraphics[width=0.33\textwidth]{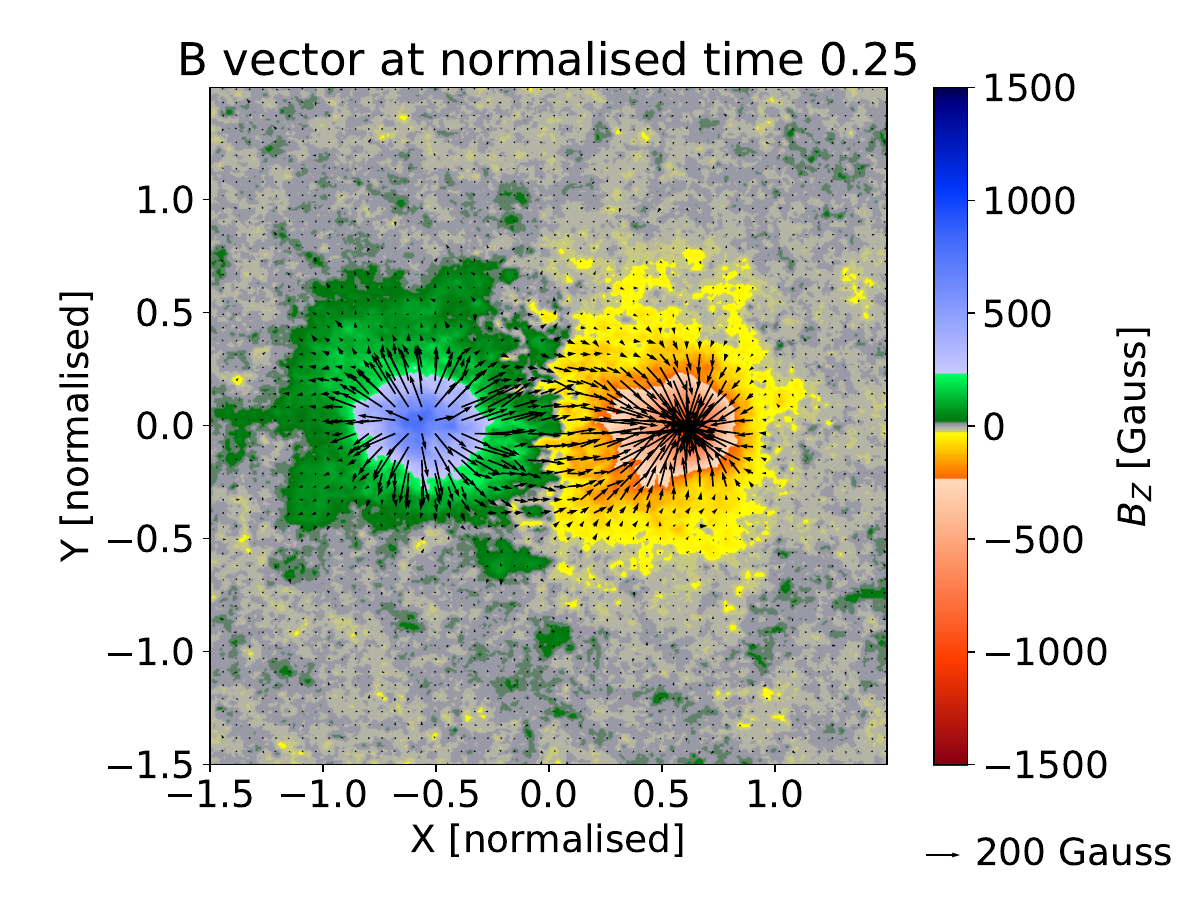} 
    \includegraphics[width=0.33\textwidth]{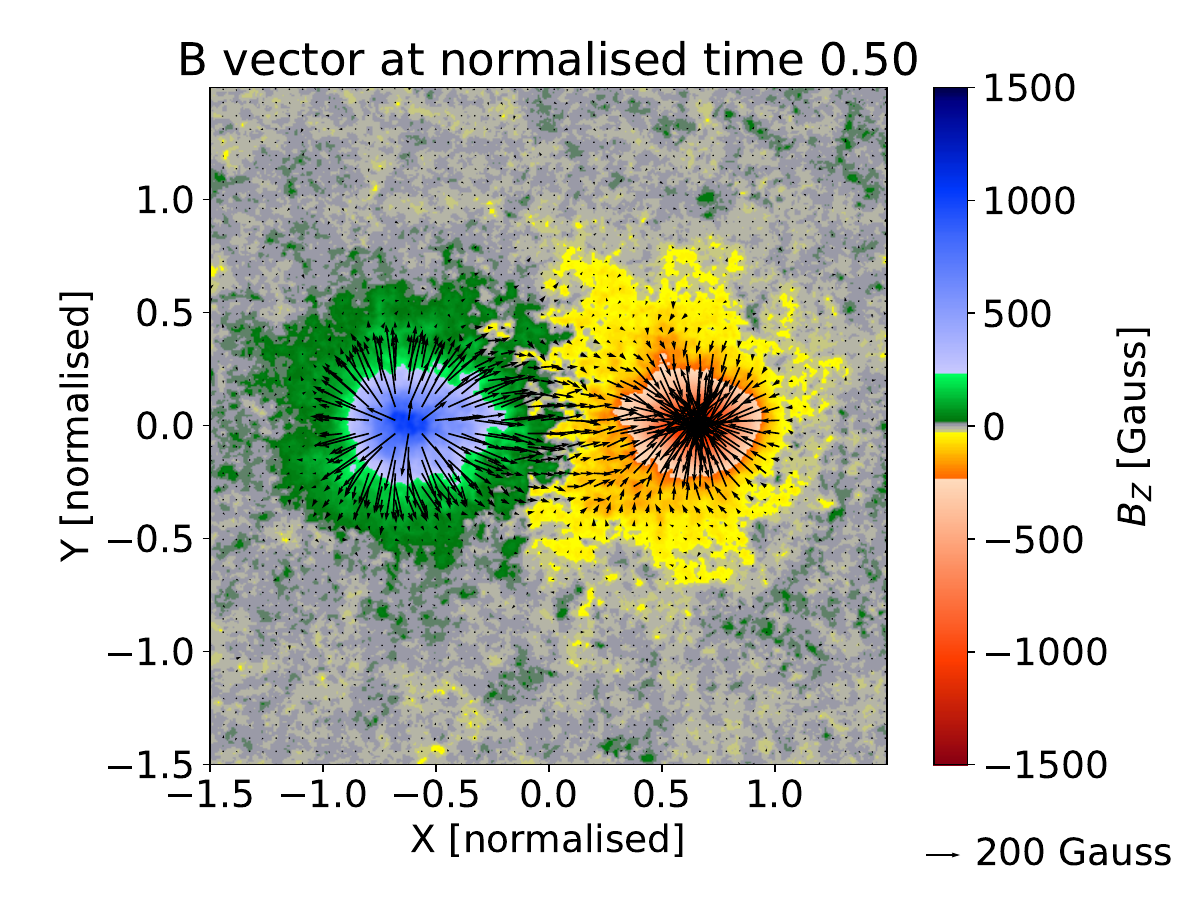} 
    \includegraphics[width=0.33\textwidth]{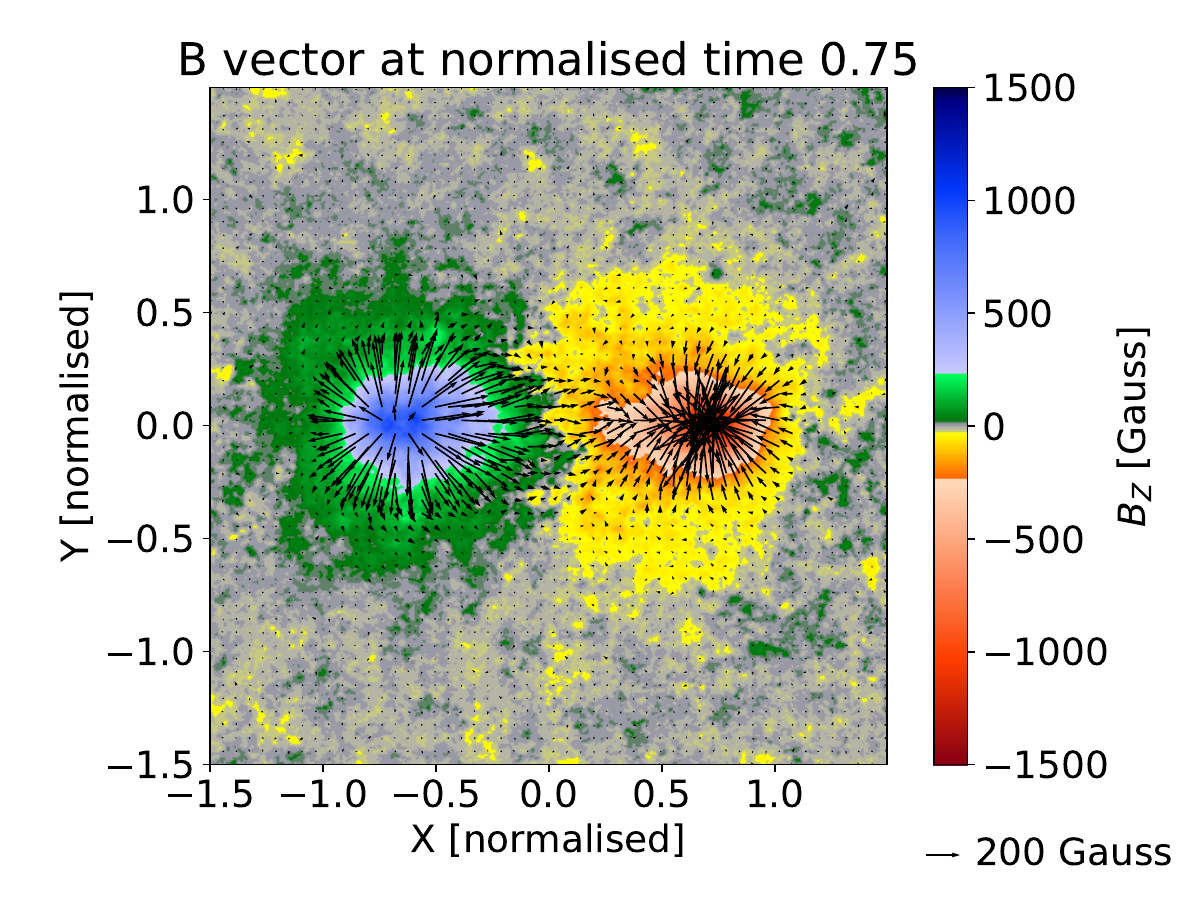} 
    \includegraphics[width=0.33\textwidth]{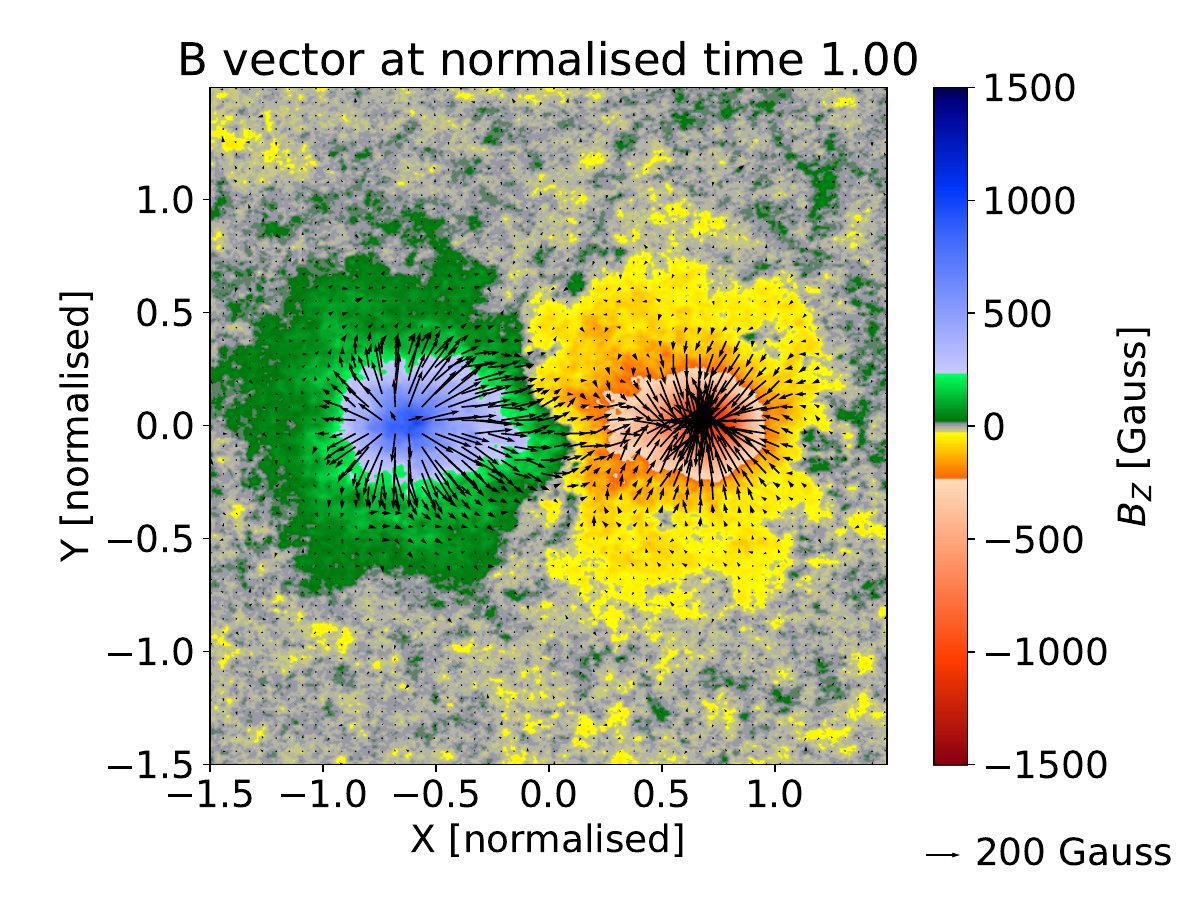} 
    \includegraphics[width=0.33\textwidth]{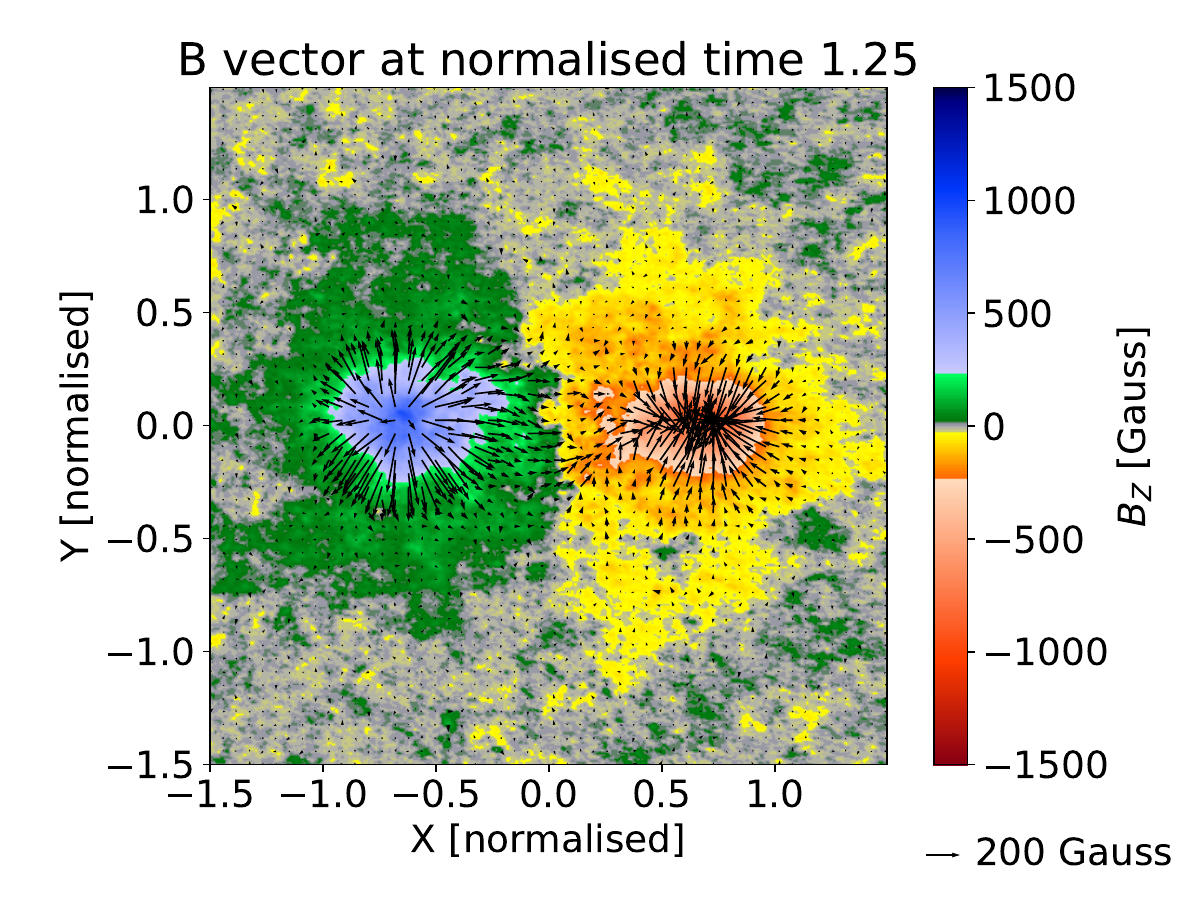} 
\caption{Same as in Fig.~\ref{fig:intensity} but only for vector magnetograms. Colours represent the vertical component of the magnetic field, whereas the arrows indicate the horizontal components. A corresponding movie of this is available online (Bvec.mp4) as well as of the line-of-sight magnetograms (magnetogram.mp4) from \url{https://zenodo.org/records/15656676}.}
\label{fig:Bvec}
\end{figure*}

First, we normalised the frames capturing various observables in the spatial domain. To do so, we used the positions of the gravity centres of polarities determined in the previous steps. Since the polarity positions were not available for each and every frame (only for those frames, where there were at least 5 pixels with the magnetic field above 500~G), the evolutionary curves were occasionally discontinuous, not to mention several jumps occurring for frames with a small number of strong magnetic pixels. Also, an extrapolation to the frames before the appearance of the strong pixels was needed. Therefore, we first formed a representative smooth curve giving positions for all frames with strong pixels. The simplification and gap filling were achieved by iteratively fitting the polynomial of the 3rd degree to the measured points (separately for each polarity and each coordinate), when removing outliers above 2$\sigma$ in each iteration ($\sigma$ was estimated from the standard deviation). The simplification ended when there were no more 2-$\sigma$ outliers in the final series. Then the cleaned polynomial fit was taken as representative of the polarity positions. Outside the ranges of interpolation, the positions were extrapolated linearly by using the first and last 100 valid points (that is about 20-hr evolution). The linear extrapolations were sewn together with the 3rd degree polynomial representing the detected positions in the middle of the common interval used for linear fitting. This sewn-together curve was continuous; however it did not have to have continuous derivatives at the sewing points. It was used as a representative robust estimate for the evolution of the position of the polarities for the normalisation. It allowed for a slow evolution and did not contain noise and jumps. 

The frames were then mapped in space using an affine transform (considering the position change, rotation, and magnification) to the predefined normalised coordinate grid $(X,Y)$, where the positions of the polarities were strictly east-west and had a fixed distance from the central pixel in the final frame. Horizontal vector fields were properly transformed as vectors so that the AR axis aligns with the $X$-direction. Directional observables such as the LCT flows were flipped in the vertical direction if the AR was located on the southern hemisphere, magnetic field (both the line-of-sight and vector magnetic field) was adjusted so that the leading polarity was always negative by flipping the sign if necessary. This way we mimicked all the ARs as if they were all located on the northern hemisphere of the Sun in cycle 25. In our sample we did not identify ARs having a ``wrong'' magnetic orientation (not respecting the Hale laws), so we did not have to deal with them. 

Then we normalised the evolution in time. A three-segment broken line was fitted to the total magnetic flux curve between the beginning (the first frame) and the maximum of the flux. The assumption was that the magnetic flux evolved somehow before the AR emergence, this was approximated by the first linear segment. Then the total flux started to rise as AR emerged and grew, and the flux reached its maximum. The growth did not have to be strictly linear, and therefore we approximated it with two linear segments. The time axis for each AR was then normalised so that the maximum in the flux corresponded to 1.0 and the first break point (that should correspond to the start of the emergence) to 0.0 by using a linear relation. For examples, see Fig.~\ref{fig:broken}. The spatially normalised frames were interpolated linearly in the time domain to the predefined normalised time axis $T$, considering the normalised time range of $-0.5$ (that is before the emergence) to $+2.0$ (after the flux maximum). 

After normalisation of each and every AR in the sample, we computed their arithmetic average of all considered observables. The normalisation naturally introduced an incomplete coverage of the target spatial and temporal coordinates across all ARs in the sample. Therefore, the average suffered from averaging of a lower number of valid observables farther from the central line in the spatial domain, and from sample incompleteness for normalised time smaller than $-0.1$ and larger than $1.1$. The incompleteness in the normalised-time domain occurred because by choice the emergence (and some time before) was captured in the AR series, whereas the decaying phases often started far on the western hemisphere or even at the western limb. Therefore our analysis focused only on normalised times between $T=0$ and $T=1.1$, where a complete coverage among the sample of ARs was secured. The coverage then decayed towards pre-emergence phases (at time $T=-0.5$ only 13 of 36 ARs were available) and to the phases after the flux maximum (at $T=1.25$ there were 19 of 36 ARs). Despite this, we also plotted some results for earlier and later normalised times $T$, where their reliability might be compromised. 

Other means of averaging are also possible. We found that the result very strongly depends on the way of normalisation in the spatial domain. If the sample is averaged on significantly different points in space (e.g. on the strongest magnetic field or on the deepest point in the intensitygrams), the resulting average is much more smeared than in the case of the use of magnetic field gravity centres. The barycentres of both polarities seem to constitute the most robust estimates for the description of the position of the magnetic field. 

On the other hand, the sensitivity to normalisation in time is much weaker. For instance, when one considers a fixed normalisation time (as an example, one could use 3.5~days, which is the mean emergence duration of our sample of ARs), the general description of the conclusions on the time evolution will be comparable, only the time axis will change accordingly. We found that in the time domain, it is critical to properly identify the beginning of the emergence and to stack the sample of ARs on this point. Since the emergence duration is then in the order of days for all ARs in our sample, it is not crucial to precisely establish the other point in the AR's evolution. This is also because of the fact that weak (and possibly quickly evolving) ARs will in the case of the fixed scaling in time contribute only weakly to the average, which will then be dominated by strong-flux ARs.

\begin{figure*}
    \includegraphics[width=0.33\textwidth]{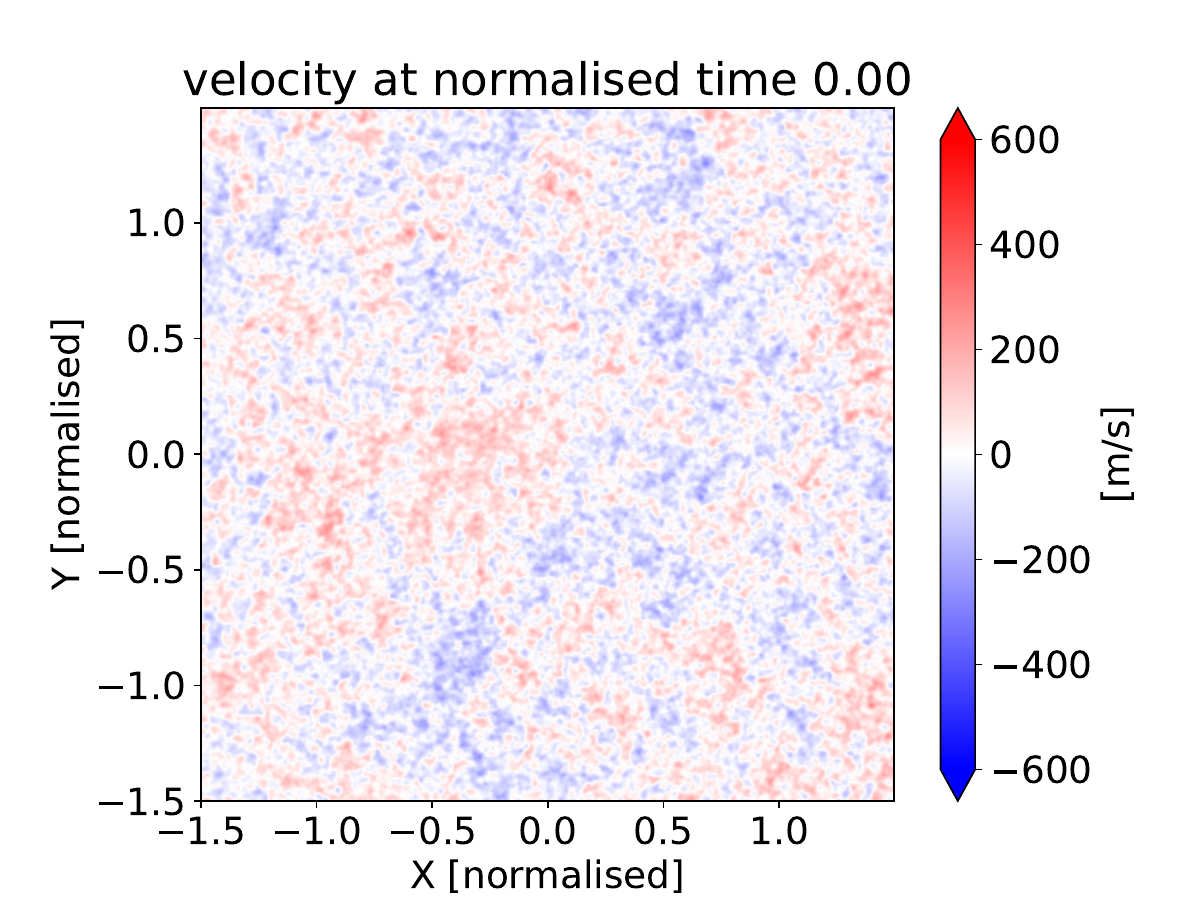} 
    \includegraphics[width=0.33\textwidth]{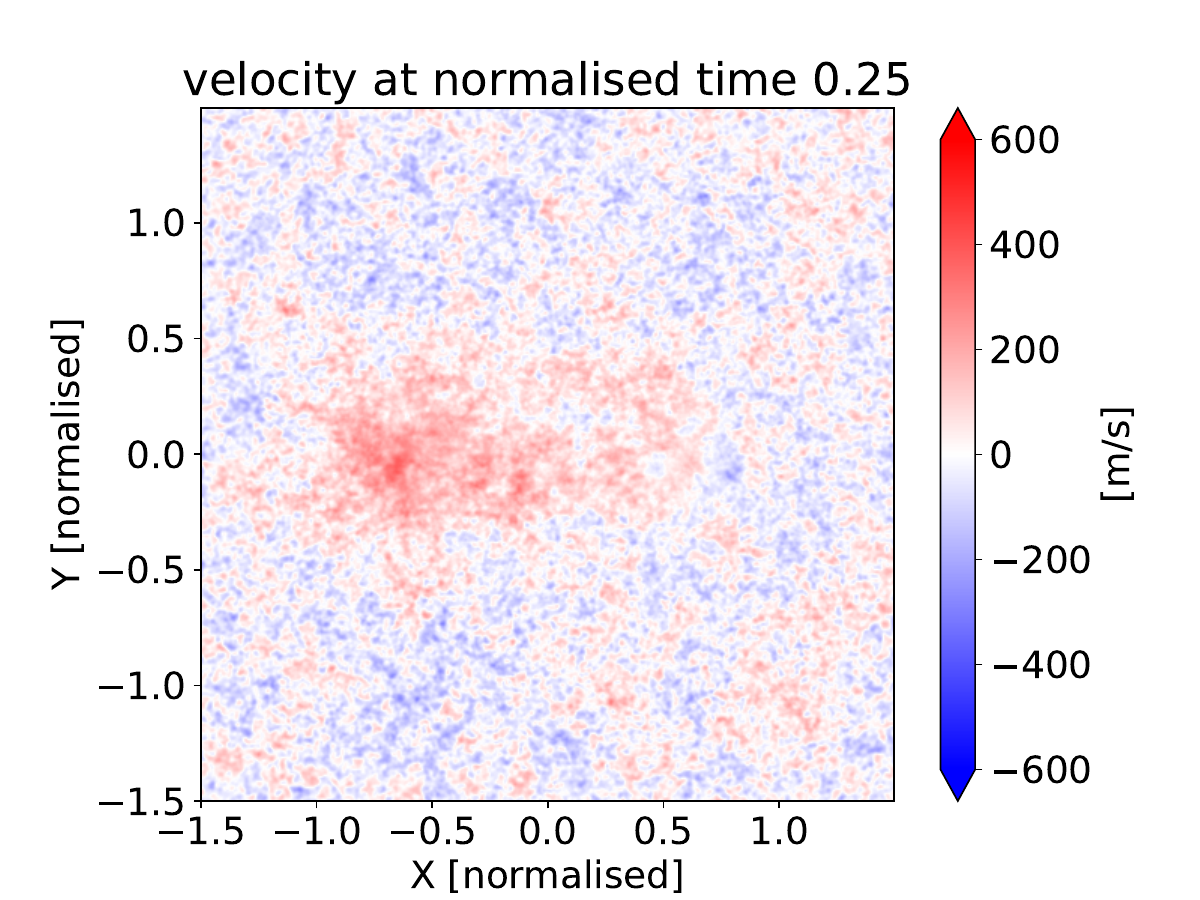} 
    \includegraphics[width=0.33\textwidth]{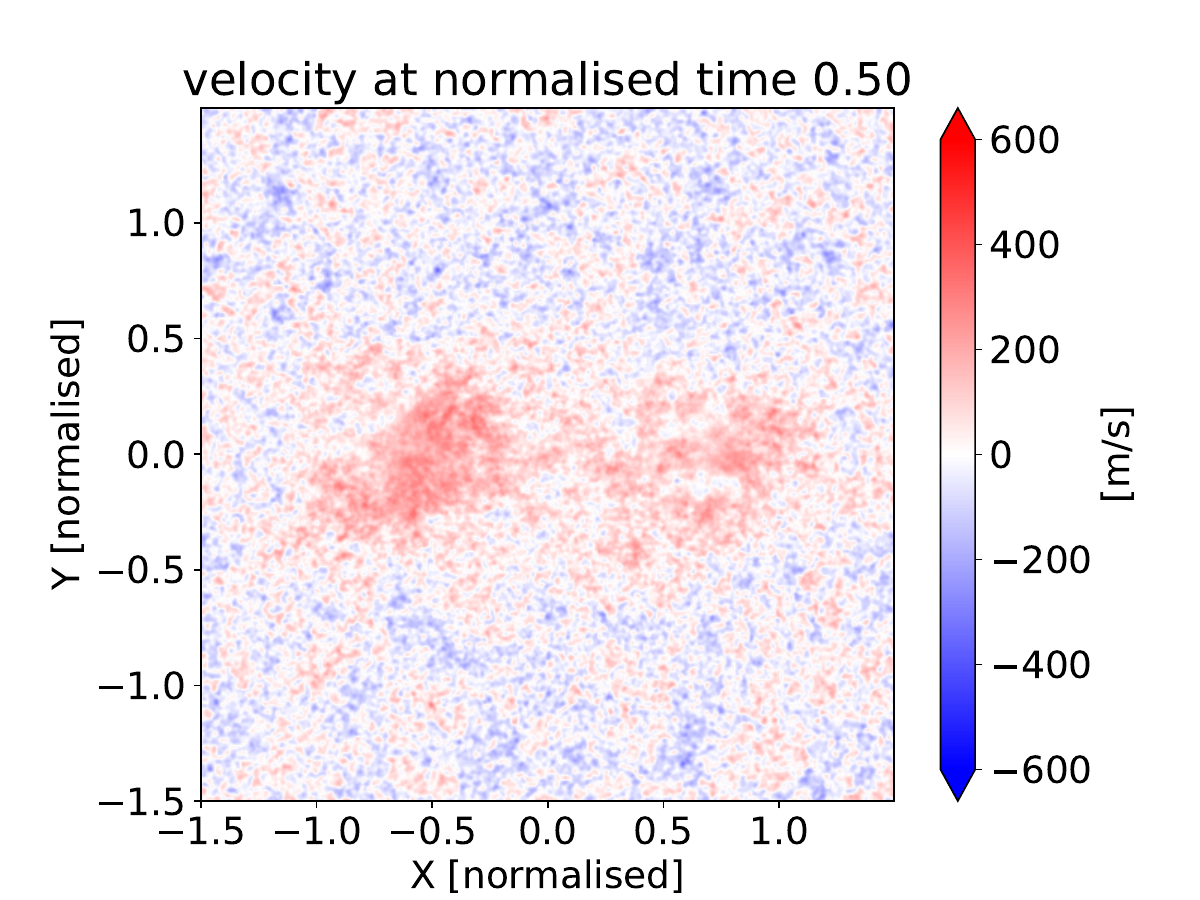} 
    \includegraphics[width=0.33\textwidth]{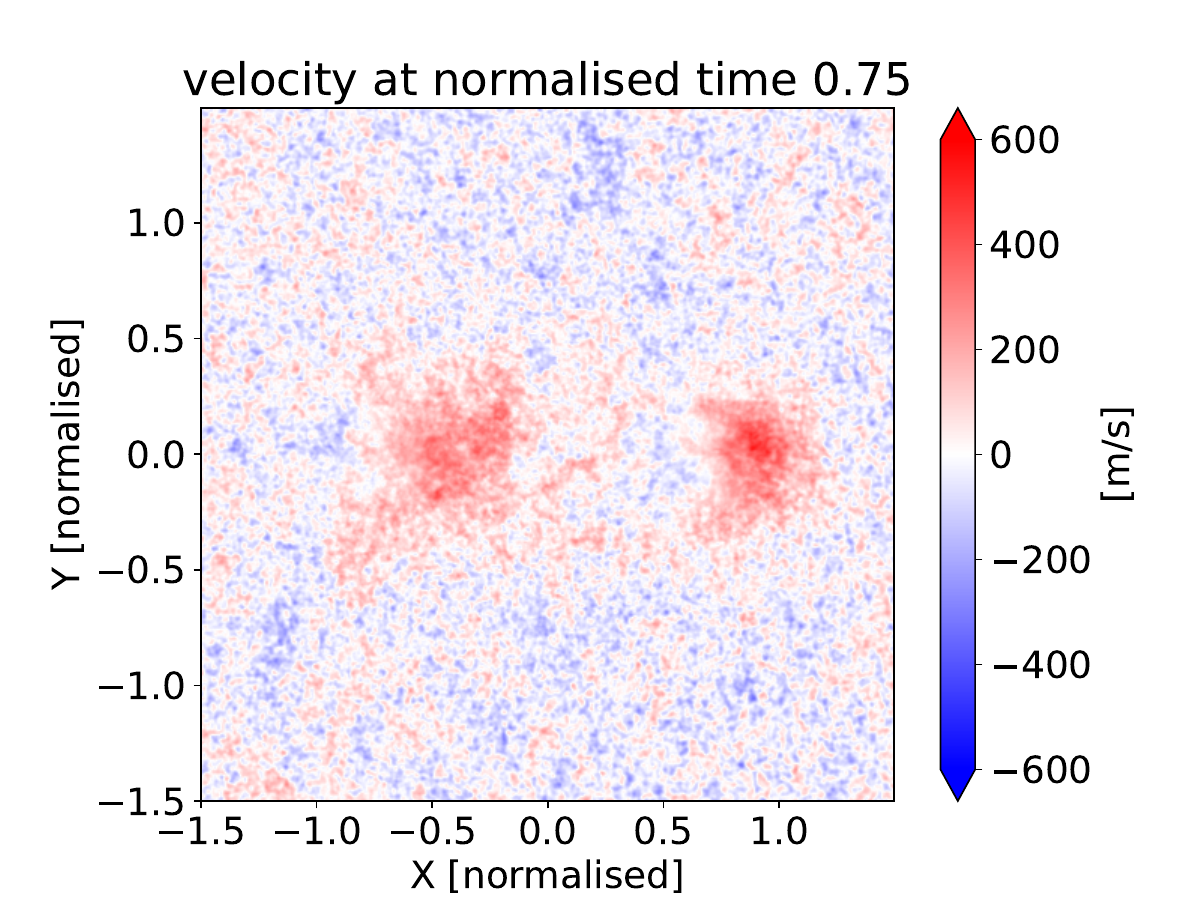} 
    \includegraphics[width=0.33\textwidth]{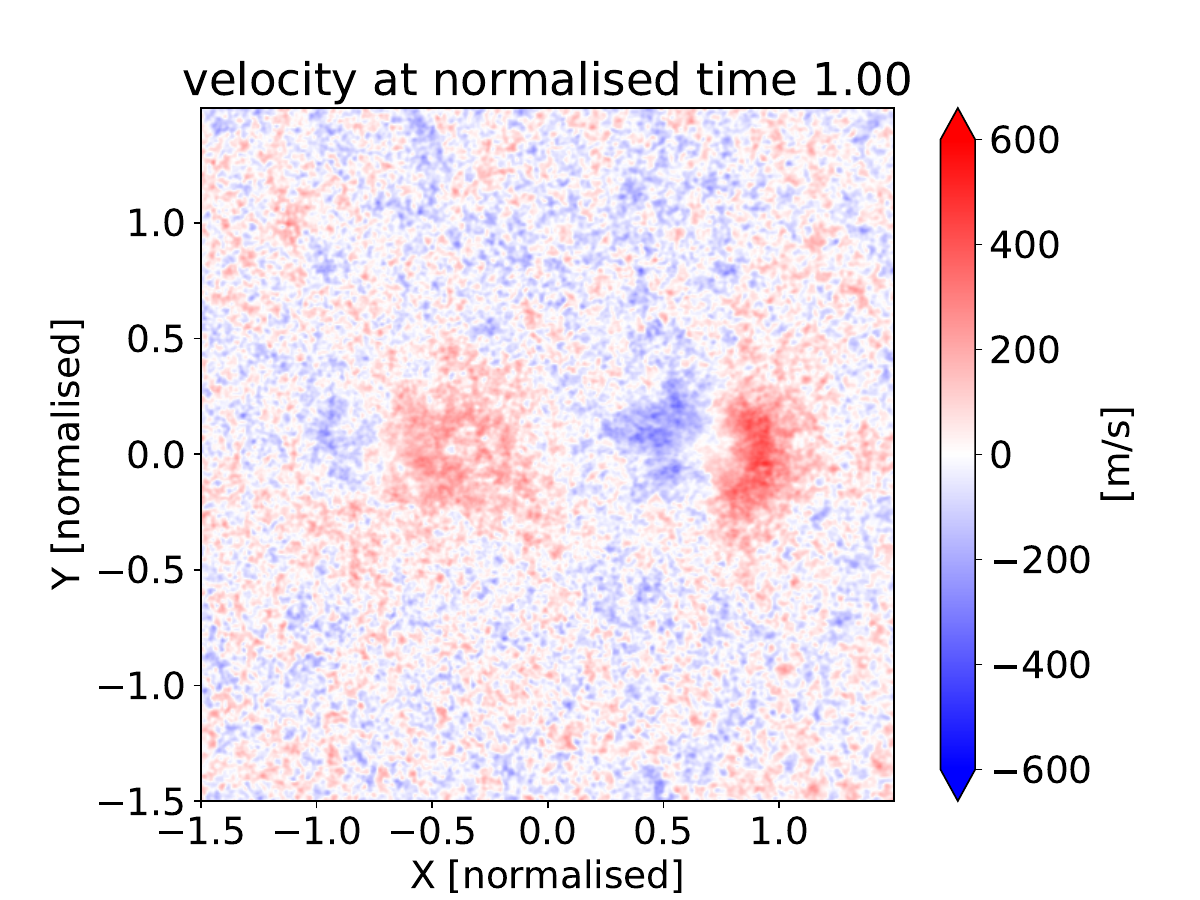} 
    \includegraphics[width=0.33\textwidth]{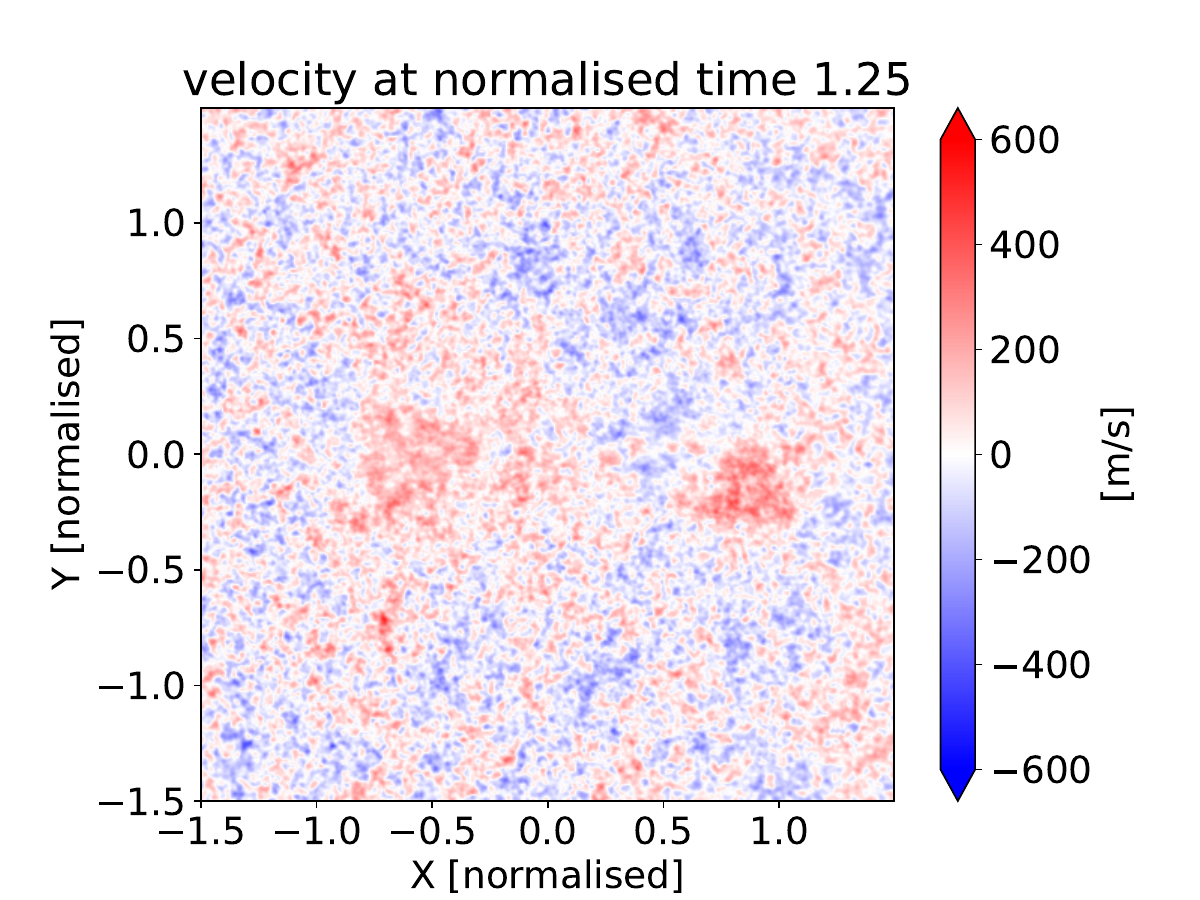} 
\caption{Same as in Fig.~\ref{fig:intensity} but only for line-of-sight velocity. A corresponding movie is available online (velocity.mp4) from \url{https://zenodo.org/records/15656676}.}
\label{fig:velocity}
\end{figure*}

\begin{figure*}
    \includegraphics[width=0.33\textwidth]{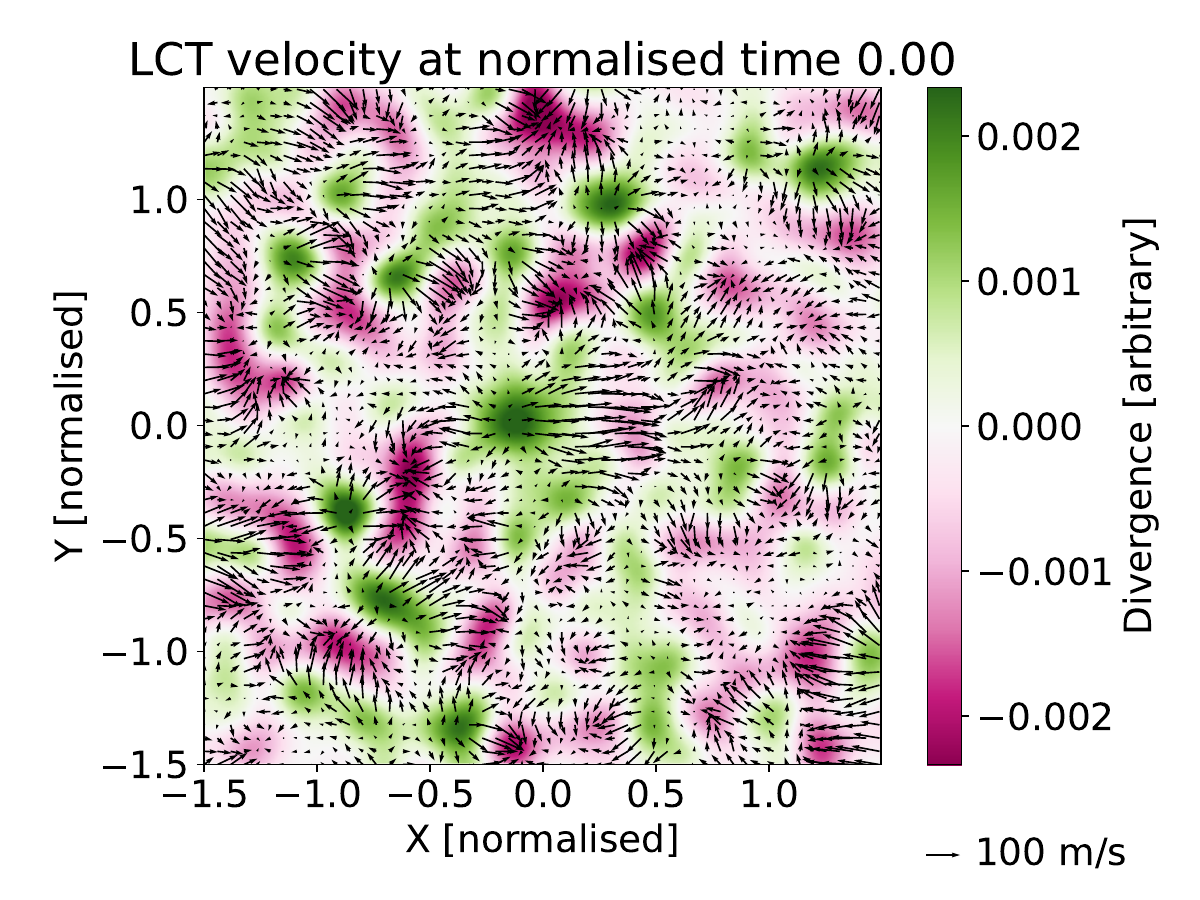} 
    \includegraphics[width=0.33\textwidth]{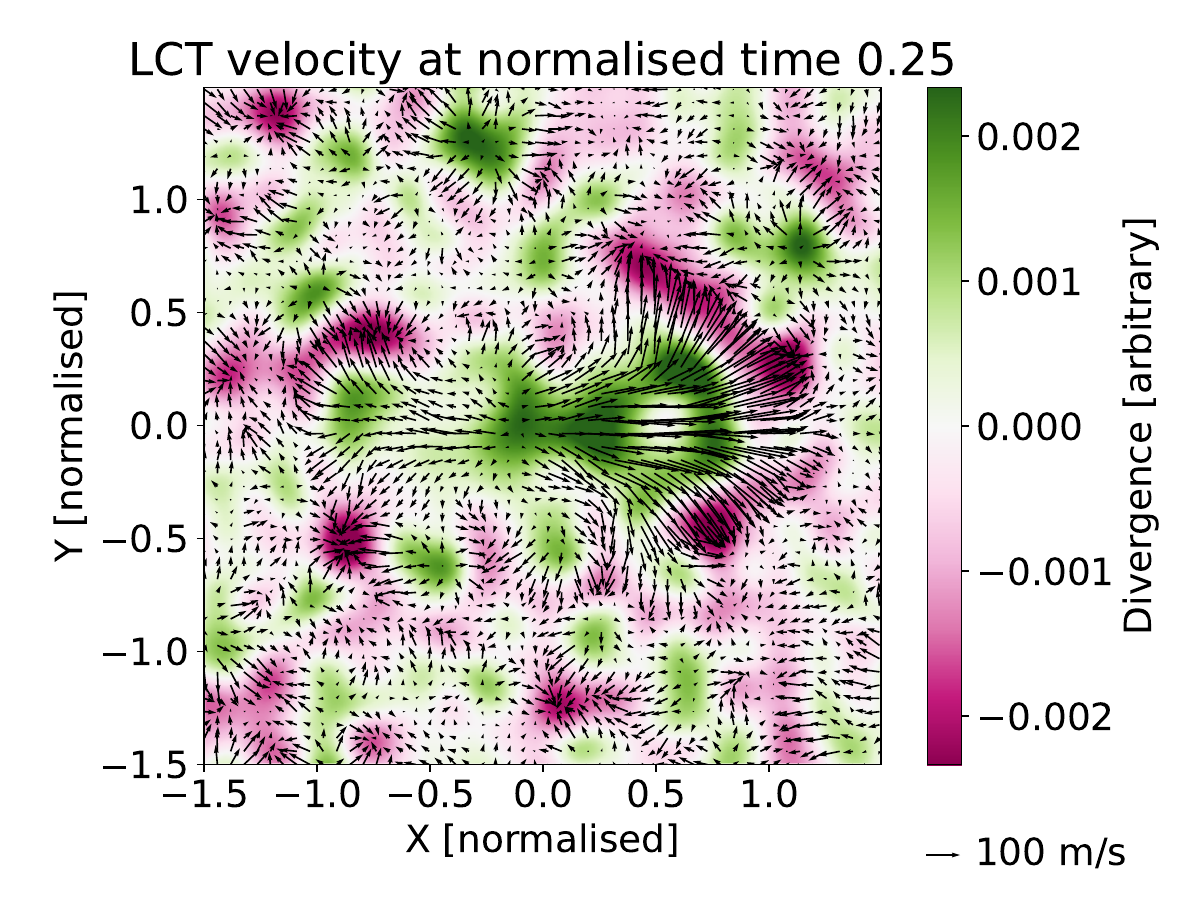} 
    \includegraphics[width=0.33\textwidth]{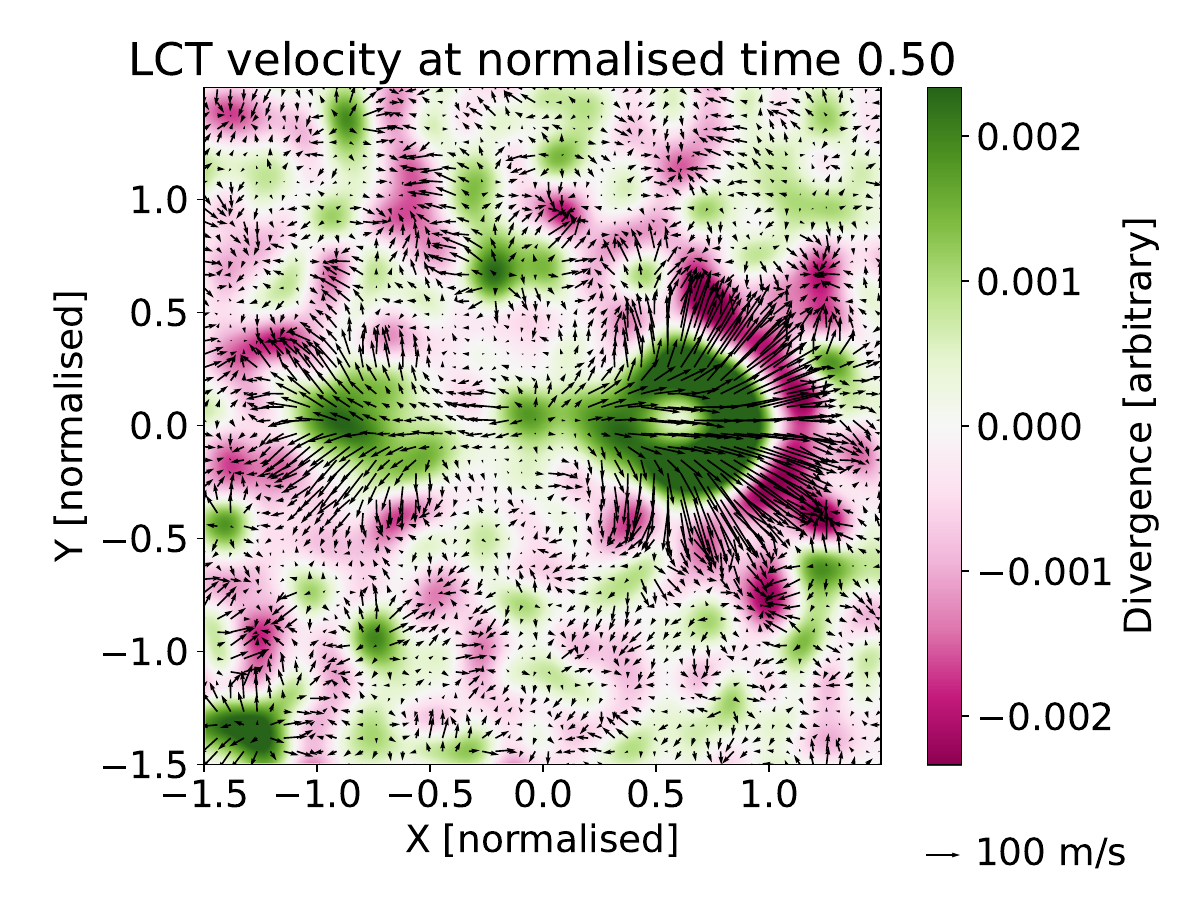} 
    \includegraphics[width=0.33\textwidth]{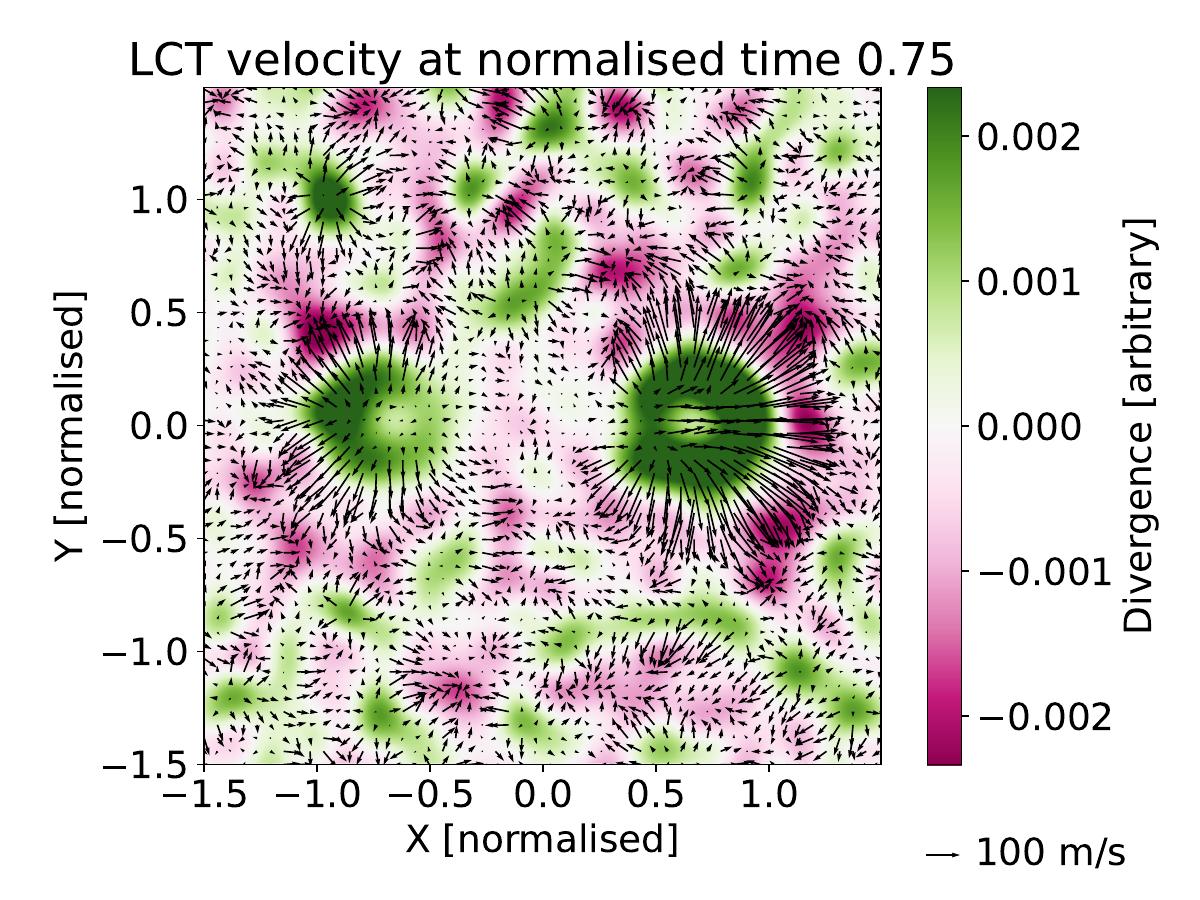} 
    \includegraphics[width=0.33\textwidth]{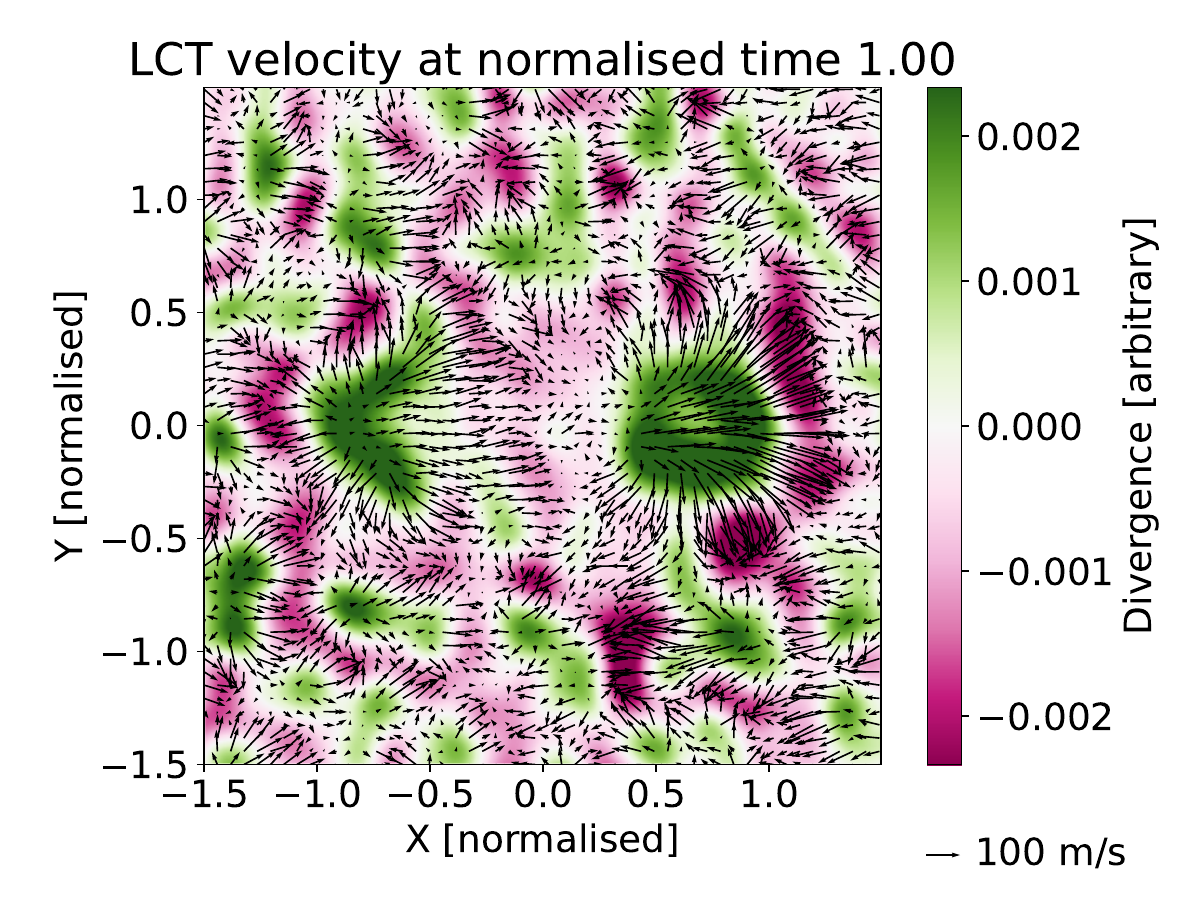} 
    \includegraphics[width=0.33\textwidth]{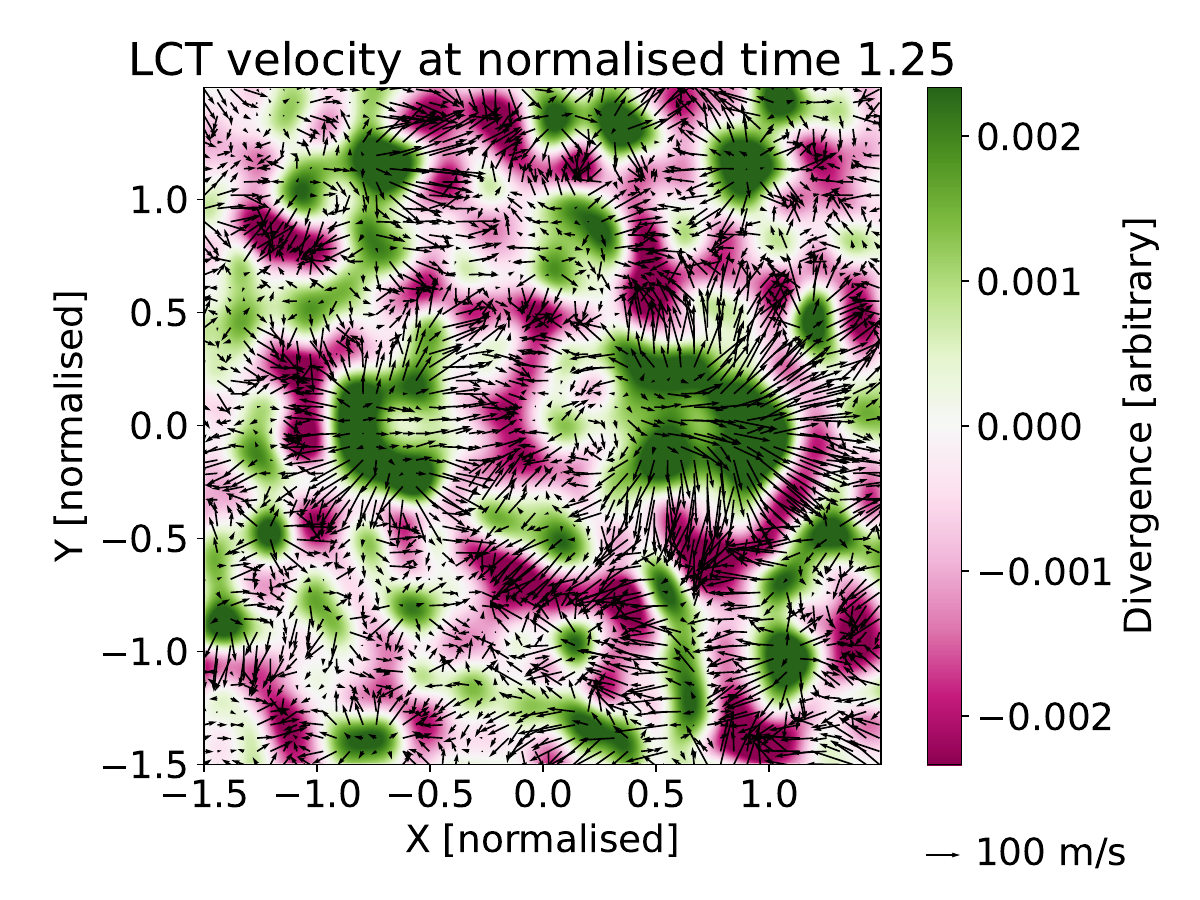} 
\caption{Same as in Fig.~\ref{fig:intensity} but only for the horizontal velocity obtained from the tracking of the HMI intensitygrams. The horizontal divergence of the velocity field is shown on the background in arbitrary units. By using the rule of thumb for the estimate of the spatial scale in the average, $0.001$ of arbitrary units roughly corresponds to $0.0036$~s$^{-1}$ in physical units. A corresponding movie is available online (LCT.mp4) from \url{https://zenodo.org/records/15656676}.}
\label{fig:LCT}
\end{figure*}

\section{Results}
We processed together 58 manually selected bipolar ARs. The total data volume of the final products is 12~TB, and almost an order of magnitude larger volume was occupied by the intermediate products (e.g. by the files downloaded from JSOC). From this sample, we selected a clean subset of 36 ARs. The remaining ARs were rejected based on visual inspection of the magnetograms averaged in time, as described in Section~\ref{sect:processing}. 

\begin{figure*}
    \includegraphics[width=0.33\textwidth]{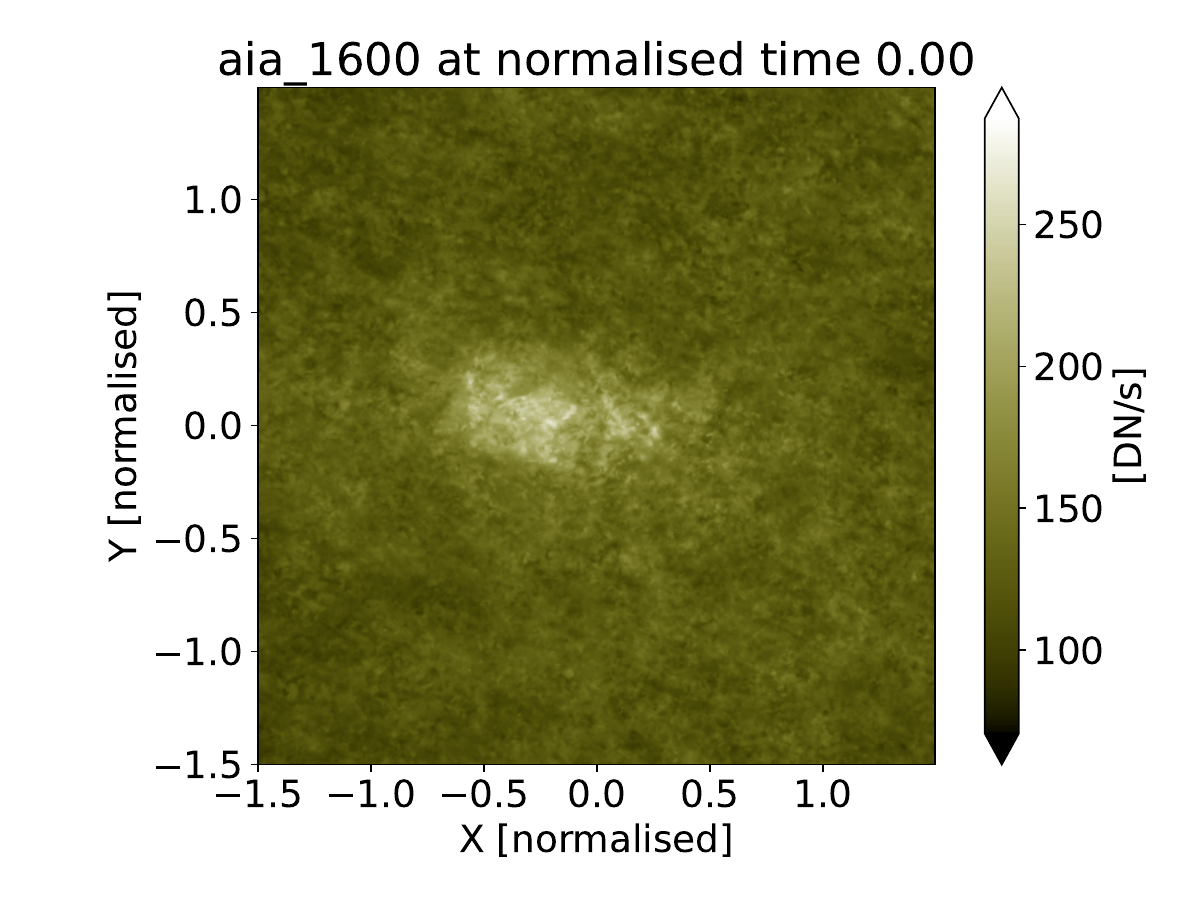} 
    \includegraphics[width=0.33\textwidth]{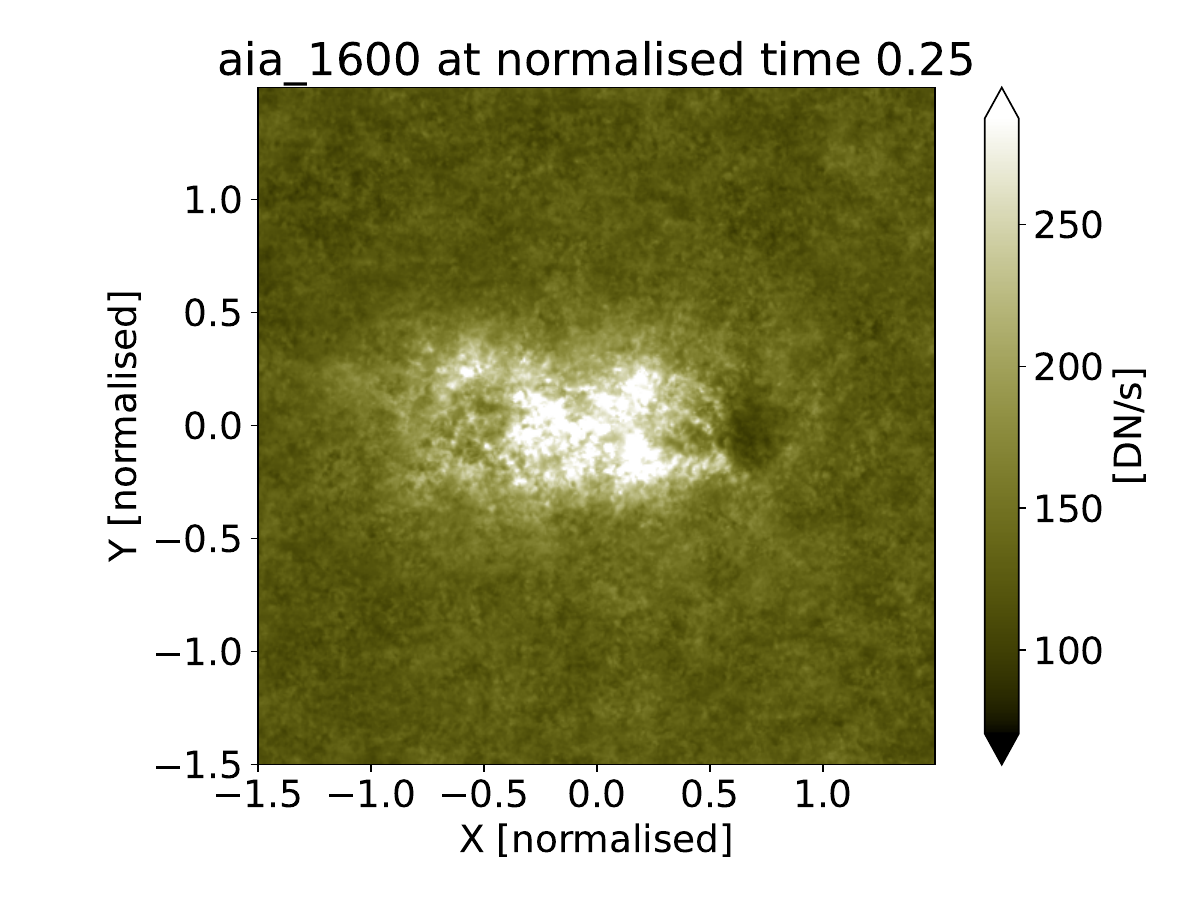} 
    \includegraphics[width=0.33\textwidth]{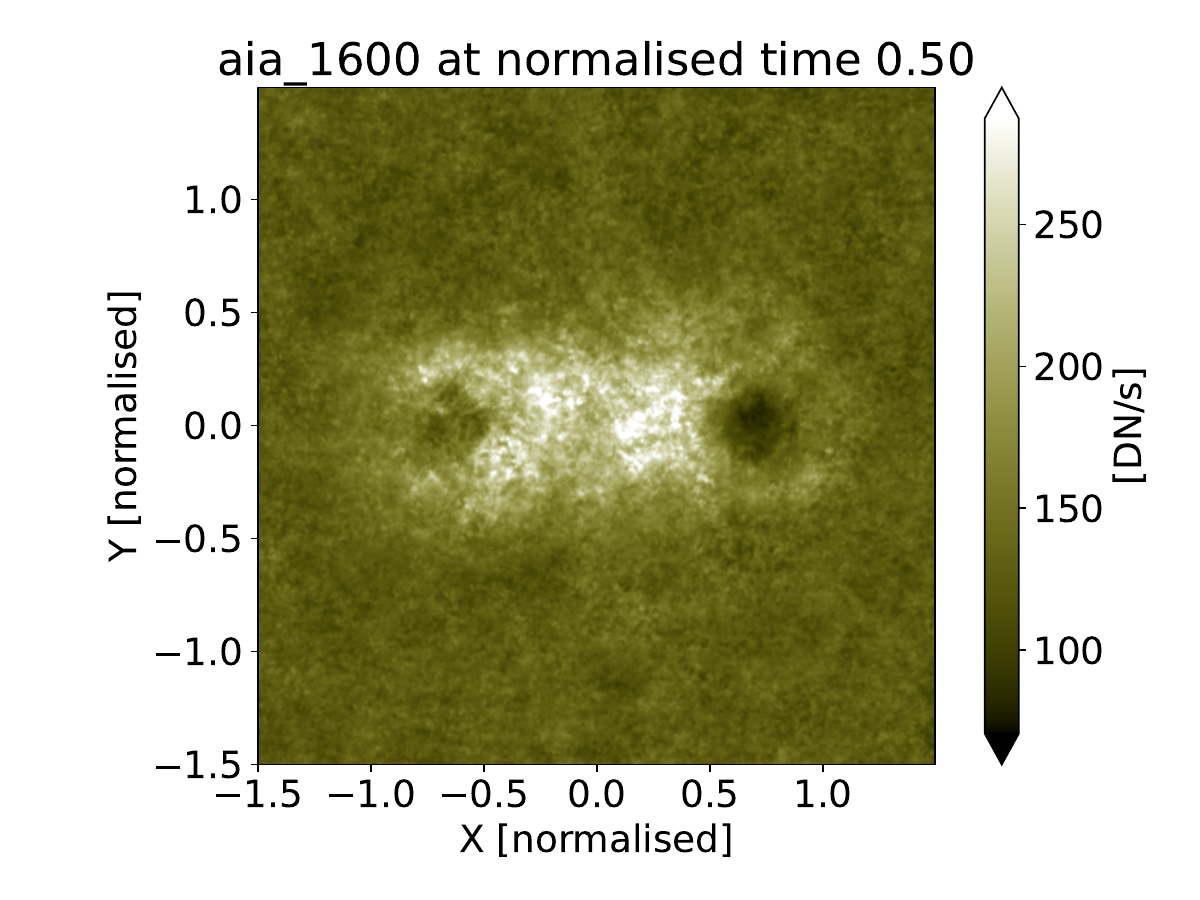} 
    \includegraphics[width=0.33\textwidth]{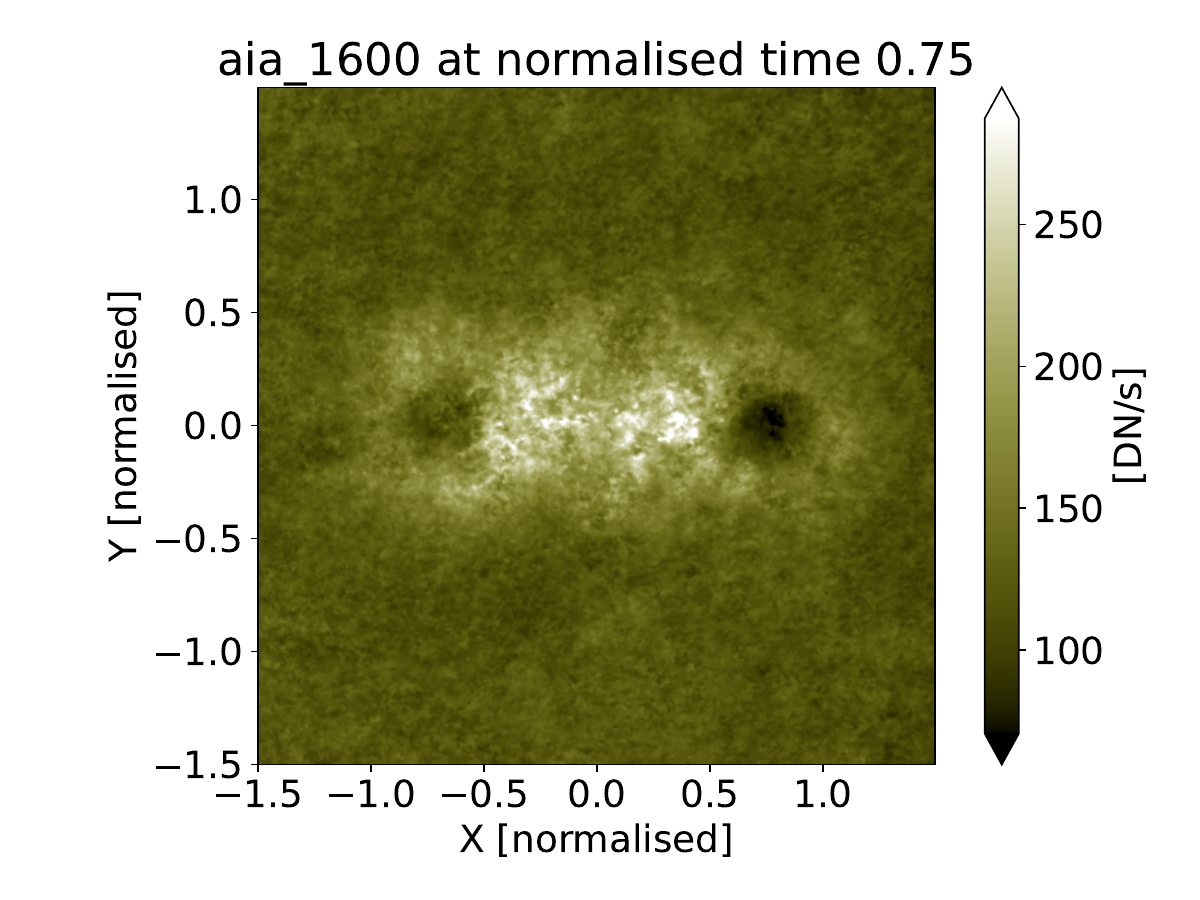} 
    \includegraphics[width=0.33\textwidth]{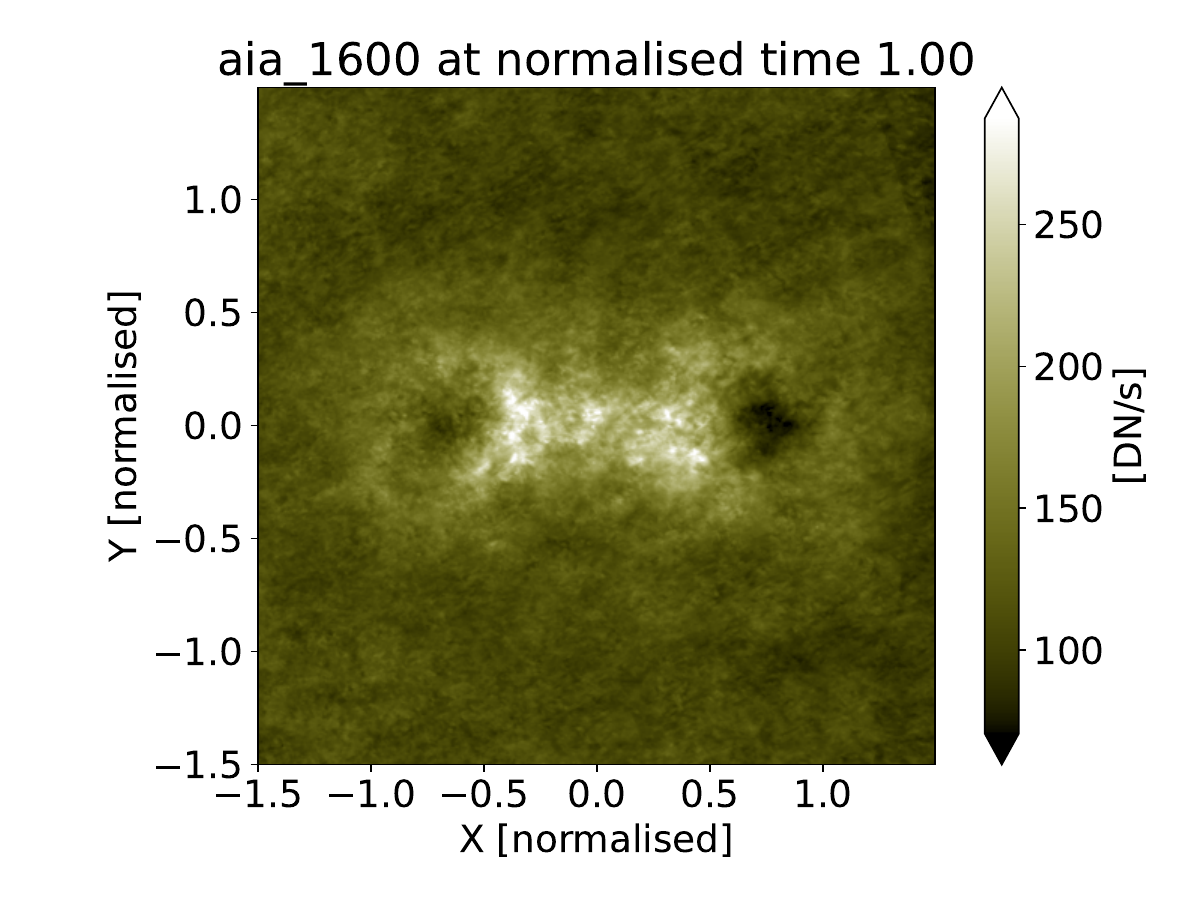} 
    \includegraphics[width=0.33\textwidth]{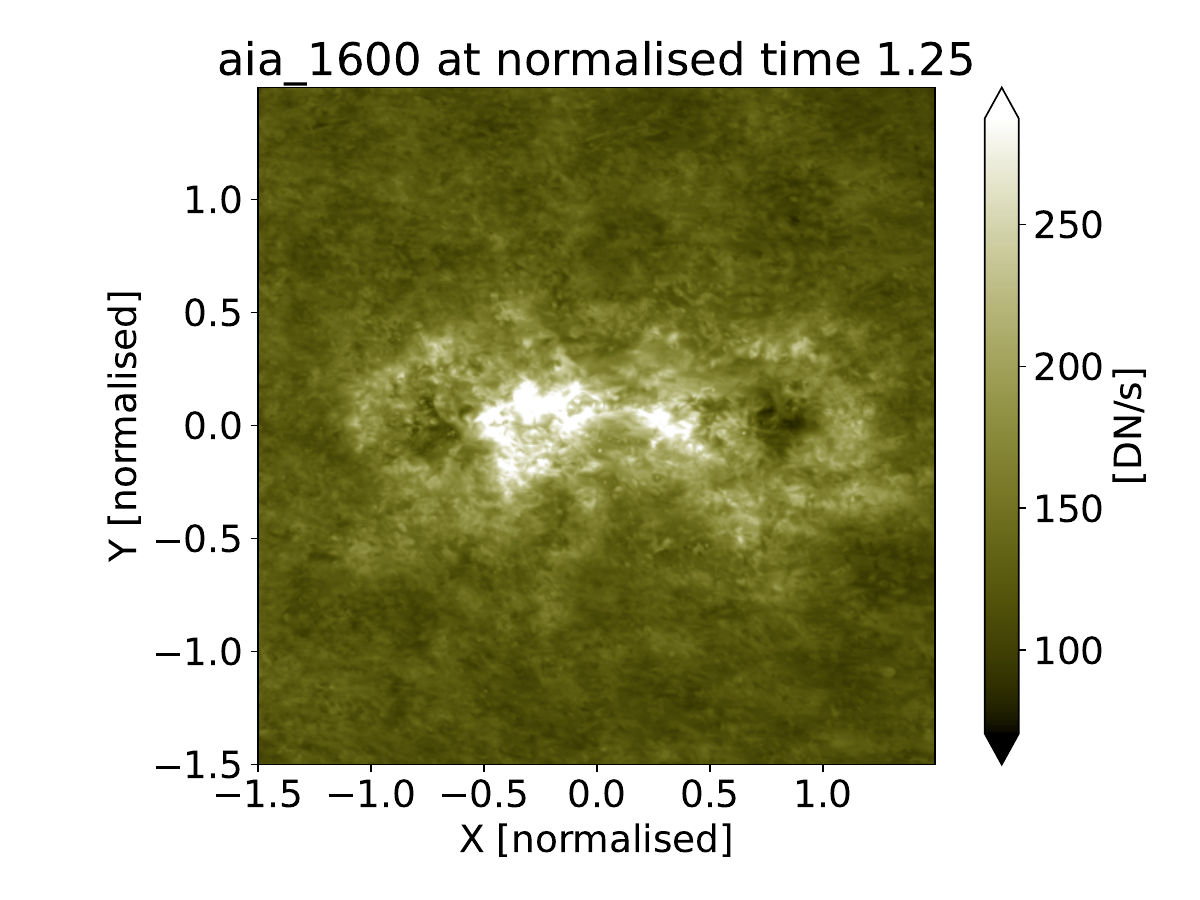} 
\caption{Same as in Fig.~\ref{fig:intensity} but only for the line-of-sight AIA 1600 channel. A corresponding movie is available online (aia\_1600.mp4) from \url{https://zenodo.org/records/15656676}.}
\label{fig:aia_1600}
\end{figure*}

All the remaining ARs had two clear points in time defined: the beginning of the magnetic field emergence and the time when the flux maximum was reached. The difference of those two instants corresponded to the duration of the emergence and growth phases. In our sample, we found that the mean emergence duration was 3.50~days with a standard deviation of 1.21~days. The minimum was 1.28~days and the maximum 6.31~days. These values are consistent with other studies, e.g. with \cite{Weber2023}. The duration of the emergence correlates with the maximum flux in the AR (Pearson's correlation coefficient of 0.55). 

These scales may be used as a rule of thumb to relate normalised time to real solar time. Given that the emergence duration is on average 3.5~days, that makes $\Delta T = 0.01$ of the normalised time equal to approximately 50 minutes and $\Delta T = 0.1$ to approximately 8.5~hours. Similarly in the spatial domain, the average distance between the polarity barycentres is about 70~Mm at the time of the maximum flux (see Fig.~\ref{fig:trends_drifts}), where the normalised distance between those points is $\Delta X=1.0$ from its definition. This makes a rule of thumb to consider the normalised distance of $\Delta X=0.1$ to approximately 7~Mm. 

During the mapping, the linear interpolation was used. The target normalised coordinates were chosen such that the frames were always upscaled, the same holds for the interpolation in time, where the number of samples was also always larger than the original. This way, we ensured that we do not downsample any information in the normalisation process. As a consequence of the upscaling, the fluxes (both radiative in the case of the filtergrams and magnetic in the case of magnetograms) are not conserved. It is therefore difficult to assign absolute physical units to quantities derived from the normalised observables that involve application of spatial or temporal differential operators (both derivatives or integrals). 

\begin{figure*}
    \includegraphics[width=0.33\textwidth]{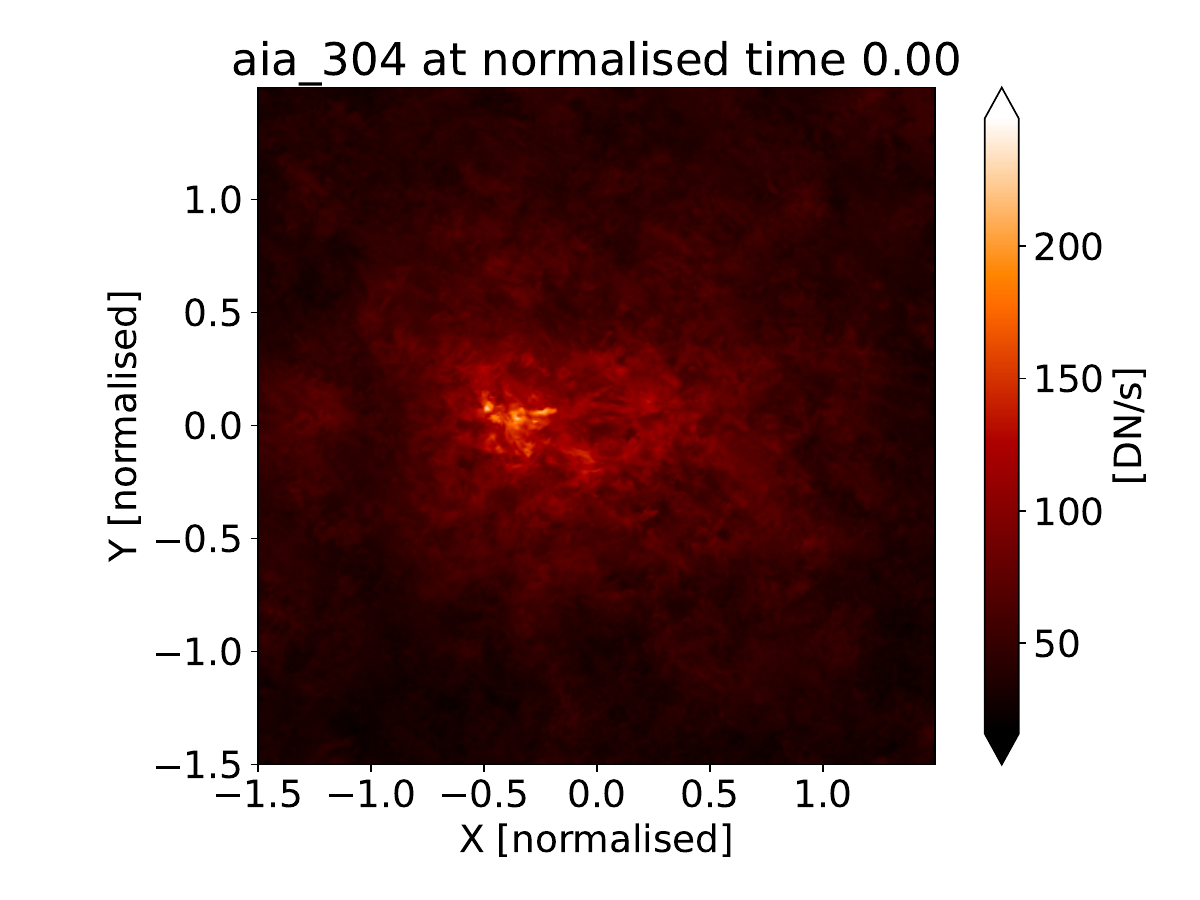} 
    \includegraphics[width=0.33\textwidth]{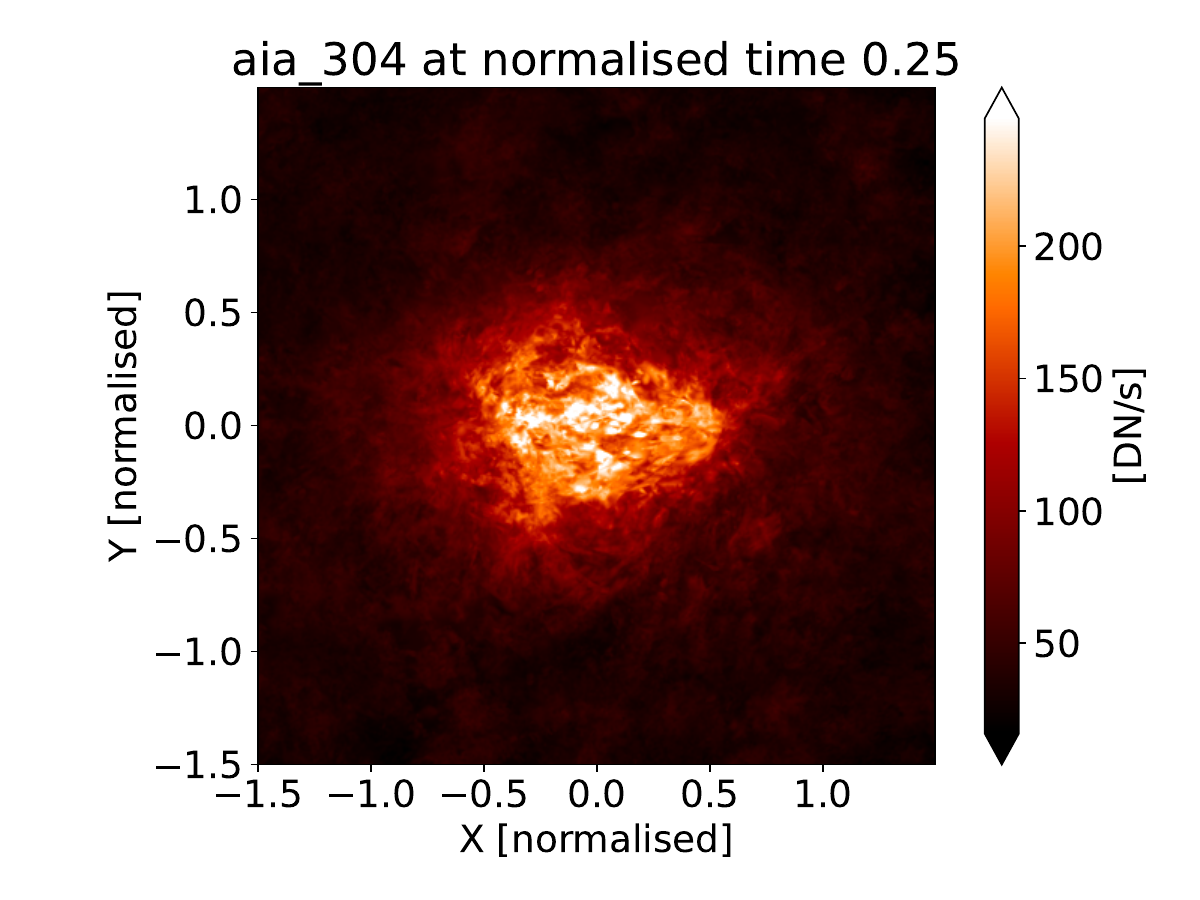} 
    \includegraphics[width=0.33\textwidth]{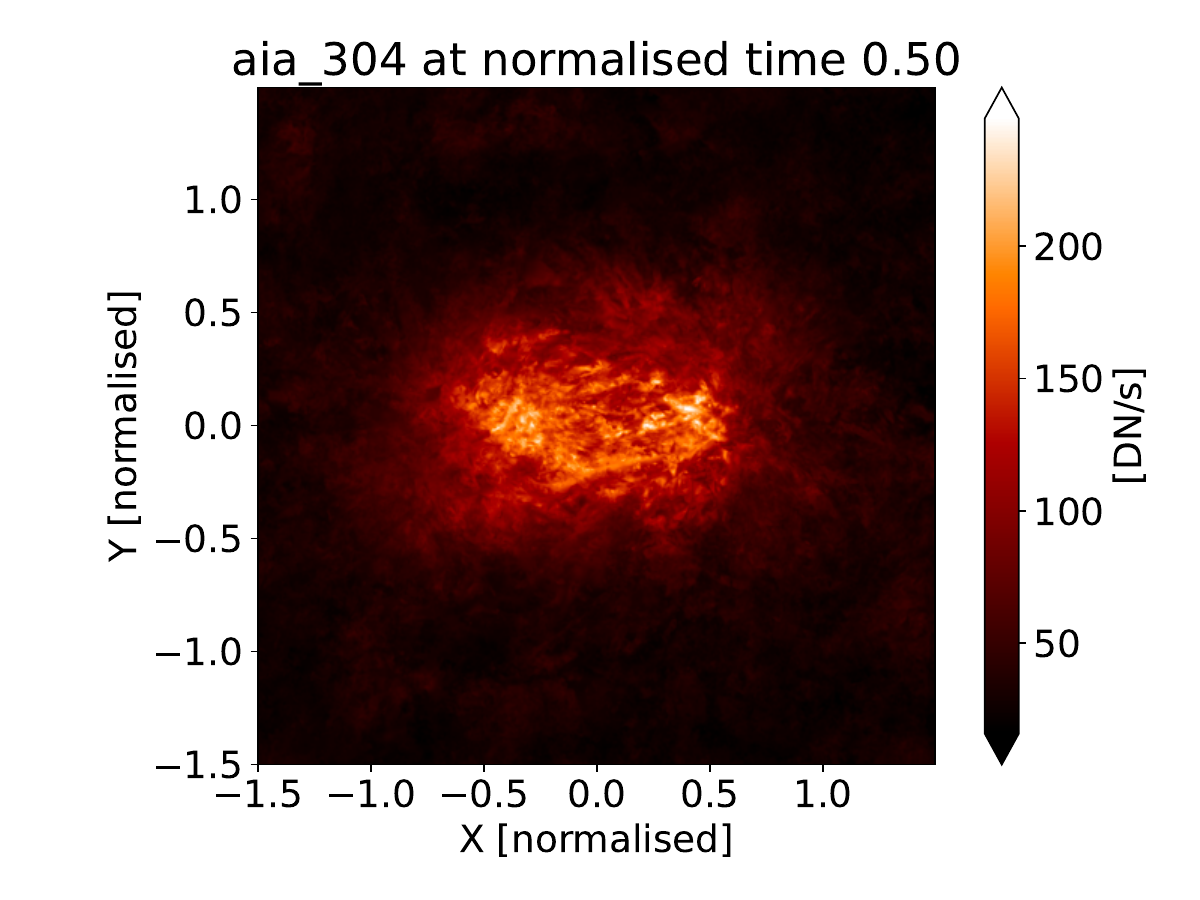} 
    \includegraphics[width=0.33\textwidth]{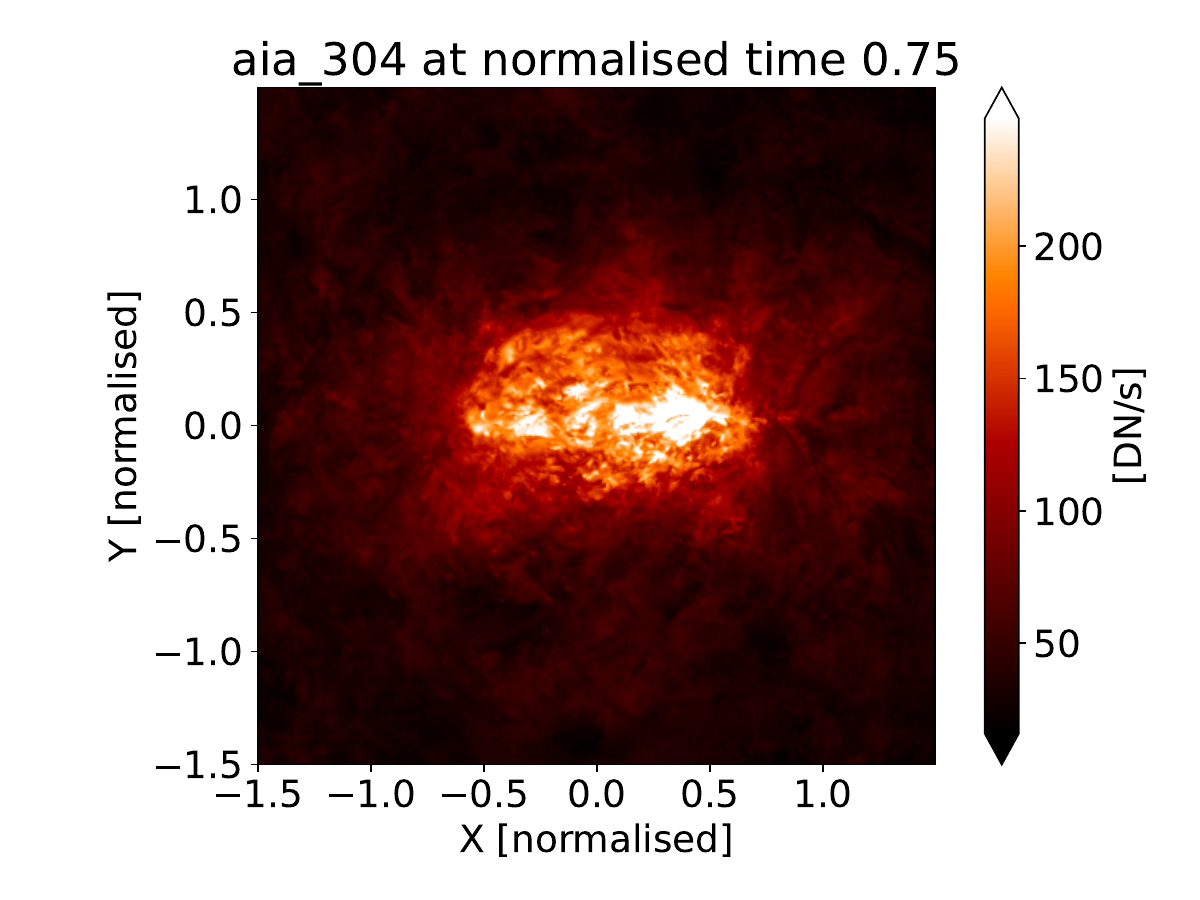} 
    \includegraphics[width=0.33\textwidth]{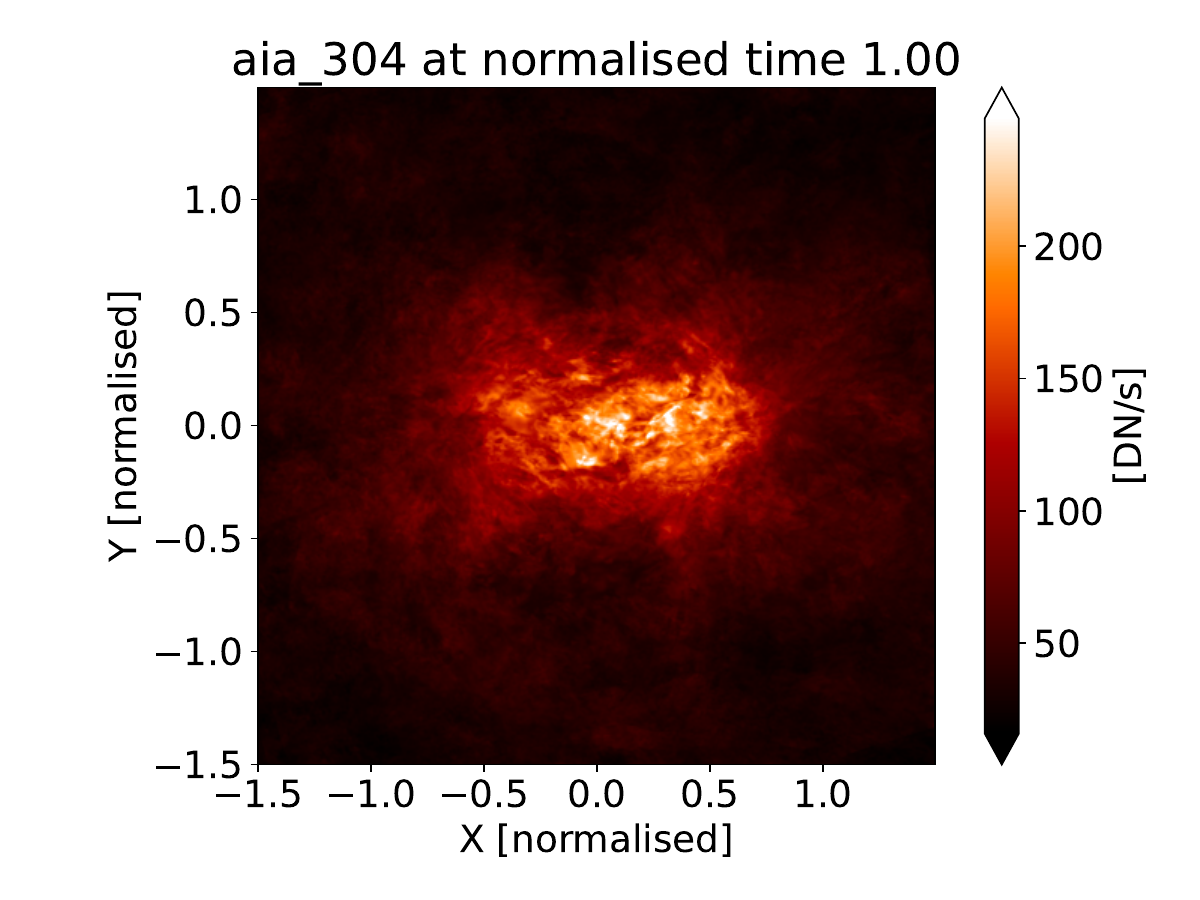} 
    \includegraphics[width=0.33\textwidth]{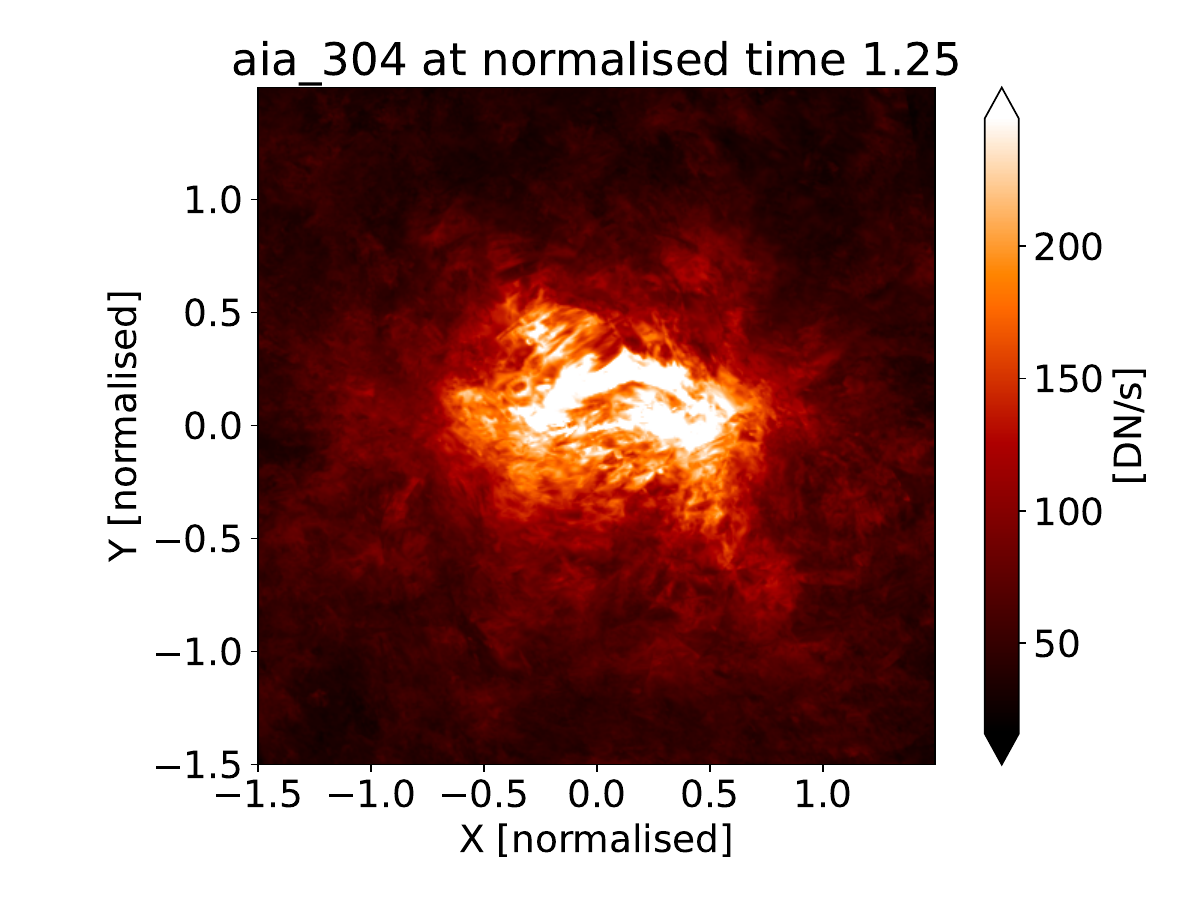} 
\caption{Same as in Fig.~\ref{fig:intensity} but only for the line-of-sight AIA 304 channel. A corresponding movie is available online (aia\_304.mp4) from \url{https://zenodo.org/records/15656676}.}
\label{fig:aia_304}
\end{figure*}

\begin{figure*}
    \includegraphics[width=0.33\textwidth]{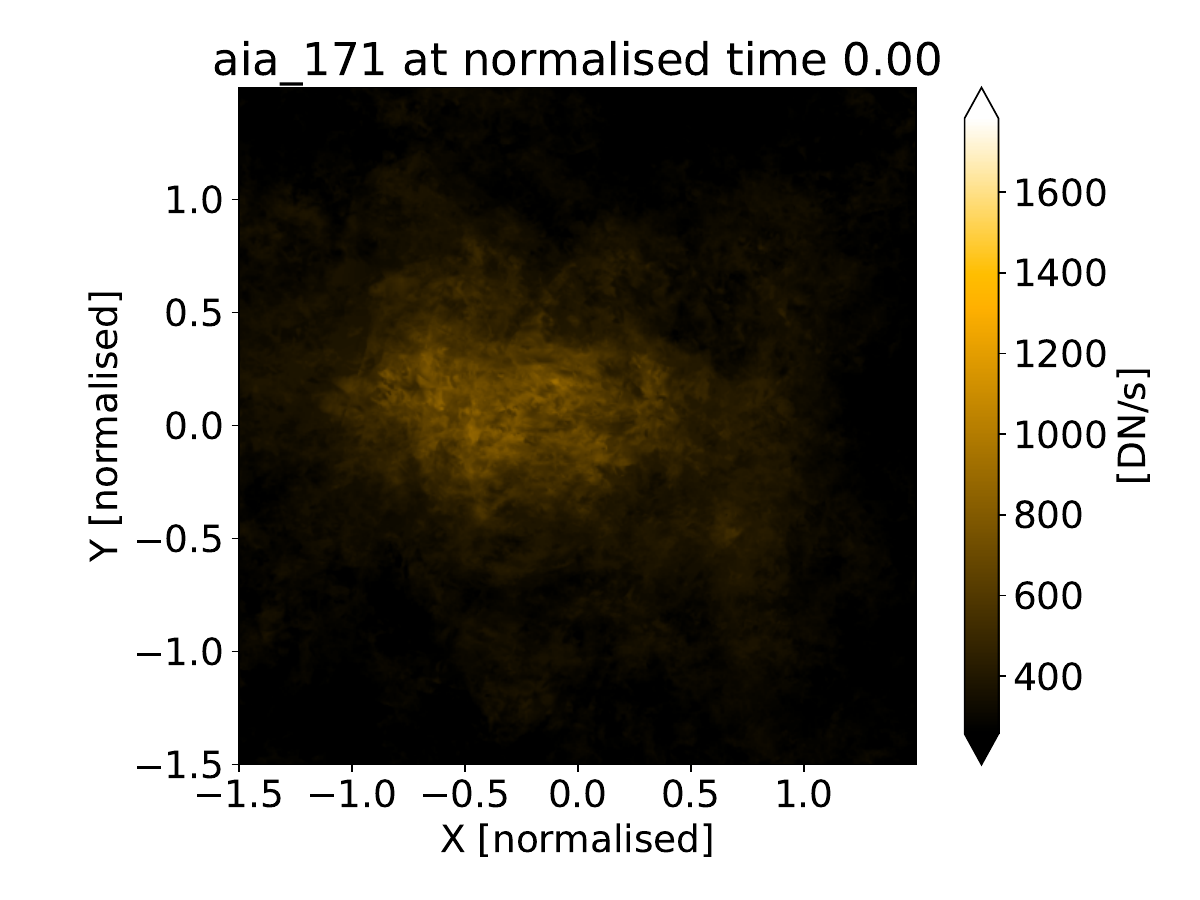} 
    \includegraphics[width=0.33\textwidth]{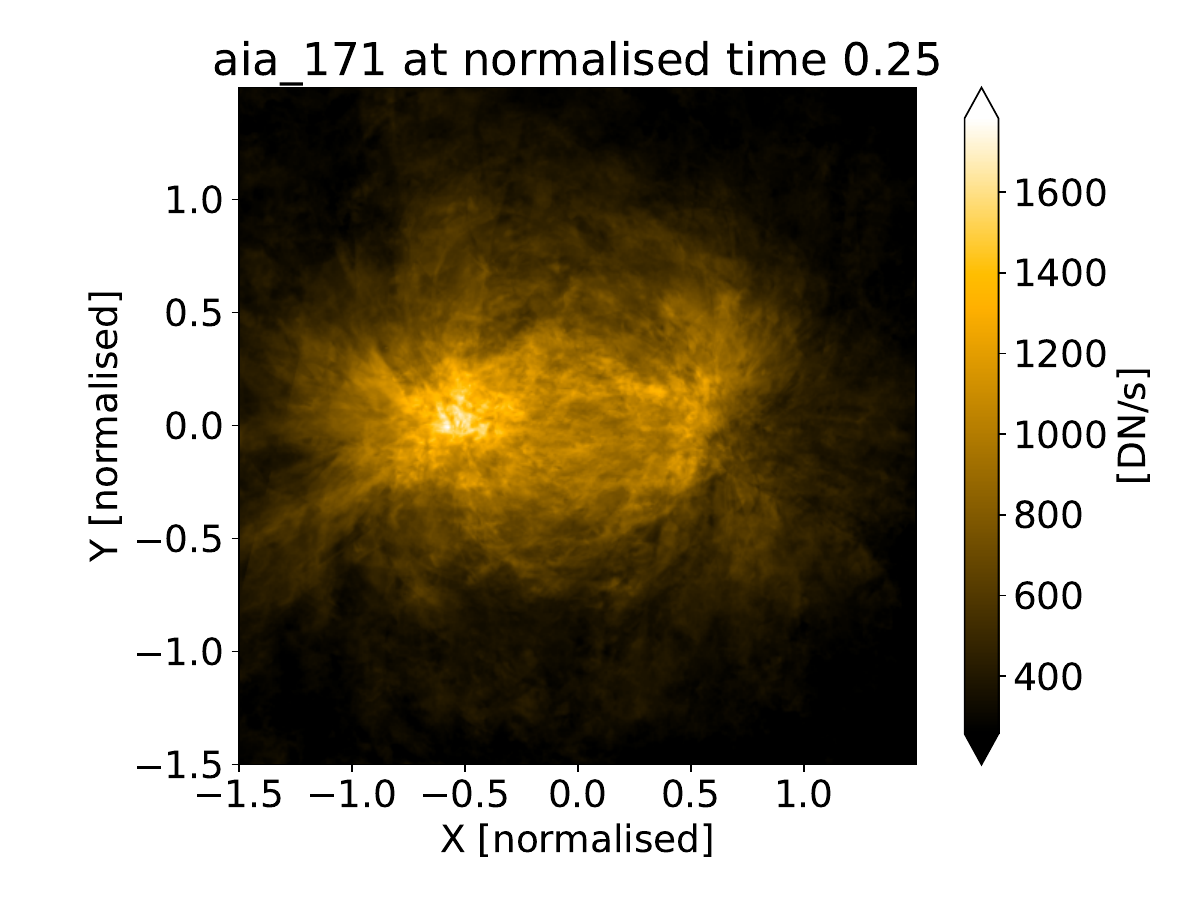} 
    \includegraphics[width=0.33\textwidth]{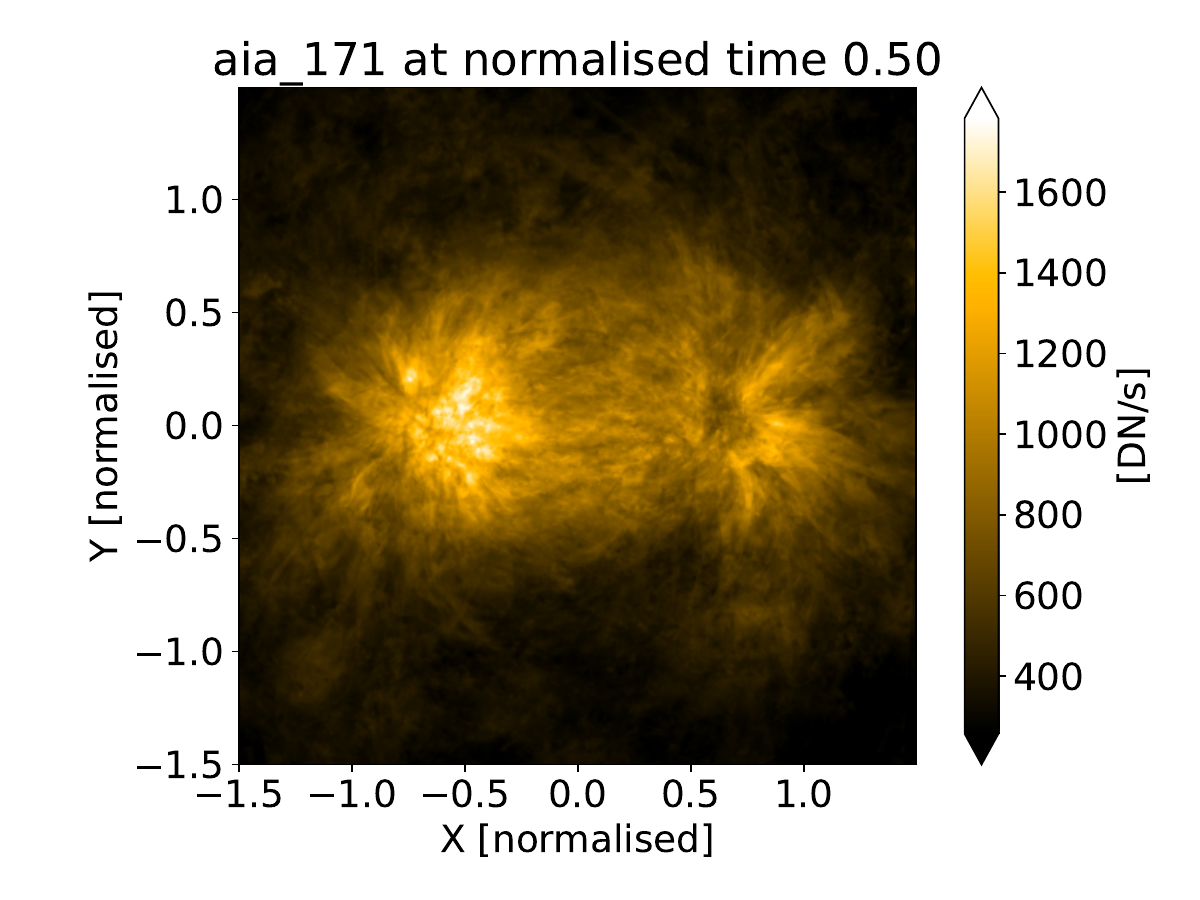} 
    \includegraphics[width=0.33\textwidth]{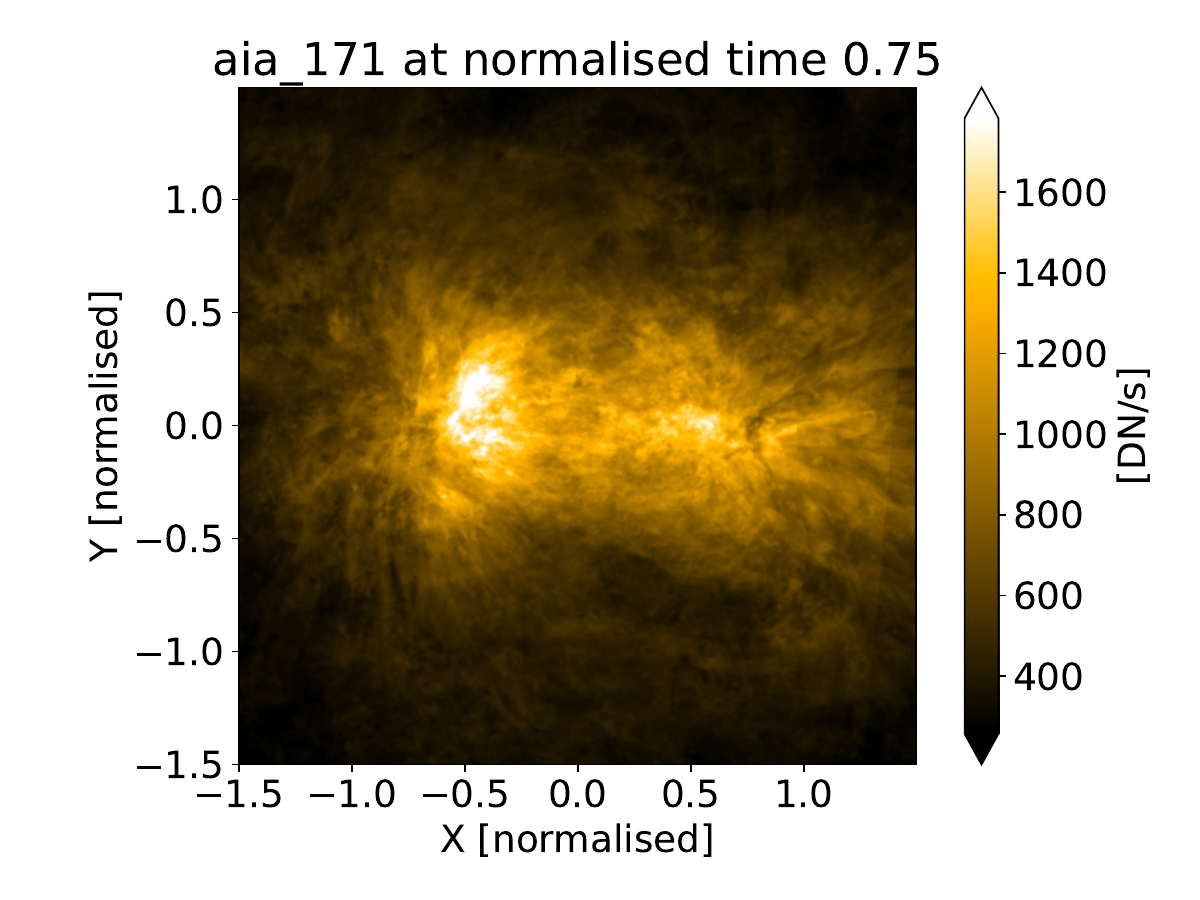} 
    \includegraphics[width=0.33\textwidth]{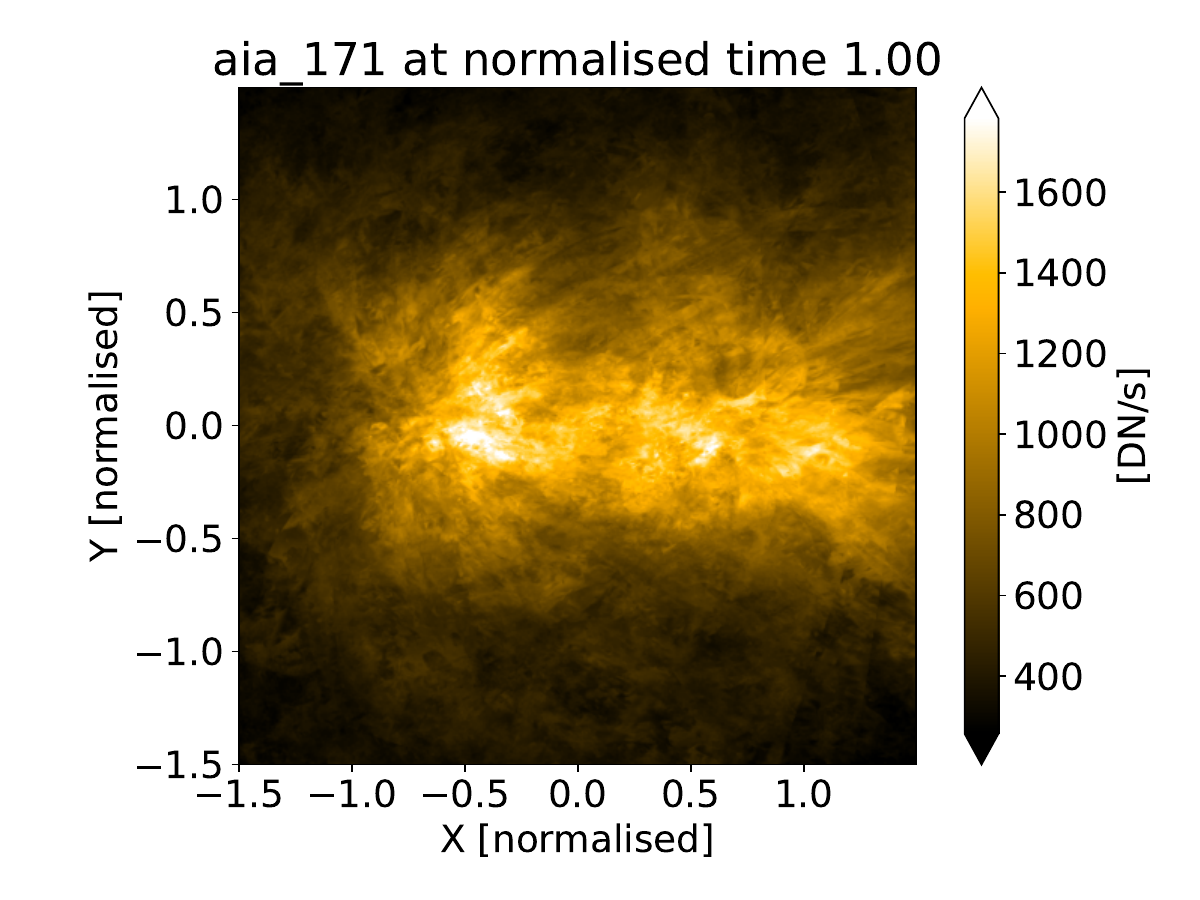} 
    \includegraphics[width=0.33\textwidth]{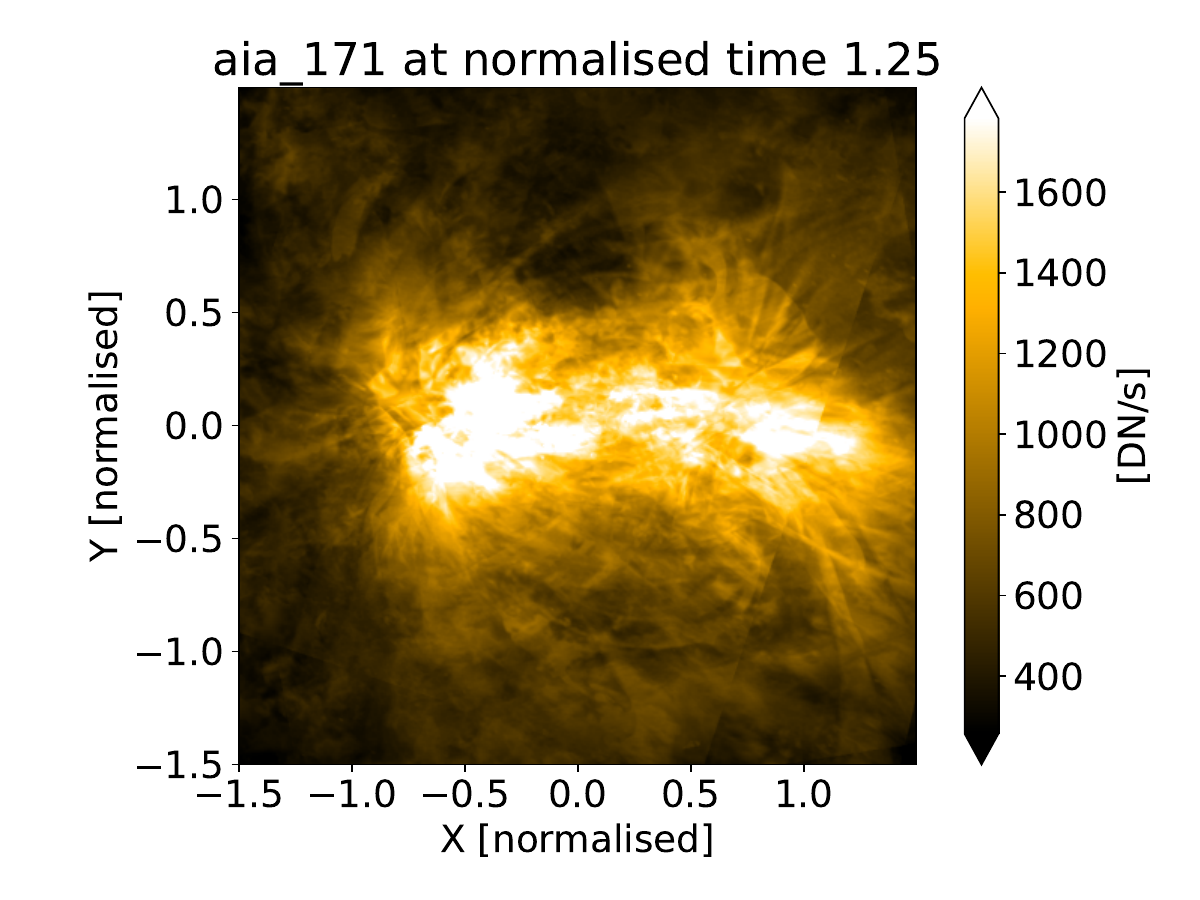} 
\caption{Same as in Fig.~\ref{fig:intensity} but only for the line-of-sight AIA 171 channel. A corresponding movie is available online (aia\_171.mp4) from \url{https://zenodo.org/records/15656676}.}
\label{fig:aia_171}
\end{figure*}

The results of the construction of the average AR are demonstrated in Figs.~\ref{fig:intensity}--\ref{fig:aia_171} for a selection of normalised times $T$, a complete evolution of the observables is available in the online movies. When looking at the evolution of the averaged intensity (Fig.~\ref{fig:intensity}), one notices two almost round sunspots with only a few weak `fragments' around. The poles very much resembled the initial simulation of the bipolar AR by \cite{2009Sci...325..171R}. It is important to note that the normalisation for stacking does not involve the intensitygrams, but was based on the spatial coincidence of the gravity centres of the magnetic field. It does not coalign the frames on the strongest point in the magnetic field nor does it on any point in the intensity. Moreover, the normalisation is set such that the stacking points lie at coordinates $X=\pm 0.5$ on the horizontal axis ($Y=0$). The locations of the centres of the averaged sunspots are obviously farther from the image centre. This systematic deviation between the average sunspot centres and the barycentres of the magnetic field is caused by the asymmetries of the magnetic polarities in the zonal direction, when the weaker field always covers a larger area towards the emergence point from the barycentres as compared to the outer side. This asymmetry is clearly visible in the magnetograms (Fig.~\ref{fig:Bvec}). The fact that, despite targeting a different point, the average forms almost round sunspots, indicates that the bipolar ARs are scale invariant. Throughout the investigated evolution, the sunspot forming the leading polarity is darker than the spot in the trailing polarity. The movies show that smaller fragments coalesce with the sunspot cores until $T\sim 1$, predominantly merging from the centre of the polarity connecting line, only rarely from the outside. This is consistent with the statistical study by \cite{2021A&A...647A.146S}. 

Even though normalised time $T=0$ is defined as the beginning of a rapid emergence in the case of individual ARs, the average shows that the vertical bipolar configuration exists before $T=0$ (Fig.~\ref{fig:Bvec}), the bipolar nature seems to be present already at $T\sim -0.5$. We note that due to the incompleteness of the AR sample at those $T$s, we cannot exclude that this is an artefact. In the individual cases the pre-existing bipole is visible only for less than 25\% of the ARs in the sample (see Appendix~\ref{app:preemergence}). The existence of a magnetic bipole for normalised times $T<0$ is confirmed also in the line-of-sight magnetograms. Therefore, it is not an artefact of the magnetic field inversion. The horizontal components are only weakly developed at the time of emergence and gain a clear bipole-like signature at the normalised time around $T=0.05$, where the divergence of the field lines is seen in the trailing and the convergence in the leading polarities. This corresponds to our choice of the average AR polarity, where the leading is chosen to have a negative vertical magnetic field. There is a strong horizontal component in the east-west direction in between the polarities, indicating an almost horizontal field around the polarity inversion line aligned with the line connecting the two polarities. 

The average Dopplergrams (Fig.~\ref{fig:velocity}) identify sunspots as downflow regions. We need to keep in mind that our Dopplergram-processing method does not provide us with absolute measurements, the method of a subtraction of a smooth background sets the Doppler velocity in a quiet Sun approximately to zero. From an absolute point of view, a convective blueshift having values between around $-400$ and $-300$~m/s at the centre of the solar disc is known from the literature \citep[see][among others]{2018A&A...617A..19L,2019A&A...622A..34S}, which evolves to zero with increasing heliocentric angle. Therefore, the terms `upflows' and `downflows' need to be understood in this context, that is, as relative terms with respect to the quiet-Sun surroundings. 

Strong magnetic concentrations are usually found to coincide with downflows, mainly during the emergence and growth phases of the AR's life, see e.g. \cite{2001ApJ...549L.139D} or Fig.~33 by \cite{2011LRSP....8....4B}, similar behaviour was seen in the state-of-the-art simulation of the AR evolution \citep[][]{Hotta2020}. In the case of the average AR, the downflows seem to appear first in the trailing polarity (the diffuse prevalence of ``red'' colour is there already at $T=-0.05$, before the rapid emergence), whereas there seems to be a weak upflow in the leading polarity. Their appearance coincides with the strong-field region. However, normalised times $T\sim0$ in the average represent stages when most of the ARs in the sample were located on the eastern hemisphere of the Sun. Therefore, the downflow-upflow configuration is likely only apparent and is a consequence of a weak horizontal outflow at the emergence point, which is projected to the line-of-sight at larger heliocentric angles. After removing the (spatial) high-frequency velocity oscillation, we estimated the magnitude of the corresponding outflow diverging from the emergence region to be around $\pm 60$~m/s. 

Later in the evolution, the downflow encompasses both polarities. Here again, we need to emphasise that this is a downflow with respect to the quiet-Sun background, which is blueshifted due to the convective blueshift. The magnitude of the sunspot downflows is of the same order as the expected magnitude of the convective blueshift, and therefore this downflow is likely only apparent. We note that the two-polarity structure at times $T>0.6$ seen in Fig.~\ref{fig:velocity} is the line-of sight effect of the outflows (Evershed flow and the moat flow) observed at larger heliocentric angles on the western hemisphere, where the majority of the ARs were located at these normalised times. 

The horizontal flows (Fig.~\ref{fig:LCT}) obtained from the tracking of the granules in the intensity frames may indicate a weak divergent flow close to the centre of the frame (slightly eastward, at $X=-0.2$) at $T=-0.07$, which turns into a clear divergent flow at $T=-0.02$ with a strong prograde component. The central divergence is strongly asymmetrical in the zonal direction initially. The divergent region is elongated in the zonal direction and later turns into a ``sausage'' shape with two centres separating from each other. These centres do not coincide with the location of the average sunspots and also do not coincide with the location of the polarity barycentres. At $T=0.65$ there remain only weak horizontal flows around the polarity inversion line, whereas the outflow centres are already located around the polarity barycentres. Almost radial outflows similar to the moat flow are formed which are slightly stronger on the outer sides. At around $T=0.85$ a clear almost symmetric moat flow is visible. Throughout the AR's growth, the outflows are stronger in the leading polarity than in the trailing one.  

\begin{figure}[!t]
    \includegraphics[width=0.49\textwidth]{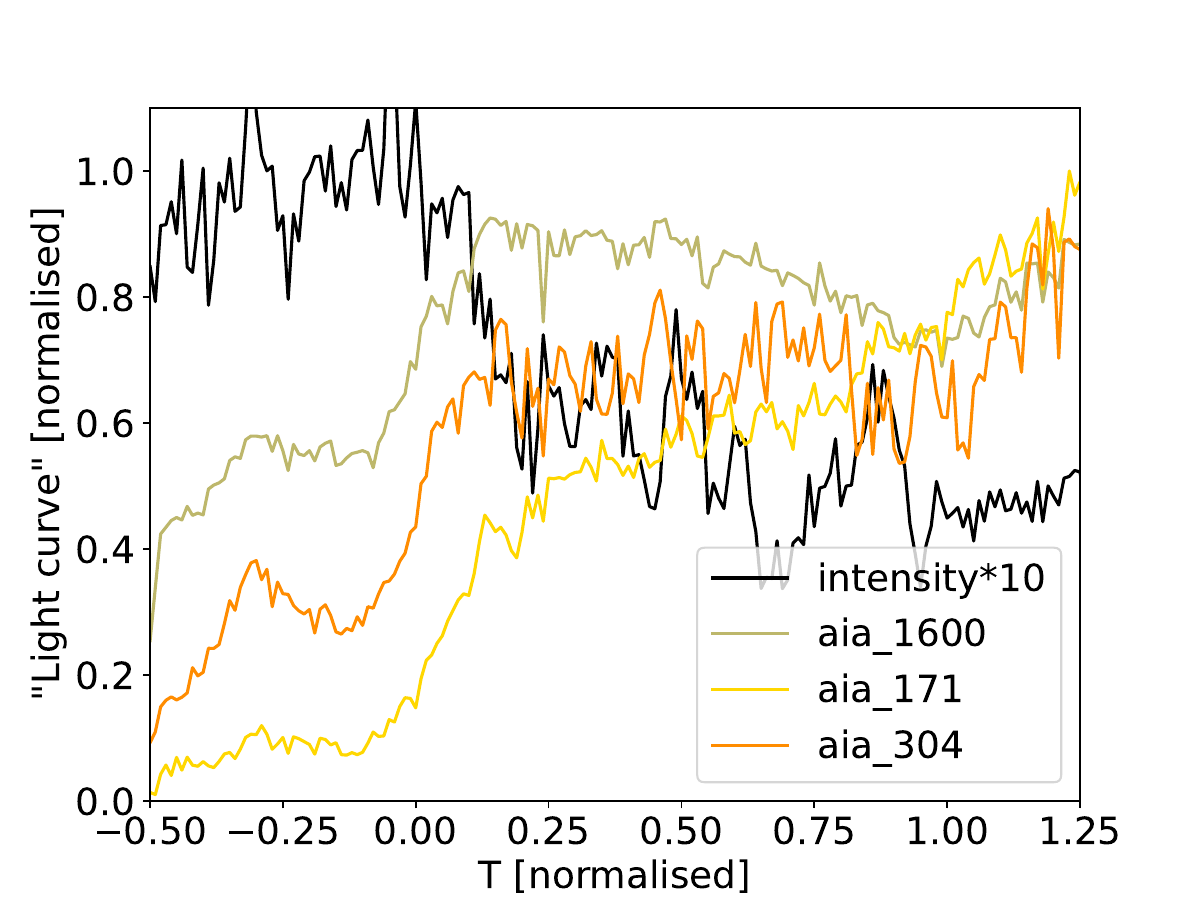}
    \caption{Light curves for the intensity channels integrated of the whole field of view as a function of the normalised time. The HMI intensity light curve was normalised to the quiet-Sun value; the AIA light curves were normalised to their maxima. For the purpose of the display, we have magnified the deviations in the HMI intensity by a factor of ten.}
    \label{fig:lightcurves}
\end{figure}

The chosen AIA channels represent rather cooler plasma, with typical temperatures below 1~MK. Those AIA observations typically capture structures in a lower corona, the characteristic formation heights of the chosen channels reach up to about 1500~km above the surface \citep[][]{2024ApJ...975..236S}. At the heliocentric angle of 60\degr{}, such a height above the surface corresponds to the projected sky-plane displacement about 750~km, which, by using a rule-of-thumb, corresponds to the displacement of about $\Delta X=0.01$ in the normalised coordinates. Given the fact that we are interested in the average behaviour, such a displacement plays only a negligible role in the construction of the average AIA flux and, therefore AIA images may be treated similarly as the HMI photospheric measurements. We postpone the analysis of hotter channels for the follow-up study. 

In the AIA 1600 (Fig.~\ref{fig:aia_1600}), brightenings are most prominent above the polarity inversion line and already start to appear at $T\sim-0.05$. Their brightness decreases from about $T=0.35$ onwards, but the region with a diffuse `glow' grows. As a consequence, the total flux integrated over the whole AIA 1600 field-of-view peaks shortly after emergence and then slowly decreases as the magnetic flux continues to emerge until its maximum. This is also visible in the integrated light curves (Fig.~\ref{fig:lightcurves}). The brightenings in AIA~1600 have mainly two origins. Some contribution comes from small-scale reconnection of the newly emerging flux into the existing one, as pointed out already in the review by \cite{vanDriel-Gesztelyi2015}. In the AIA~1600 channel, emission from the small-scale magnetic elements is visible in the weaker-flux regions, such as the surroundings of the strong-flux regions and at the lanes between supergranular cells. The effect of small-scale reconnections effect also contributes to the observed emission in the averaged AR between the polarities, where a weaker flux is abundant. 

The AIA 304 (Fig.~\ref{fig:aia_304}) behaves similarly at first, but in the later phases (for $T>0.75$) the most intensive brightening moves towards the leading polarity. The integrated light curve is similar to that of AIA 1600 (Fig.~\ref{fig:lightcurves}) first, but the integrated flux does not decrease after $T=0$, because dark spots are not visible in AIA~304. The emissions in AIA~304 channel consist of a mixture of both emissions at the loop footpoints and ribbons of flares of all kinds. Cold absorbing filaments may also affect the radiation flux recorded by AIA. The increase of the total emission for $T>1$ is likely dominated by the magnetic field reconnections in the weak flares. 

The brightness of an average AIA 171 (Fig.~\ref{fig:aia_171}) increases gradually (Fig.~\ref{fig:lightcurves}) as coronal loops develop in the lower corona. The brightenings appear above both polarities, demonstrably stronger above the trailing one. At later phases, radiation from the hot loops dominates. The loops may be heated by the heating events during the dissipation of the coronal magnetic field. Coronal rain may also contribute to the emission in the AIA~171 and AIA~304 channels. 

\begin{figure}[tb]
    \includegraphics[width=0.49\textwidth]{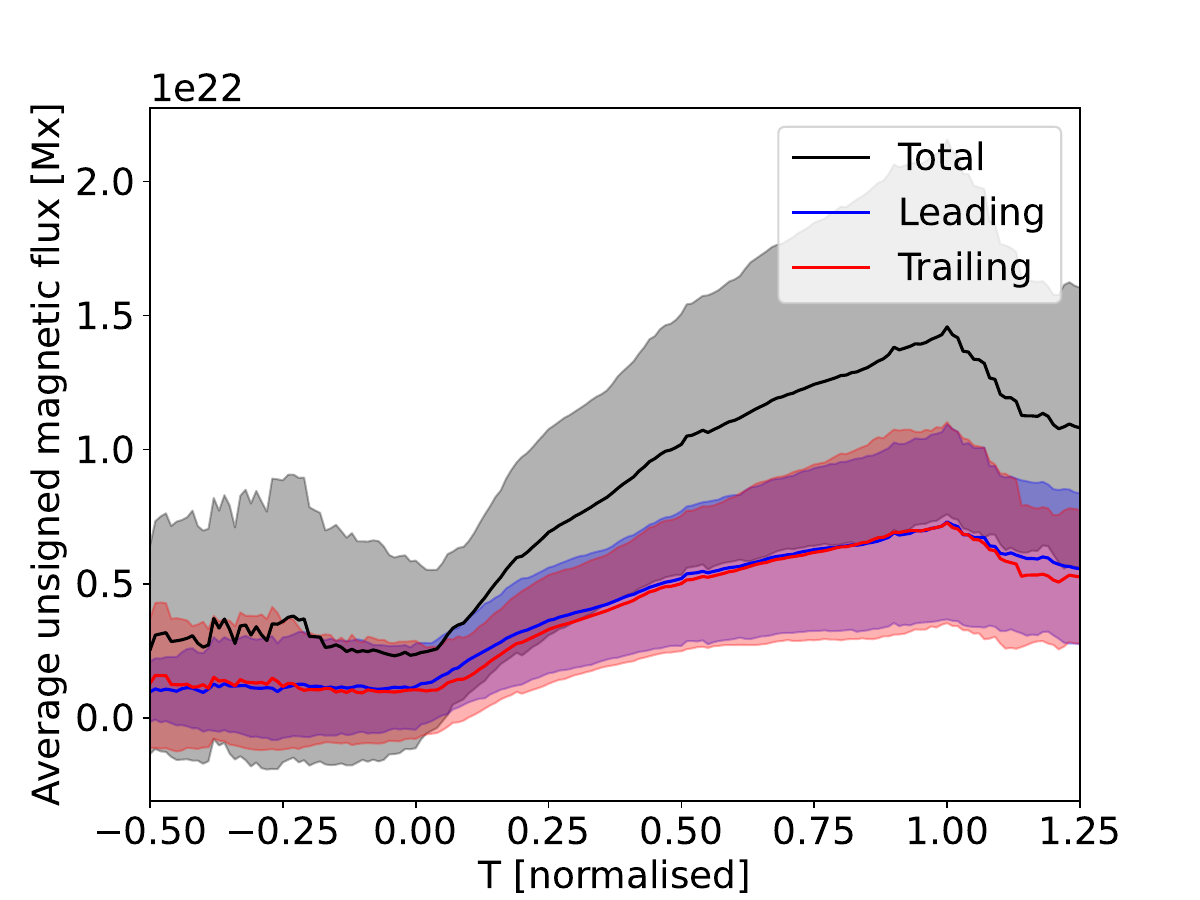} 
 \caption{Average unsigned magnetic flux as a function of normalised time, $T$, in the polarities and a total in the ARs. The solid lines indicate the mean over the sample of ARs, whereas the filled regions represent $\pm \sigma$ bounds.  }
\label{fig:trends_fluxes}
\end{figure}

\begin{figure}[!ht]
    \includegraphics[width=0.49\textwidth]{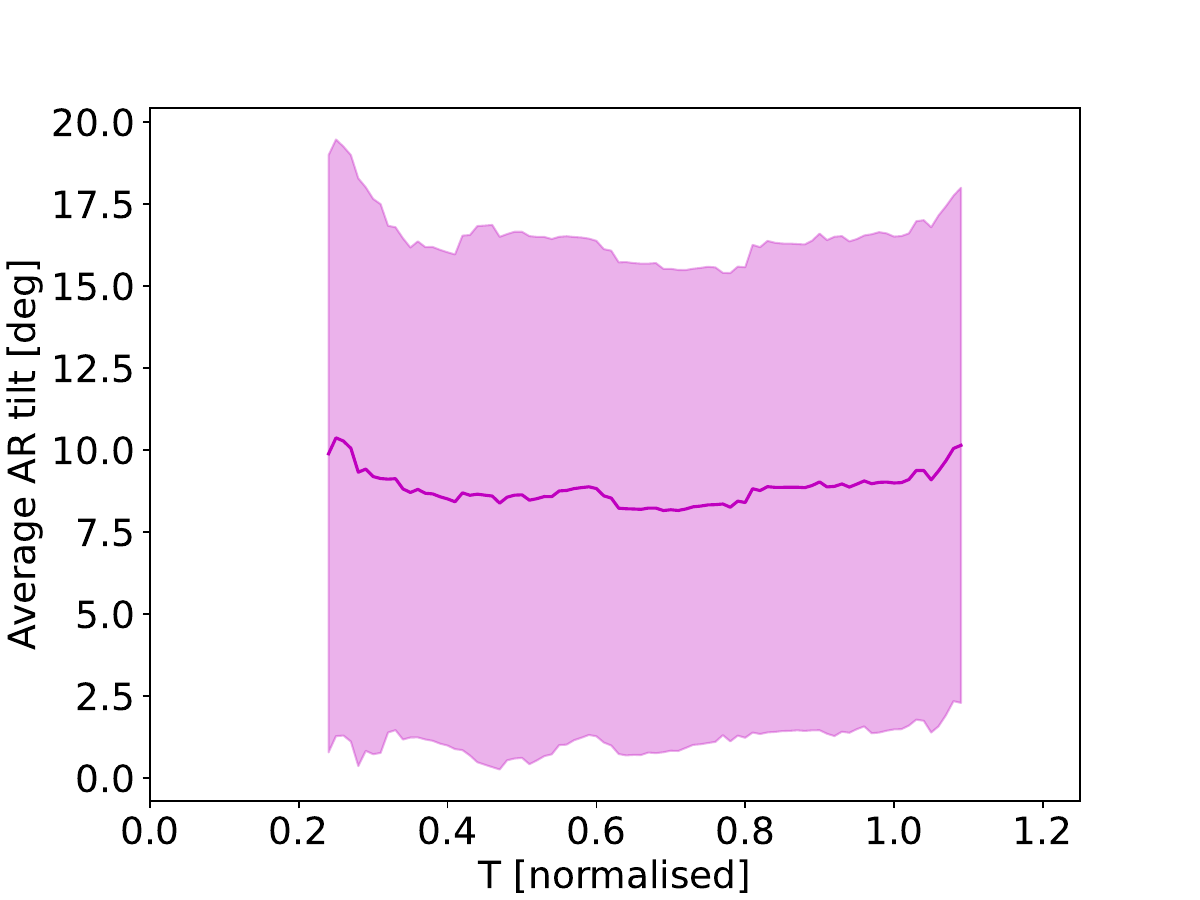} 
 \caption{Similar to Fig.~\ref{fig:trends_fluxes} but only for the AR tilts measured by polarity positions.}
\label{fig:trends_tilts}
\end{figure}

\begin{figure}[!bht]
    \includegraphics[width=0.49\textwidth]{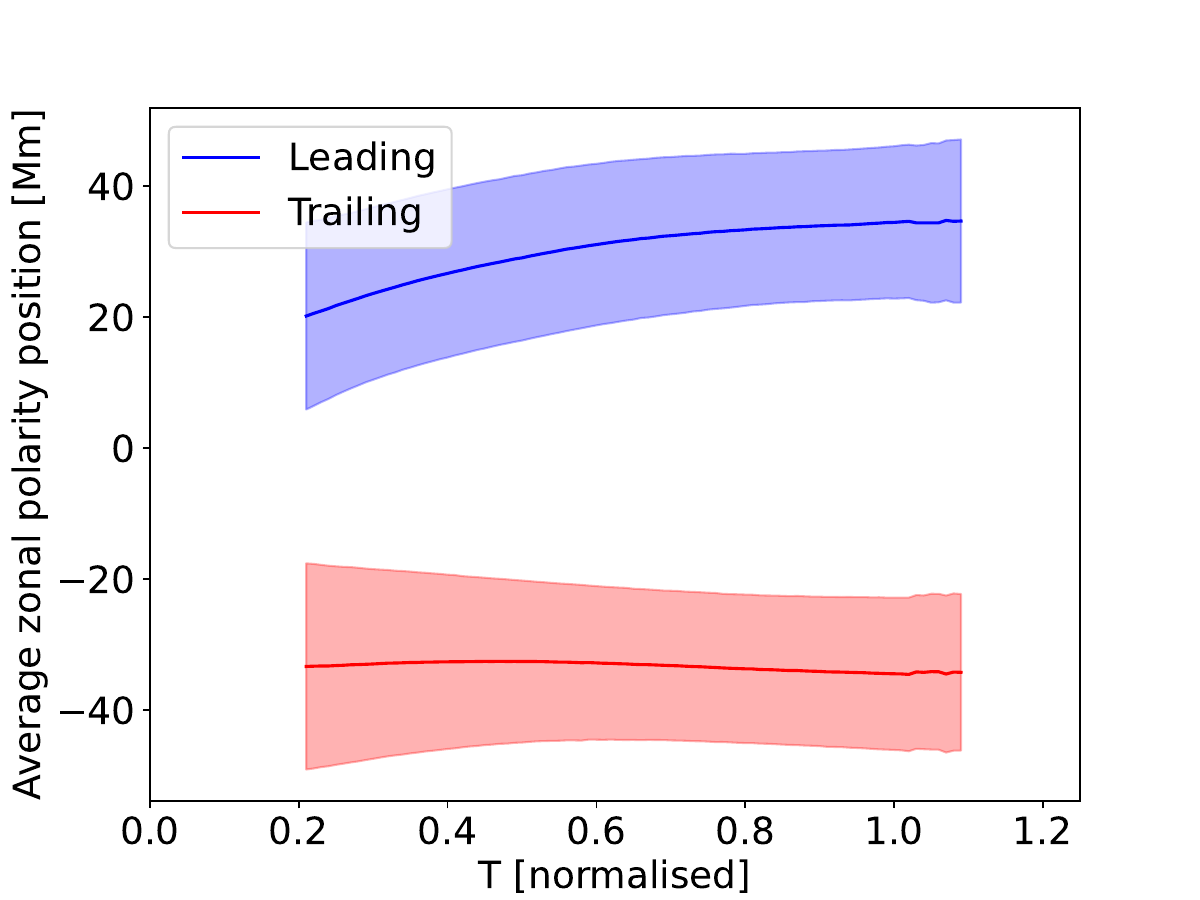}\\ 
    \includegraphics[width=0.49\textwidth]{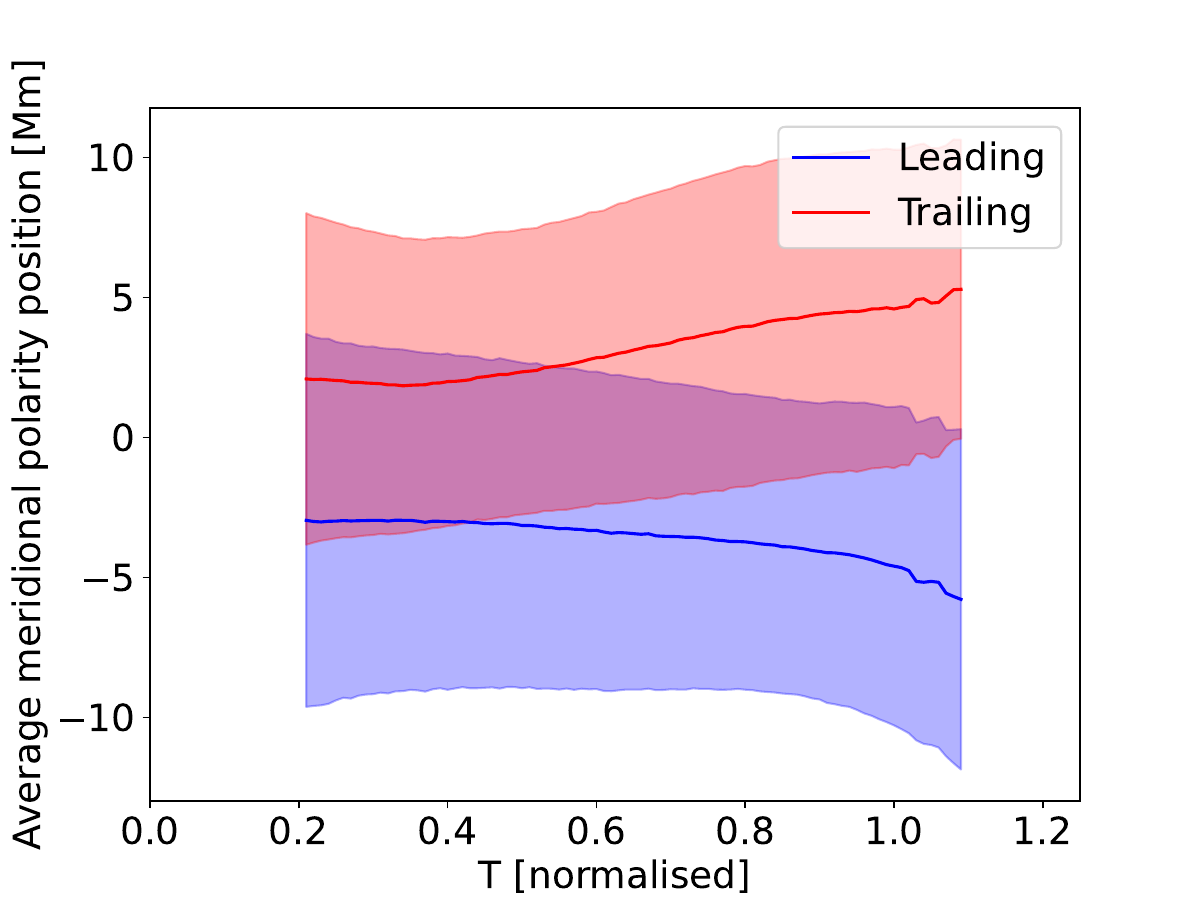} 
 \caption{Similar to Fig.~\ref{fig:trends_fluxes} but only for the positions of the polarity barycentres in the zonal direction (upper) and in the meridional direction (lower). The drifts are measured with respect to the middle point between the polarity positions at $T=1$.}
\label{fig:trends_drifts}
\end{figure}

The construction of the average AR as a normalised ensemble average also allows for investigation of the average properties of the individual ARs by means of averaging these physical quantities. In other words, in the following, we do not discuss the properties of the average AR but the average properties of individual ARs. 

For instance, we may investigate the average evolution of the magnetic fluxes in the ARs' polarities. This is displayed in Fig.~\ref{fig:trends_fluxes}. At the beginning, the magnetic flux in the leading polarity grows faster. Later on, however, there is almost a perfect balance of the flux between these two polarities. After reaching the maximum, the flux in the trailing polarity seems to disperse faster, which is also well known from literature. Here we need to remind that before $T=-0.1$ and after $T=1.1$ the coverage over the sample of ARs drops, so the results are less trustworthy than in the times in between.  

The average AR tilts are positive (Fig.~\ref{fig:trends_tilts}), meaning that on average, the leading polarity is closer to the equator than the trailing polarity, as expected. The spread of the average is quite large, and there does not seem to be a significant evolutionary trend throughout the growing phase of the AR. The tilts are further discussed in Appendix \ref{app:tilts}. 

\begin{figure}[!tb]
    \includegraphics[width=0.49\textwidth]{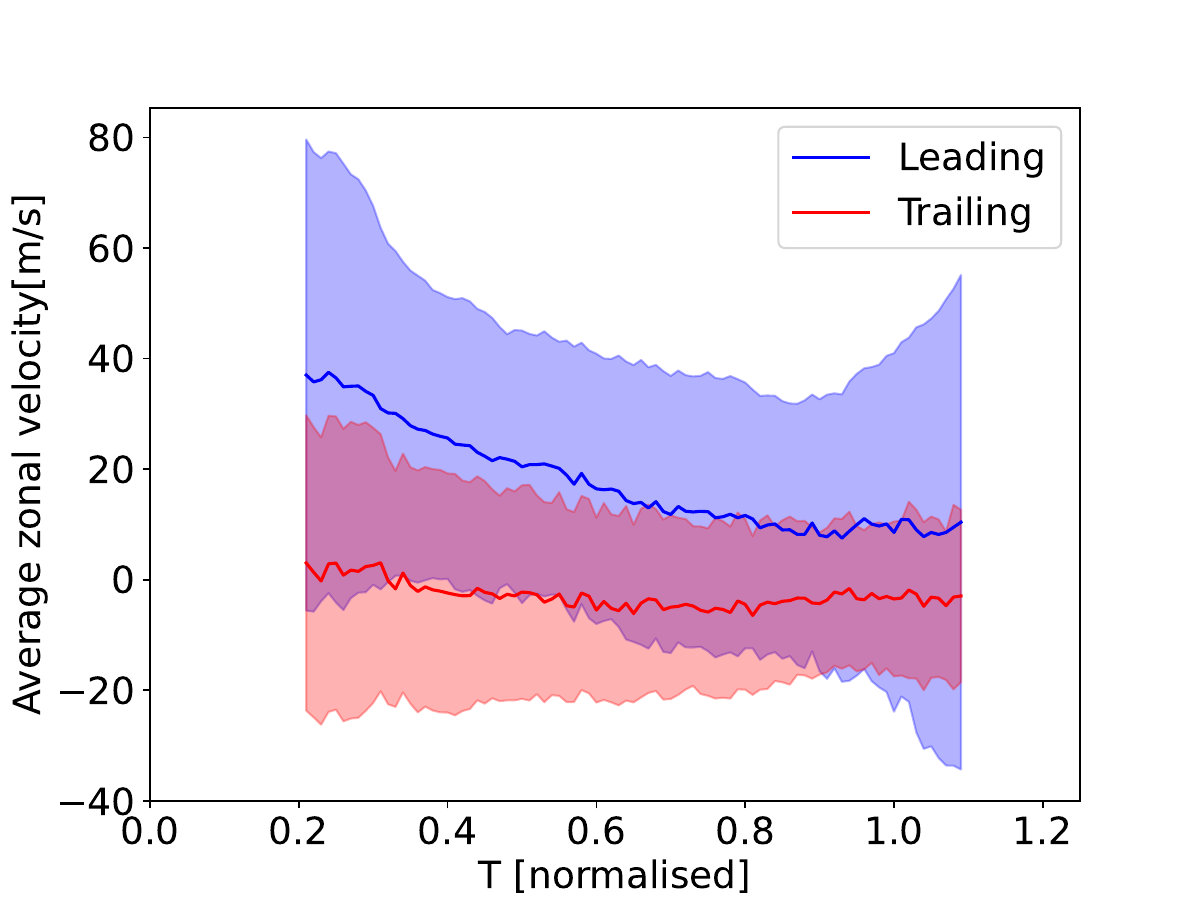}\\ 
    \includegraphics[width=0.49\textwidth]{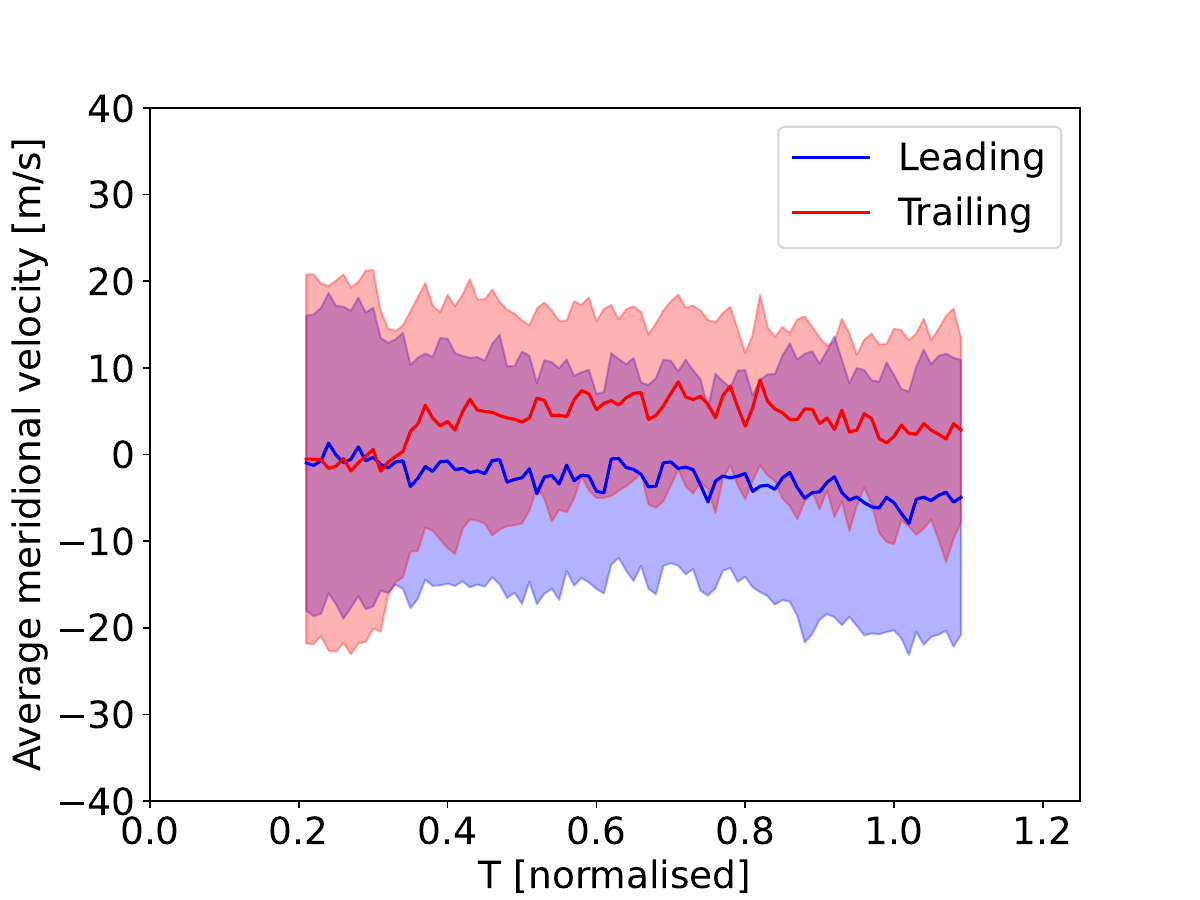} 
 \caption{Similar to Fig.~\ref{fig:trends_fluxes} but only for the average velocity in the zonal direction (upper) and in the meridional direction (bottom) with respect to the Carrington coordinate system. }
\label{fig:trends_velocities}
\end{figure}

\cite{2016A&A...595A.107S} reported that while there is an apparent asymmetry of the leading and following polarity motion of the AR with respect to the Carrington rotation rate, this zonal motion is actually symmetric with respect to the local differential rotation speed. Our average results are apparently not consistent with the conclusions by \cite{2016A&A...595A.107S}. On average, it seems that the proper motion of the trailing polarity with respect to the Carrington rotation is very slow and remains almost constant throughout the evolution of AR, whereas the leading polarity moves on average about 40~m/s faster than the Carrington rotation and the gradually slows down to about 10~m/s at $T=1.0$. This is consistent with a long-term increase of the average tilt. This asymmetry is inherently resulting from the construction of the average. Individual ARs may reside in plasma rotating faster or slower than the Carrington reference frame. By aligning around a fixed central point without accounting for these individual rotation differences the asymmetry becomes embedded in the averaged result. 

\begin{figure}[!t]
    \includegraphics[width=0.49\textwidth]{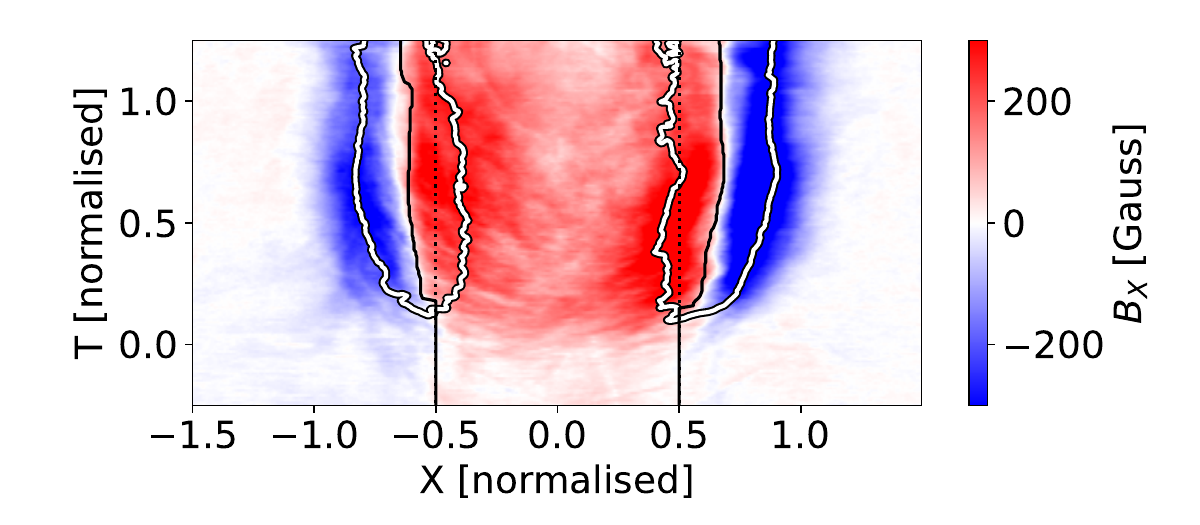}\\ 
    \includegraphics[width=0.49\textwidth]{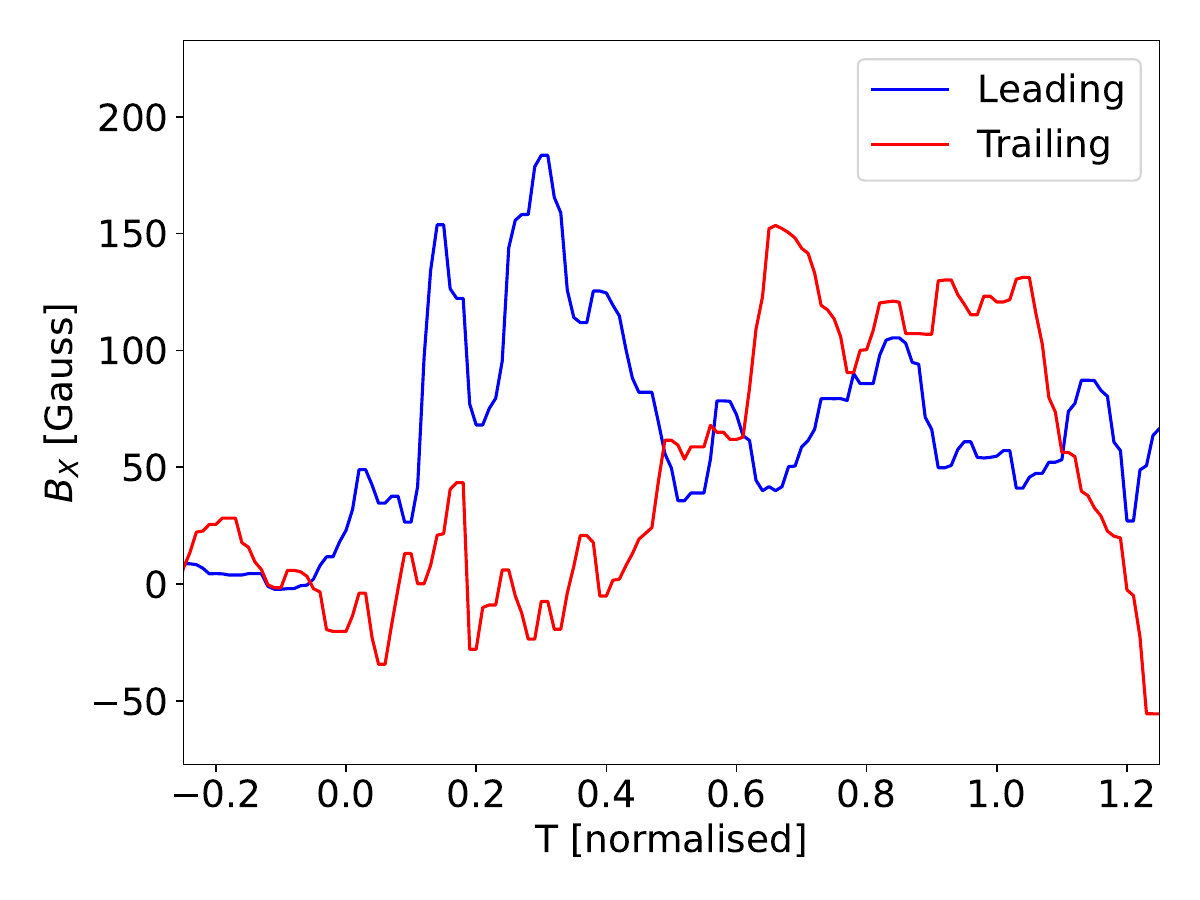} 
 \caption{Distance--time diagrams of the zonal component of the magnetic field ($B_X$) around the line connecting the polarities (upper panel). The black solid lines indicate the positions of the strongest vertical field, and the dotted line represents the position of the polarity barycentres. The outlined lines represent the position of $\pm 500$~G vertical field iso-contour. In the bottom panel, we plot the time evolution of the same quantity evaluated along the solid black lines. A corresponding movie is available online (Bvec\_X.mp4) from \url{https://zenodo.org/records/15656676}.}
\label{fig:Bx}
\end{figure}

Representative trends may also be obtained for the speeds measured in the Carrington rotation system (Fig.~\ref{fig:trends_velocities}). In general, the motions in both directions gradually slow down with time. At around $T=0.9$ it seems that the velocities saturate for both zonal and meridional motions, after that both remain constant until the end of the investigated interval. The zonal velocity difference between the motion of the leading polarity and the trailing polarity converges to about 15~m/s. Given that the meridional position difference between the leading and the trailing polarity at that time is less than 10~Mm (Fig.~\ref{fig:trends_drifts} bottom), the shear due to the differential rotation following the \cite{Snodgrass1983} rate is 12~m/s for the latitude of 10~degrees and 25 m/s for the latitude of 20~degrees. Therefore, the increasing east-west separation at normalised times around $T=1$ may be explained by the action of the differential rotation alone. The velocity difference between the polarities in the meridional direction saturates at about 10~m/s and contributes to the continuous tilting of the ARs on average. We note that the corresponding panels in Fig.~\ref{fig:trends_drifts} and Fig.~\ref{fig:trends_velocities} cannot be directly compared, because the real time axis involved in the speed calculation is different for each and every AR in the sample. As a consequence, the average of the individual velocities does not correspond to the time derivative of the average velocity with respect to the normalised time. 

\begin{figure}[!t]
    \includegraphics[width=0.49\textwidth]{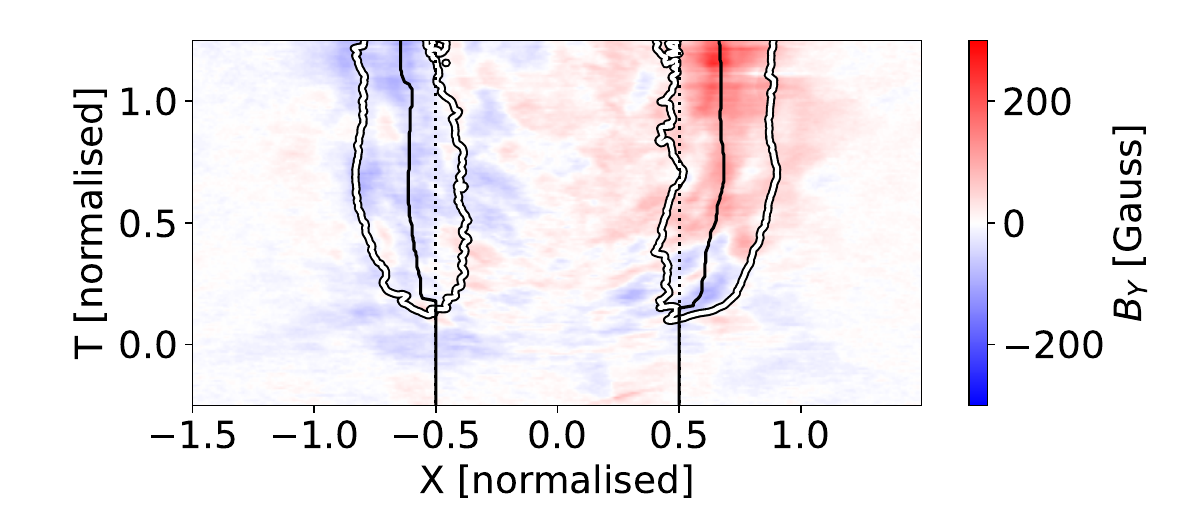}\\ 
    \includegraphics[width=0.49\textwidth]{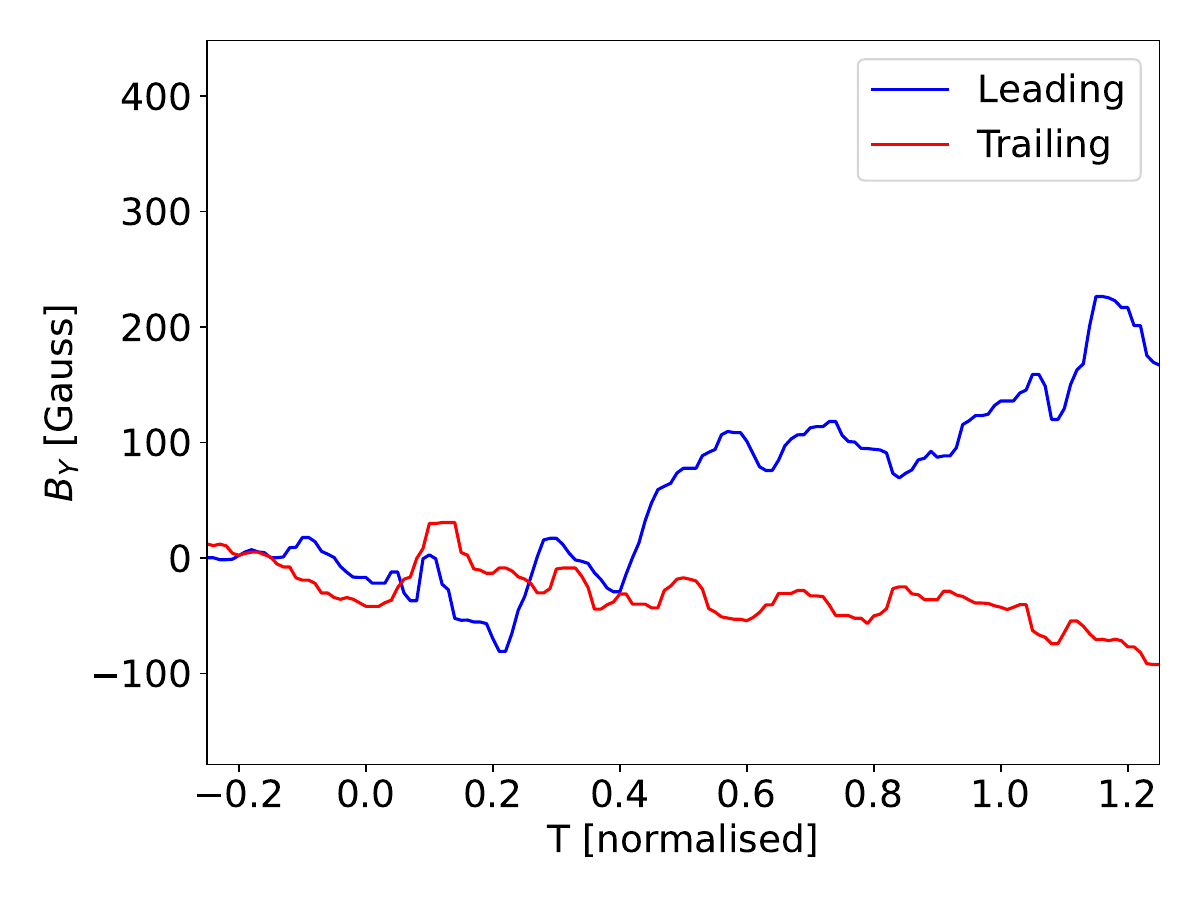} 
 \caption{Same plot as in Fig.~\ref{fig:Bx} but only for the meridional component of the magnetic field ($B_Y$). A corresponding movie is available online (Bvec\_Y.mp4) from \url{https://zenodo.org/records/15656676}.}
\label{fig:By}
\end{figure}

Last but not least, we also have a full magnetic vector at the photosphere resulting from the Milne-Eddington inversion of the polarimetric measurements at hand. The existence of such a data product warrants its use for also studying the magnetic field in the higher atmosphere by means of extrapolation, which we leave for a follow-up study. At the moment, we focus solely on its description at the photospheric level. 

The overall appearance (Fig.~\ref{fig:Bvec}) indicates the typical magnetic field diverging in the positive polarity and converging back in the negative polarity. The horizontal field in between the polarities is also clearly visible. 

\begin{figure}[!t]
    \includegraphics[width=0.49\textwidth]{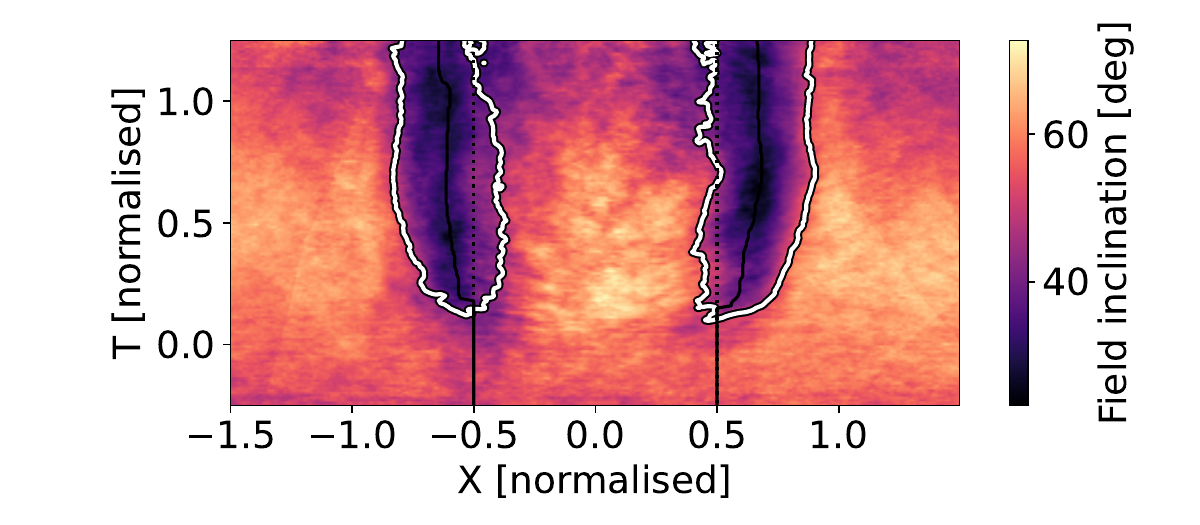}\\ 
    \includegraphics[width=0.49\textwidth]{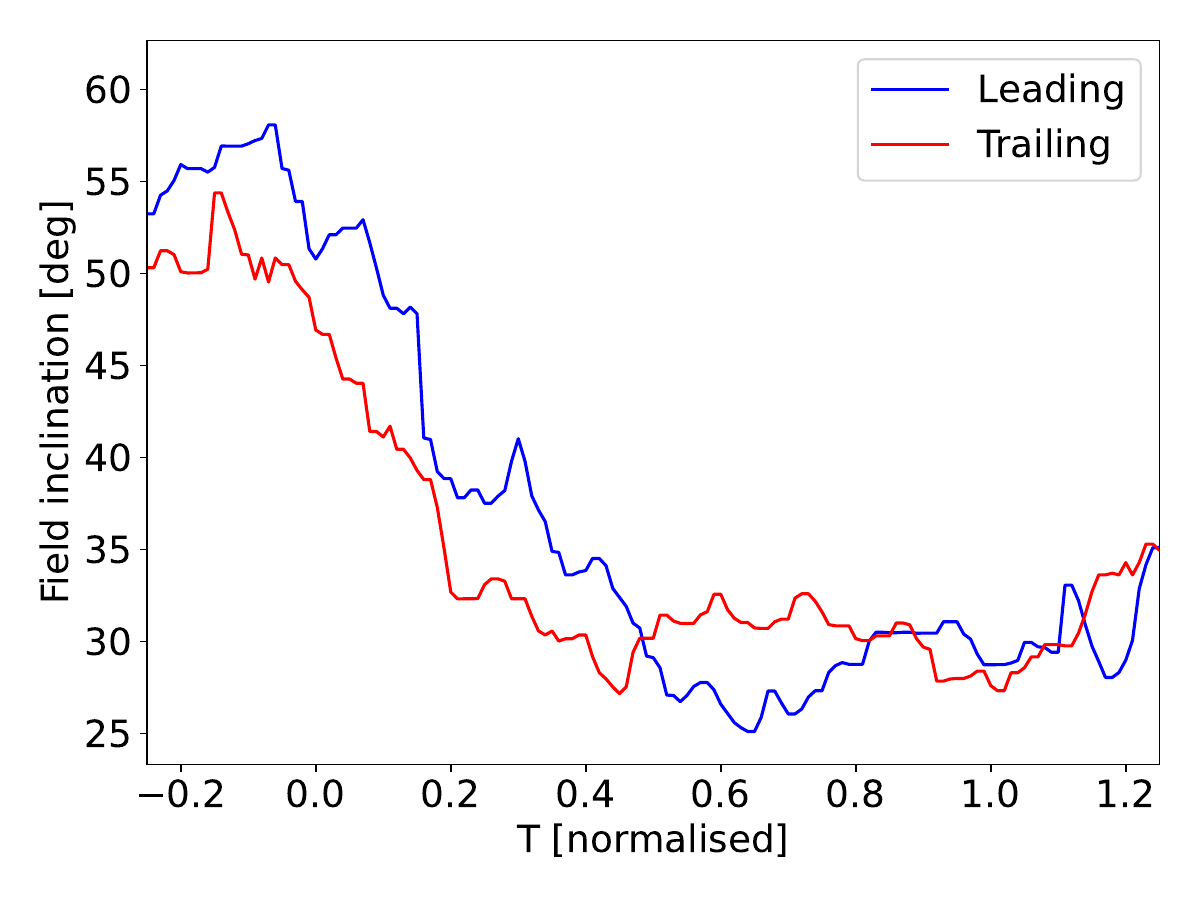} 
 \caption{Same plot as in Fig.~\ref{fig:Bx} but only for the inclination of the magnetic field. The average inclinations were obtained as an average of the normalised inclinations computed for individual ARs. A corresponding movie is available online (Bvec\_inc.mp4) from \url{https://zenodo.org/records/15656676}.}
\label{fig:Binc}
\end{figure}

\begin{figure}[!t]
    \includegraphics[width=0.49\textwidth]{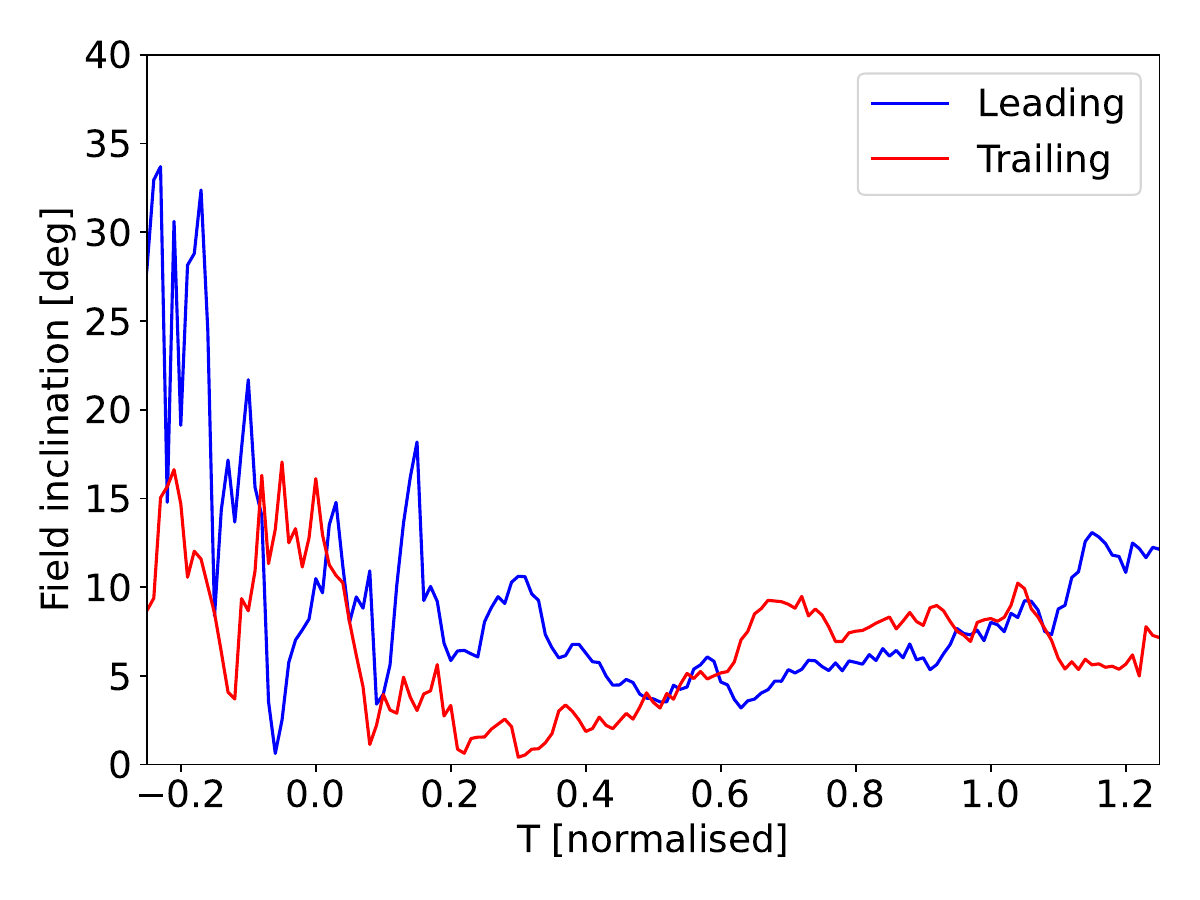} 
 \caption{Same plot as in the lower panel of Fig.~\ref{fig:Binc} but only for the inclinations computed from the averaged vector magnetic field. }
\label{fig:Binc_from_average}
\end{figure}

We investigated the overall behaviour of the vector magnetic field components along the line connecting the two polarities. To obtain more robust results, we averaged the magnetic field properties over 10 pixels along the connecting line and displayed the results in Figs.~\ref{fig:Bx}--\ref{fig:Binc_from_average}. 

In Fig.~\ref{fig:Bx} we show the horizontal magnetic field component aligned with the AR axis, $B_X$. We call $B_X$ ($B_Y$) zonal (meridional) for ease of orientation as if the AR axis were parallel to the equator. It resembles a bipolar nature around the polarities with the polarity inversion occurring near the location of the strongest vertical magnetic component. At the location of the strongest vertical field, however, positive (westward) polarity prevails for most of the evolution in both polarities. The deviation is stronger for the leading polarity for $T<0.5$ period, then the spatial displacement between the $B_X$ inversion and the location of the strongest vertical-field pixel appears in the trailing polarity. This observation may be interpreted in such a way that the field is slightly slanted towards the west in the photosphere.

A different trend is seen in the meridional $B_Y$ component (Fig.~\ref{fig:By}), which shows a gradual increase (the field is becoming northward) in the leading polarity and a decrease (the field becomes more southward) in the trailing polarity. For $T<0.4$ both polarities depict a slightly negative (southward) orientation, as if the loop legs were pushed towards the equator with the field systematically deviating from the radial direction. Later on, the $B_Y$ increases in the AR poles in the opposite directions, which indicates an increasing writhe of the loops in the meridional direction with time. 

The overall field inclination (Fig.~\ref{fig:Binc}) evolves over time. The average inclination values were obtained by averaging normalised inclination maps for each AR in the sample. Due to ensemble averaging, only the trend or change in inclination should be analysed, not the absolute values. The stacking was based on polarity barycenters rather than the pixels with the strongest vertical field.

The average inclination around the strongest vertical field \textit{in the average} decreases over time, likely due to averaging effects and misalignment in the ensemble. As ARs evolve and grow, scale invariance dominates, causing the strongest vertical pixels in individual regions to align better, effectively reducing inclination spread. For normalised times $T>0.4$, the average inclination stabilises at approximately 30 degrees.

A similar analysis can be performed for the vector magnetic field in the averaged AR (Fig.~\ref{fig:Binc_from_average}). Around the strongest vertical pixel, the field initially emerges almost vertically. The oscillations for $T<0.2$ are mainly due to the low average field strength, making inclination calculations sensitive to small variations. For $T>0.4$, we observe a gradual increase in average inclination. Since the values are averaged over a $10\times 10$ pixel box around the strongest vertical pixel, they are also influenced by the spread within this region. This gradual increase is consistent with the slow transformation of the field from an almost vertical configuration to an opening funnel. A more detailed analysis of the vector magnetic field evolution is left for future work.

\section{Concluding remarks}
By utilising an ensemble averaging method, we constructed a time sequence capturing the evolution of an average bipolar AR in a handful of observables available from instruments aboard SDO involving a normalisation in both the spatial and temporal domain. Our results are consistent with a scenario of a rising and emerging flux tube, the shape of which is strongly influenced by the near-surface turbulence in the case of the individual ARs. In the constructed average, these particular appearances behave similar to random noise and are suppressed by the averaging. 

We note that our average was constructed from only 36 representatives, whereas there are very likely many more available in the archive of observations. In the course of working on the project, we noticed a gradual convergence to the presented state as more ARs were added to the sample. One could assume that if even more ARs were added to the sample, the results would improve. Technical limitations stopped us from using more, but we also observed that the convergence to a clean average slowed down with a larger number of ARs. 

The evolution of our average AR very much agrees with what has already been observed by large-scale studies of individual ARs \citep[see e.g.][]{vanDriel-Gesztelyi2015}. Our construction of the average AR observationally indicates that at least bipolar ARs with clearly separated polarities are scale invariant. Both idealised flux‐emergence dynamo models and detailed magnetoconvection simulations naturally produce spot‐forming magnetic structures whose size statistics are essentially power laws (i.e. spatially scale invariant).  In the classic picture \citep{1955ApJ...122..293P}, buoyant toroidal flux tubes rising through a highly stratified convection zone fragment under aerodynamic drag and convective buffeting.  Each break‐off event seeds daughter flux concentrations over a broad range of scales so that by the time the field reaches the photosphere, it appears as a hierarchy of $\Omega$- and $U$-loops of many sizes, with no single preferred length scale, which is exactly the condition for a power-law spot-size distribution \citep{2004A&A...426.1047A}. Advances in fully compressible radiative magnetohydrodynamic magnetoconvection simulations have confirmed and quantified this. \cite{2012ApJ...753L..13S} showed that turbulent downflows shred emerging flux into a filamentary network whose loop‐length distribution follows a near–power‐law over more than two decades in scale; when those loops coalesce at the surface, they naturally form sunspot umbrae and pores whose individual sizes inherit this scale‐invariance. Thus, both the fundamental buoyant‐flux‐tube paradigm and the state-of-the-art magnetoconvective modelling predict that sunspot groups should exhibit spatial scale invariance.

A matter for discussion may be the phases just before the flux emergence. For instance, we found an indication for the diverging outflow at the emergence site with an amplitude of about $\pm 60$~m/s already at $T=-0.05$, which corresponds to around 4~hours of real solar time. This finding is apparently not consistent with the simulation results of \cite{Hotta2020}, where no divergent outflows were detected even 3 hours before flux emergence. In agreement with \cite{2014ApJ...786...19B} or \cite{2023PASA...40...59A}, we observed hints of a pre-existing magnetic bipole field before a rapid flux emergence was detected. Thus, the indication for the outflow is present at times when the field already started to slowly emerge. These outflows only appeared after the averaging. No indications are visible in the Dopplergram series nor are they in the maps of horizontal divergences of individual ARs (see Fig.~\ref{fig:kolaz_preemergence_div} and compare to the first panel of Fig.~\ref{fig:LCT}). \cite{Birch2016}, using non-normalised averaging over 70 ARs, did not find conclusive indications for radial outflows 3 hours before rapid emergence, and their estimate for the radial outflow was $-8 \pm 50$~m/s (the quiet-Sun control regions indicated $-5 \pm 40$~m/s). However, we point out that our results are not directly comparable to \cite{Birch2016}, for instance because our definition of the emergence time is different. 

We believe that our methodology of construction of an average bipolar AR creates a new playground for studies of their typical behaviour. In follow-up studies, we would like to focus on related aspects occurring both in the atmosphere above such an average AR and below. 

\section*{Data availability}
The normalised datacubes of individual ARs and the final averaged datacubes are available within the Czech National Data Repository (\url{https://data.narodni-repozitar.cz/}) with a record ID v5fq7-mvc15 \citep{published_data}. The primary observations were retrieved from the Joint Science Operations Center (\url{http://jsoc.stanford.edu/}). Processing codes are available upon request from the corresponding author. The indicated movies are available from \url{https://zenodo.org/records/15656676}.

\begin{acknowledgements}
This research is supported by the Czech--German bilateral grant, funded by the Czech Science Foundation under the project 23-07633K and by the Deutsche Forschungsgemeinschaft under the project BE 5771/3-1 (eBer-23 13412). M\v{S} and JJ were further supported by the institutional support RVO:67985815. The authors also acknowledge the help of Tatiana V\'ybo\v{s}\v{t}okov\'a and Lucia Mravcov\'a, who performed a trial visual search for the active regions following particular requirements, including the bipoles. The sample used in this study built up on their initial work. We thank Alena Zemanov\'a and Jana Ka\v{s}parov\'a for their inputs regarding the discussion of the AIA observations. We also thank to Art Amezcua from Stanford University for a making it possible to download the Postel's mapped vector magnetograms transformed to local Cartesian system. The authors would like to thank the anonymous referee for their valuable comments and suggestions that greatly improved the quality of the paper.  
\end{acknowledgements}

\bibliographystyle{aa} 
\bibliography{aa54440-25}

\onecolumn
\appendix
\section{Investigated active regions}
\begin{table*}[!h]
\caption{Complete list of the ARs used in the construction of an average AR.}
\label{tab:noaas}
\centering
\begin{tabular}{l|ccccc}
\hline \hline \rule{0pt}{12pt}
NOAA & Time range & Longitude & Latitude & Emergence & Flux maximum \\
\hline
\rule{0pt}{12pt}11072 & 2010.05.19\,--\,2010.05.29 & $312$ & $-16$ & 2010.05.20 17:00 & 2010.05.23 16:00 \\
11076 & 2010.05.30\,--\,2010.06.07 & $193$ & $-20$ & 2010.05.31 09:12 & 2010.06.03 07:36 \\
11124 & 2010.11.09\,--\,2010.11.18 & $171$ & $14$ & 2010.11.10 10:00 & 2010.11.15 15:12 \\
11130 & 2010.11.26\,--\,2010.12.05 & $330$ & $13$ & 2010.11.27 17:24 & 2010.12.01 02:12 \\
11142 & 2010.12.29\,--\,2011.01.08 & $208$ & $-12$ & 2010.12.30 10:24 & 2011.01.04 07:48 \\
11157 & 2011.02.10\,--\,2011.02.18 & $65$ & $18$ & 2011.02.11 21:48 & 2011.02.14 00:00 \\
11199 & 2011.04.24\,--\,2011.05.01 & $189$ & $21$ & 2011.04.25 10:12 & 2011.04.29 09:00 \\
11204 & 2011.04.30\,--\,2011.05.11 & $48$ & $17$ & 2011.04.30 17:48 & 2011.05.06 16:36 \\
11208 & 2011.05.06\,--\,2011.05.16 & $321$ & $12$ & 2011.05.07 06:00 & 2011.05.11 10:00 \\
11214 & 2011.05.12\,--\,2011.05.21 & $270$ & $-25$ & 2011.05.13 10:00 & 2011.05.17 20:24 \\
11250 & 2011.07.08\,--\,2011.07.19 & $209$ & $-27$ & 2011.07.09 22:12 & 2011.07.14 06:00 \\
11294 & 2011.09.10\,--\,2011.09.20 & $103$ & $-17$ & 2011.09.10 10:36 & 2011.09.15 20:36 \\
11316 & 2011.10.10\,--\,2011.10.20 & $49$ & $-12$ & 2011.10.10 17:48 & 2011.10.14 03:48 \\
11327 & 2011.10.18\,--\,2011.10.26 & $336$ & $-21$ & 2011.10.19 12:24 & 2011.10.23 06:24 \\
11345 & 2011.11.07\,--\,2011.11.14 & $84$ & $-25$ & 2011.11.08 03:00 & 2011.11.10 19:24 \\
11640 & 2012.12.28\,--\,2013.01.07 & $320$ & $28$ & 2012.12.28 17:09 & 2013.01.04 15:03 \\
11670 & 2013.02.04\,--\,2013.02.16 & $161$ & $19$ & 2013.02.05 20:48 & 2013.02.10 09:36 \\
11805 & 2013.07.21\,--\,2013.07.31 & $128$ & $-7$ & 2013.07.24 04:48 & 2013.07.30 07:24 \\
12273 & 2015.01.24\,--\,2015.02.02 & $67$ & $-3$ & 2015.01.25 14:06 & 2015.01.27 15:42 \\
12353 & 2015.05.19\,--\,2015.05.28 & $341$ & $7$ & 2015.05.20 21:12 & 2015.05.24 14:48 \\
12414 & 2015.09.08\,--\,2015.09.17 & $317$ & $-10$ & 2015.09.08 09:00 & 2015.09.12 19:24 \\
12614 & 2016.11.26\,--\,2016.12.04 & $164$ & $5$ & 2016.11.26 09:12 & 2016.11.29 06:48 \\
12615 & 2016.11.26\,--\,2016.12.09 & $139$ & $-7$ & 2016.11.27 20:36 & 2016.12.04 10:48 \\
13214 & 2023.02.04\,--\,2023.02.14 & $206$ & $16$ & 2023.02.05 15:12 & 2023.02.12 08:36 \\
13315 & 2023.05.22\,--\,2023.06.01 & $235$ & $-17$ & 2023.05.22 19:24 & 2023.05.28 09:48 \\
13367 & 2023.07.06\,--\,2023.07.13 & $16$ & $10$ & 2023.07.06 16:12 & 2023.07.11 11:24 \\
13403 & 2023.08.11\,--\,2023.08.21 & $242$ & $26$ & 2023.08.11 19:36 & 2023.08.17 09:00 \\
13407 & 2023.08.12\,--\,2023.08.22 & $225$ & $-16$ & 2023.08.14 10:00 & 2023.08.17 04:36 \\
13486 & 2023.11.11\,--\,2023.11.17 & $161$ & $-9$ & 2023.11.11 19:48 & 2023.11.14 22:12 \\
13583 & 2024.02.09\,--\,2024.02.19 & $7$ & $9$ & 2024.02.10 07:12 & 2024.02.14 09:24 \\
13595 & 2024.02.24\,--\,2024.03.05 & $165$ & $20$ & 2024.02.24 12:48 & 2024.03.01 10:00 \\
13613 & 2024.03.14\,--\,2024.03.21 & $323$ & $-23$ & 2024.03.15 20:12 & 2024.03.17 18:00 \\
13641 & 2024.04.11\,--\,2024.04.20 & $272$ & $11$ & 2024.04.13 01:12 & 2024.04.15 09:00 \\
13662 & 2024.04.27\,--\,2024.05.06 & $60$ & $30$ & 2024.04.27 22:24 & 2024.05.01 22:12 \\
13703 & 2024.05.31\,--\,2024.06.10 & $326$ & $-7$ & 2024.05.31 20:12 & 2024.06.06 10:48 \\
13759 & 2024.07.15\,--\,2024.07.23 & $116$ & $-5$ & 2024.07.17 02:00 & 2024.07.19 20:12 \\
\hline
\end{tabular}
\tablefoot{We state the NOAA identifiers of each considered active region, time range covered by observations (both dates are at midnight TAI), Carrington longitude and latitude of the central pixel of the tracked region, and times of the detected emergence and the time when the total magnetic flux reached its maximum. All regions are mapped to Postel's coordinates with a pixel size of $0.0301$ heliographic degrees and each subframe has a size of $768\times 768$ pixels.}
\end{table*}

\begin{figure*}[!th]
    \includegraphics[width=\textwidth]{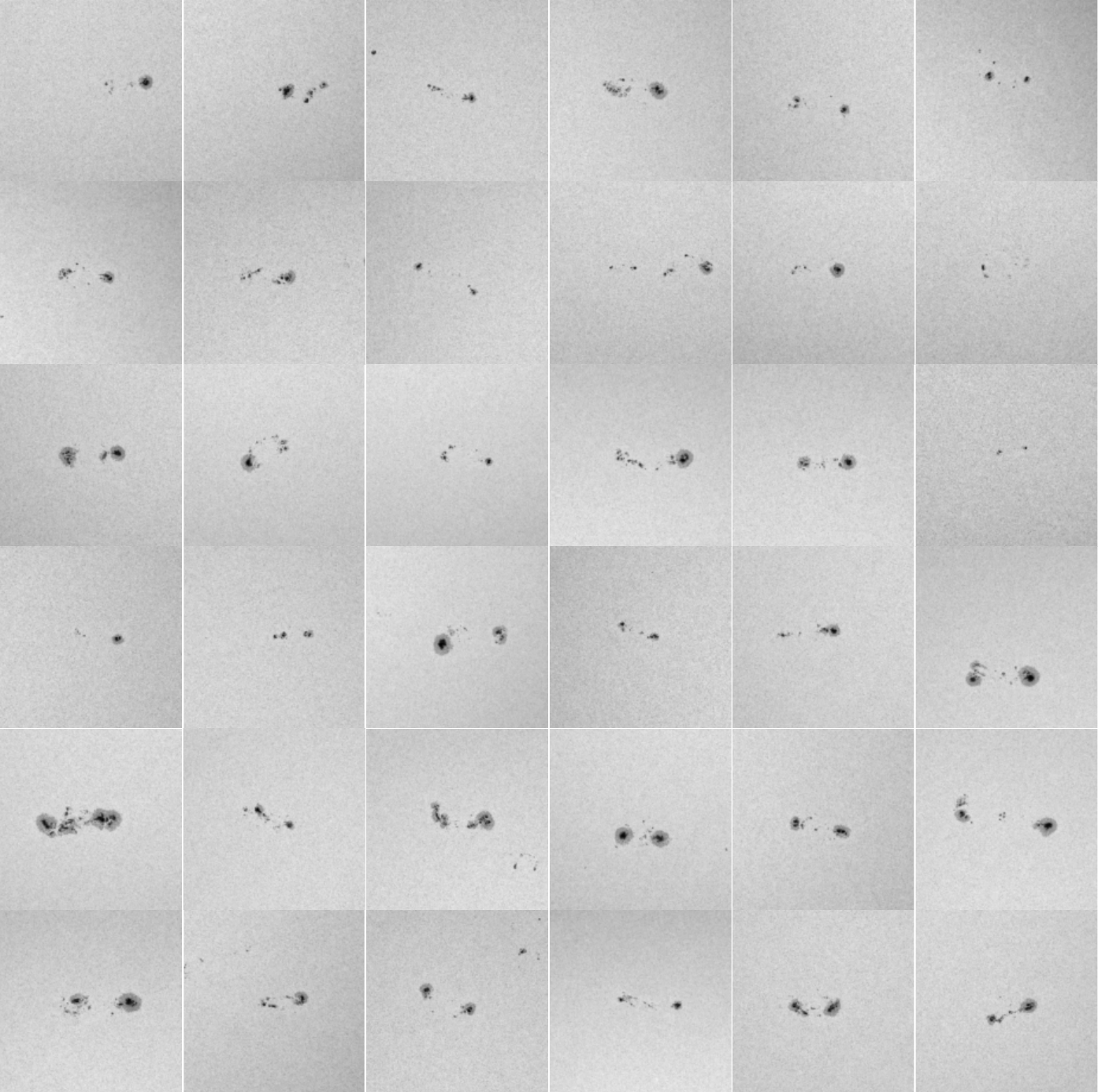}
    \caption{Considered ARs in their initial tracked frames before normalisation and averaging. Here we show intensitygrams before the removal of the limb darkening.}
    \label{fig:kolaz_int}
\end{figure*}

\begin{figure*}[!th]
    \includegraphics[width=\textwidth]{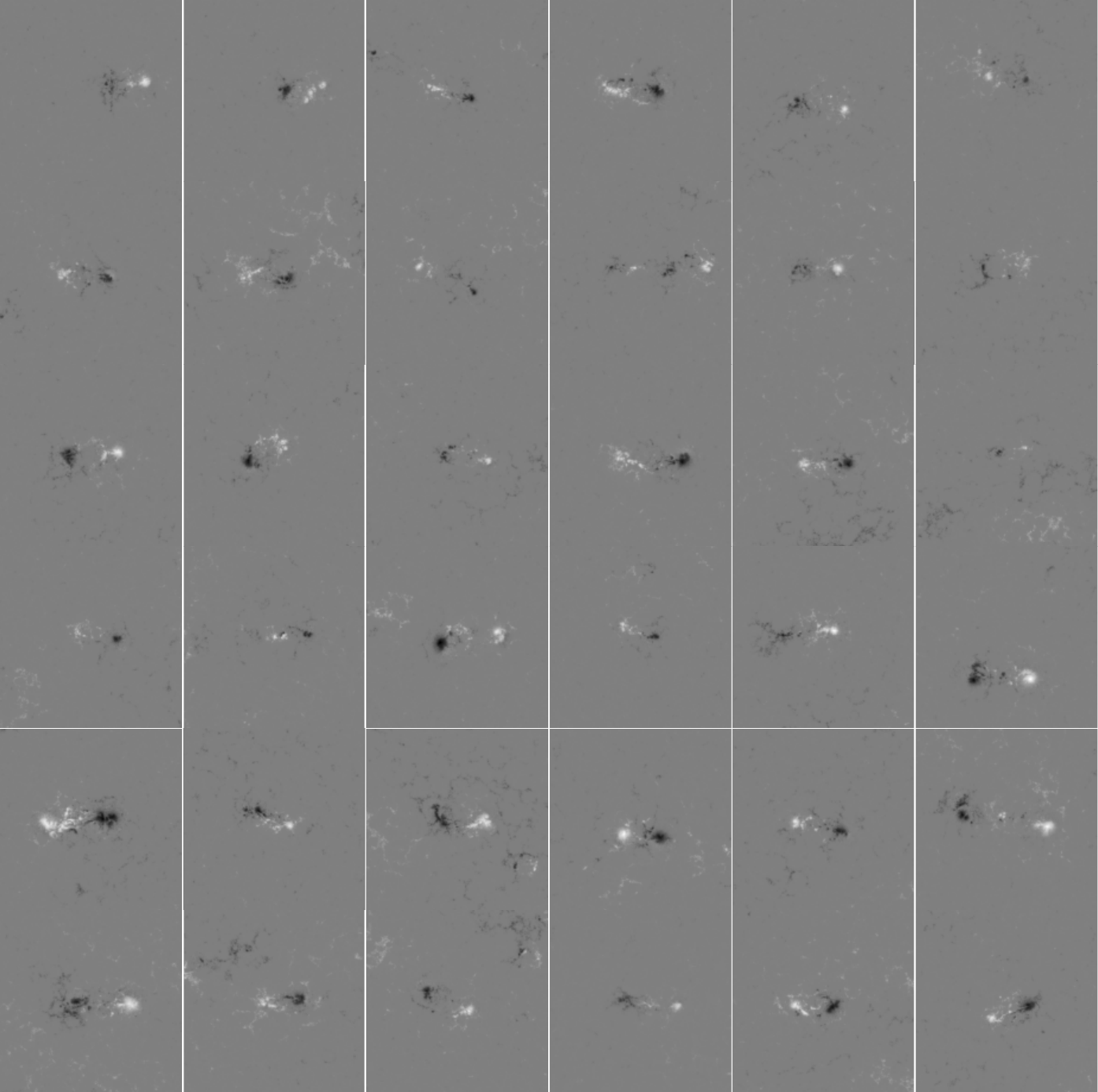}
    \caption{Similar to Fig.~\ref{fig:kolaz_int} but only with line-of-sight magnetograms, $B_\mathrm{LOS}$. All the subplots were saturated at 2000~Gauss. The snapshots represent the ARs appearance around the flux maximum. }
    \label{fig:kolaz_mag}
\end{figure*}

Example fields of view of the finally considered sample of ARs are given in Fig.~\ref{fig:kolaz_int} and~\ref{fig:kolaz_mag} in their original coordinate frames. 

\clearpage
\section{Pre-emergence bipoles}
\label{app:preemergence}

\begin{figure*}[!th]
    \includegraphics[width=\textwidth]{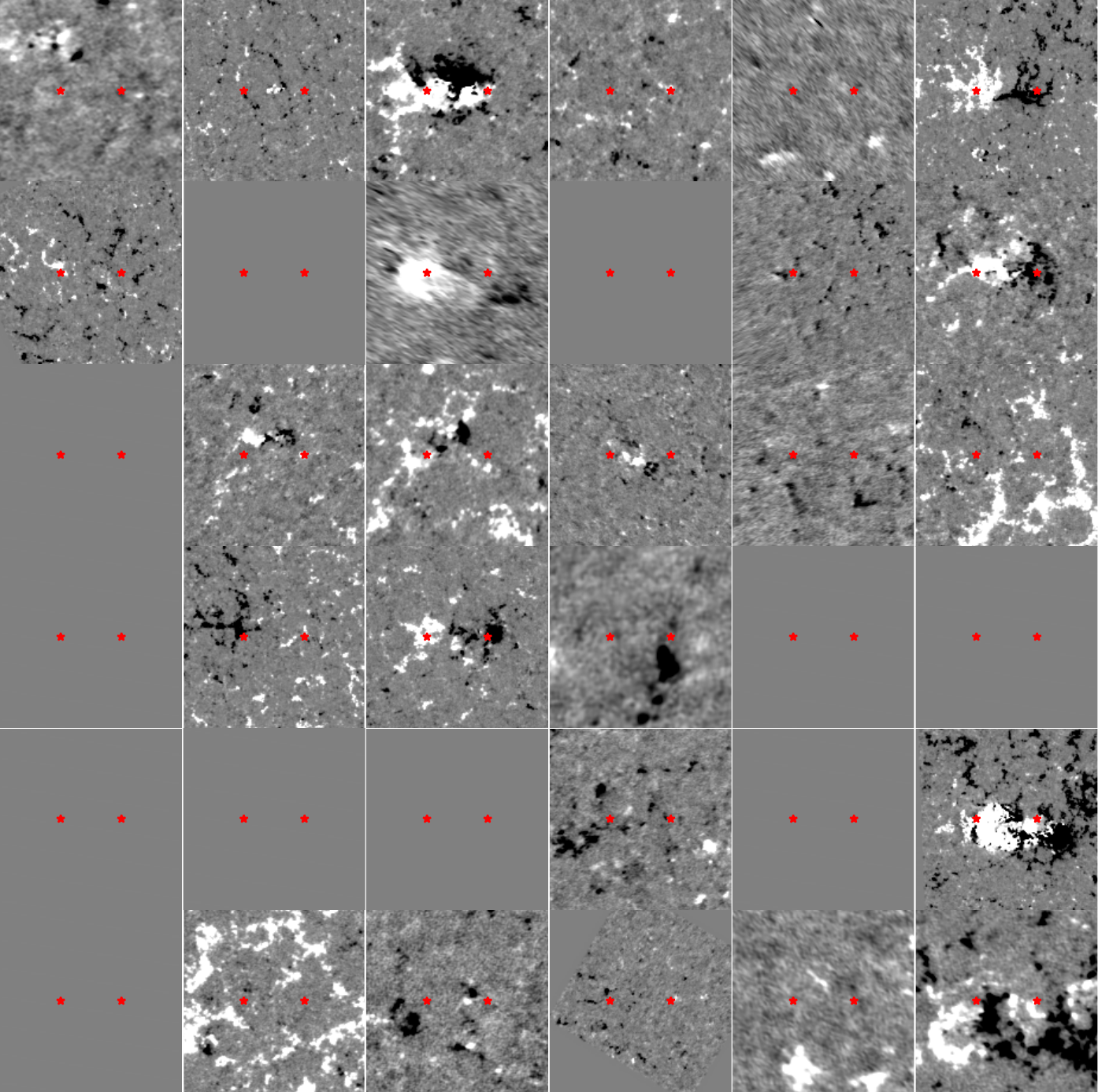}
    \caption{Line-of-sight magnetograms of the considered ARs before their emergence (at $T=-0.3$) in the normalised coordinates. Red stars indicate the determined positions of the polarities. We note that frames are not available for all of the ARs (some were too close to the limb to be considered). Also, the polarity positions were extrapolated from their positions after the emergence. All the subplots were saturated at $\pm$100~Gauss.}
    \label{fig:kolaz_preemergence}
\end{figure*}

\begin{figure*}[!th]
    \includegraphics[width=\textwidth]{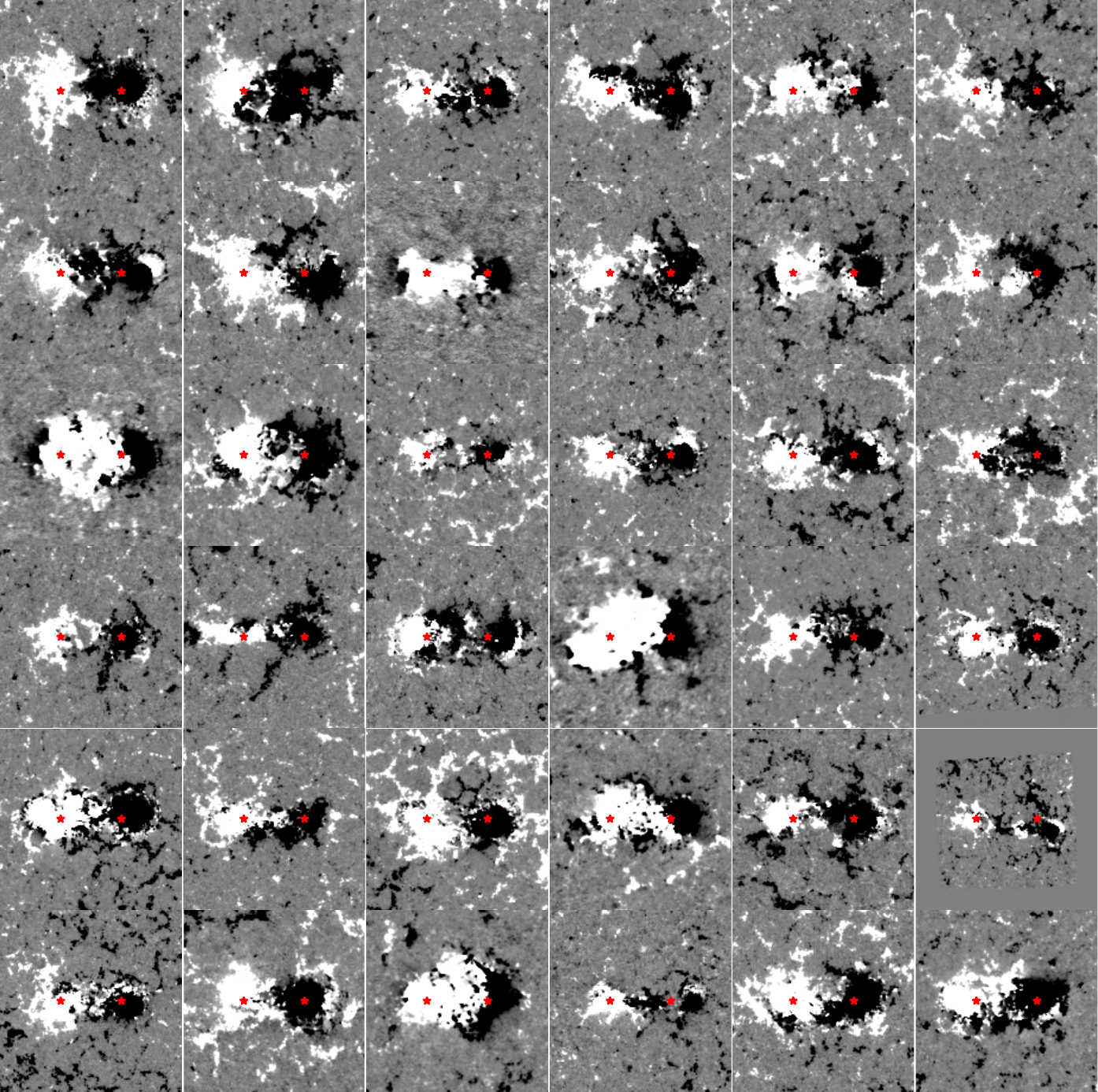}
    \caption{Same as Fig.~\ref{fig:kolaz_preemergence} but only shortly after emergence (at $T=0.05$). For most ARs at this stage, sunspots were not visible at both AR poles in the intensitygrams. Hence, in this case, polarity positions indicated by red stars were extrapolated from their positions determined for later stages of their evolution. }
    \label{fig:kolaz_postemergence}
\end{figure*}

Some of the studied physical quantities seem to be buried in the realisation noise and are revealed only after averaging. For instance, the magnetic bipole seems present only for a small fraction of AR before the evolution of the total magnetic flux registered a rapid emergence (see Fig.~\ref{fig:kolaz_preemergence}), whereas such a bipole is clearly present in all studied active regions after beginning of the emergence (Fig.~\ref{fig:kolaz_postemergence}). 

\begin{figure*}[!th]
    \includegraphics[width=\textwidth]{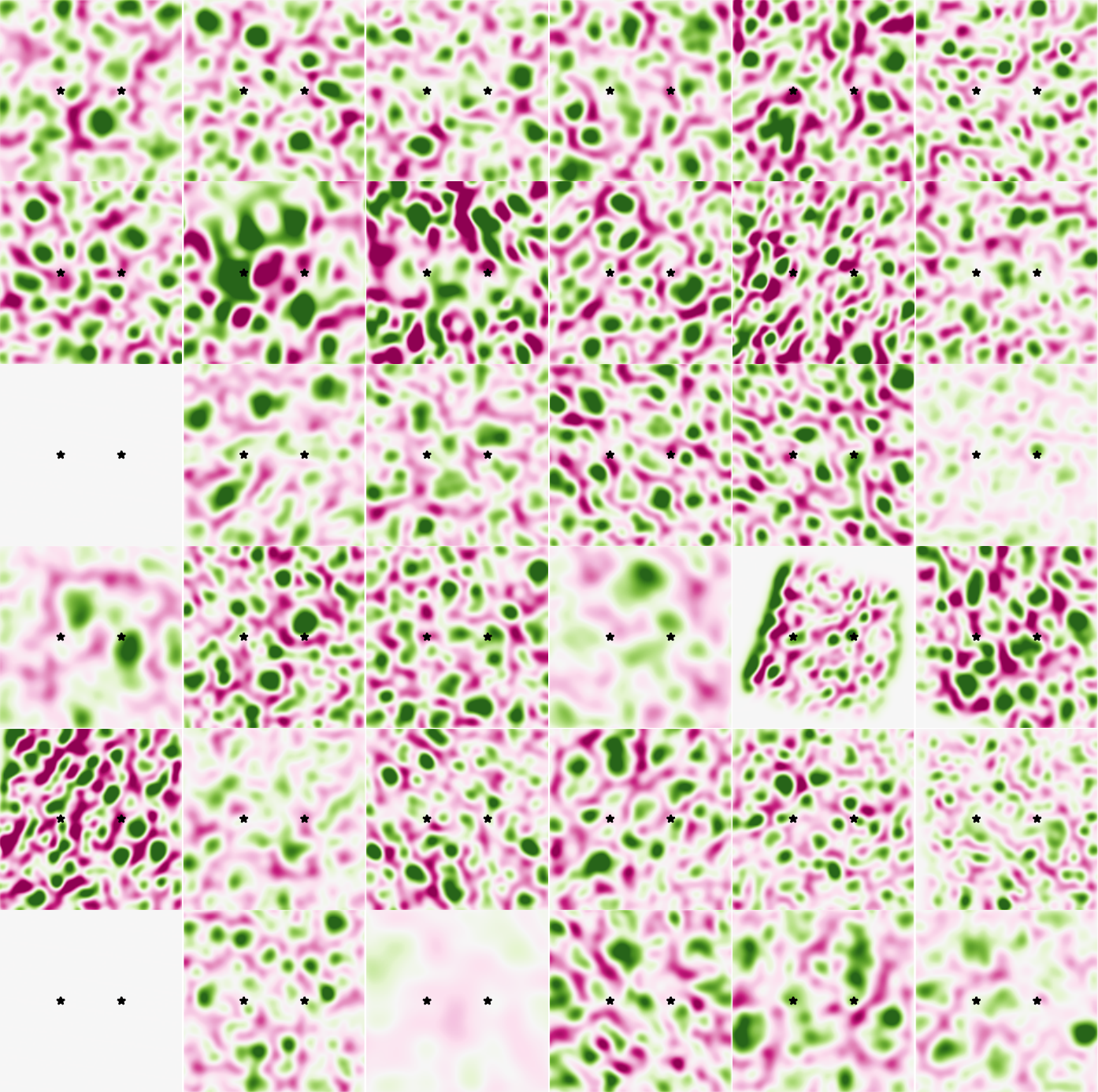}
    \caption{Horizontal flow divergences of the considered active regions before their emergence (at $T=-0.05$) in the normalised coordinates. Black stars indicate the determined positions of the polarities. We note that not for all ARs the frames are available (they were too close to the limb to be considered). All the subplots were saturated at $\pm$0.01, which roughly corresponds to $\pm0.036$~s$^{-1}$. Green colour indicates a positive horizontal divergence, purple the negative one (a convergence). Over the fields of view, the divergence pattern is dominated by supergranular flows. }
    \label{fig:kolaz_preemergence_div}
\end{figure*}

Similarly, there are no clear indications of a systematic flow divergence before beginning of the emergence in the maps of individual active regions (Fig.~\ref{fig:kolaz_preemergence_div}). 

\clearpage
\newpage\section{The tilts}
\label{app:tilts}
The tilt of the bipolar active region may be measured by several method. In the main text, we only discussed the tilt obtained from the polarity barycentres for the frames, where a sufficient amount of pixels with the vertical magnetic field larger than 500~G was present. Such measurements are not available for all frames in the series. 

\begin{figure}[!h]
\sidecaption
    \includegraphics[width=0.49\textwidth]{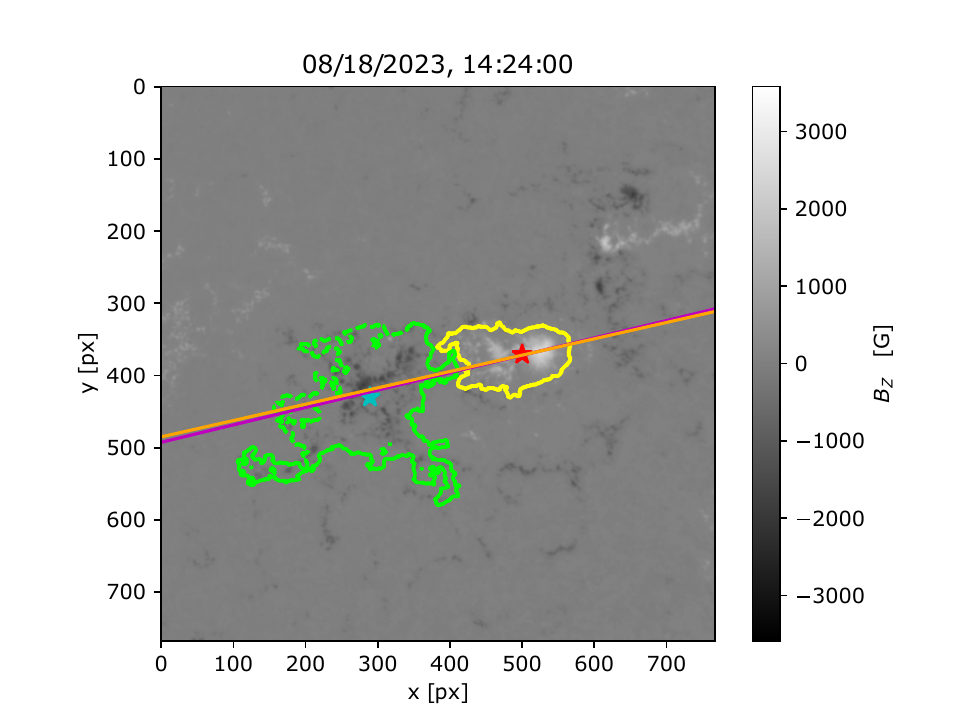} 
\caption{Demonstrative frame showing the performance of the automatic procedure to detect the magnetic polarities. The isolines outline compact regions of the positive (yellow) and negative (lime) polarity. The red star indicates the center of mass of the positive polarity within the outlined region, the blue star then the same for the negative polarity. The purple line then represents the linear fit through the unsigned magnetic field that represents the axis of the active region at the given time, the orange line has the same meaning, only the fit was obtained for vertical field stronger than 500~G.}
\label{fig:frame-tilts}
\end{figure}

From the frame magnetograms, where we applied masking of the leading and trailing polarity, we also measured the active region tilt alternatively by fitting the straight line through the magnetised pixels when each pixel was weighted by its absolute value of the vertical magnetic field. The tilts obtained by this method were available for every frame in the sequence. We repeated the same procedure of tilt determination by fitting when keeping only pixels with magnetic induction larger than 500~G. This way we separated the magnetic ``cores'' from the weaker dispersed flux. The comparison of the three methods is displayed in Fig.~\ref{fig:frame-tilts}.

\begin{figure}[!ht]
\sidecaption
    \includegraphics[width=0.49\textwidth]{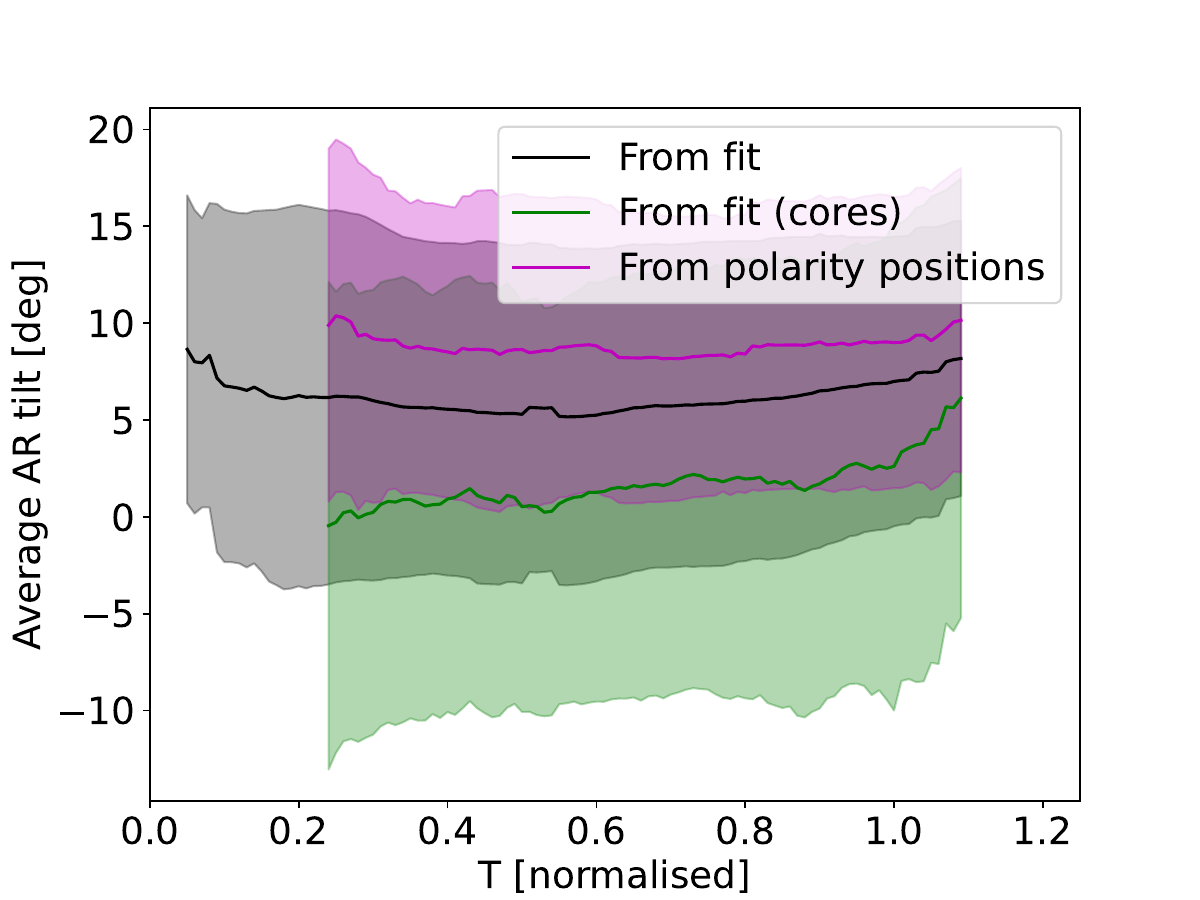} 
 \caption{Similar to Fig.~\ref{fig:trends_fluxes} but only for the AR tilts measured by three independent methods: Linear fit through the unsigned magnetic field (black), linear fit through the unsigned magnetic field of the cores (where $B_Z>500$\,G, green), and from polarity positions (purple).}
\label{fig:trends_all_tilts}
\end{figure}

The AR tilts (Fig.~\ref{fig:trends_all_tilts}) measured from fitting the straight line through magnetised pixels are systematically smaller than those calculated from the positions of the barycentres. This is caused by the asymmetry of the magnetic field in the meridional direction. There is usually a larger amount of the flux on the poleward side of the trailing polarity and the equatorial side of the leading polarity, an effect known as magnetic tongues \citep{2020ApJ...894..131P}. The presence of the tongues affects the gravity centre calculation. The strong field cores present even a smaller tilt systematically, which is again consistent with the effect of the meridional flux asymmetry. 

\end{document}